# Sr-Doped Molecular Hydrogen: Synthesis and Properties of $SrH_{22}$

Dmitrii V. Semenok,[1] Wuhao Chen,[2] Xiaoli Huang,[2,*] Di Zhou,[1] Ivan A. Kruglov,[3,4] Arslan B. Mazitov,[3,4] Michele Galasso,[1] Christian Tantardini,[1,7] Xavier Gonze,[1,5] Alexander G. Kvashnin,[1] Artem R. Oganov,[1,*] and Tian Cui[2,6,*]

[1] Skolkovo Institute of Science and Technology, Skolkovo Innovation Center, 30 Bolshoy Boulevard, Moscow 121205, Russia
[2] State Key Laboratory of Superhard Materials, College of Physics, Jilin University, Changchun 130012, China
[3] Dukhov Research Institute of Automatics (VNIIA), Moscow 127055, Russia
[4] Moscow Institute of Physics and Technology, 9 Institutsky Lane, Dolgoprudny 141700, Russia
[5] European Theoretical Spectroscopy Facility, Institute of Condensed Matter and Nanosciences, Université catholique de Louvain, Chemin des étoiles 8, bte L07.03.01, B-1348 Louvain-la-Neuve, Belgium
[6] School of Physical Science and Technology, Ningbo University, Ningbo 315211, China
[7] Institute of Solid State Chemistry and Mechanochemistry SB RAS, 630128, Russia

* Corresponding authors
E-mails: huangxiaoli@jlu.edu.cn (X.H.); a.oganov@skoltech.ru (A.R.O.); cuitian@jlu.edu.cn (T.C.)

## Abstract

Recently, several research groups announced reaching the point of metallization of hydrogen above 400 GPa. Following the mainstream of extensive investigations of compressed polyhydrides, in this work we demonstrate that small (4 atom %) doping of molecular hydrogen by strontium leads to a dramatic reduction in the metallization pressure to about 200 GPa. Studying the high-pressure chemistry of the Sr–H system at 56–180 GPa, we observed the formation of several previously unknown compounds: $C2/m$-$Sr_3H_{13}$, pseudocubic $SrH_6$, $SrH_9$ with cubic $F\bar{4}3m$ Sr sublattice, and pseudotetragonal $P1$-$SrH_{22}$, the metal hydride with the highest hydrogen content discovered so far. Unlike Ca and Y, strontium forms molecular semiconducting polyhydrides, whereas calcium and yttrium polyhydrides are high-$T_C$ superconductors with an atomic H sublattice. The latter phase, $SrH_{22}$ or $Sr_{0.04}H_{0.96}$, may be considered as a convenient model of the consistent bandgap closure and metallization of hydrogen. Using the impedance measurements in diamond anvil cells at 300–440 K, we estimated the direct bandgap of the $Pm\bar{3}n$-like compound $P1$-$SrH_6$ to be 0.44–0.51 eV at 150 GPa, and its metallization pressure to be 220 GPa. Together with the machine learning interatomic potentials, the impedance spectroscopy allowed us to estimate the diffusion coefficients of hydrogen $D_H$ = 1-2.8×$10^{-10}$ $m^2/s$ in $SrH_6$ and 1.2-2.1×$10^{-9}$ $m^2/s$ in $P1$-$SrH_{22}$ at 500–600 K.

## I. Introduction

Reaching the metallic state of pure molecular hydrogen by compression is one of the most spectacular challenges in high-pressure physics and chemistry. Studies of hydrogen metallization[1–3] have been in the focus since the 1990s, with the experimental metallization limit moving step by step from 150 GPa[4] to 430–500 GPa.[1,5] A consistent increase in pressure leads to a series of phase transitions in solid hydrogen (phases I–V[1,5]), gradual quenching of the Raman signals, and darkening of the sample down to a complete loss of transparency. Despite a significant progress in achieving ultrahigh pressures in the last five years, detecting superconductivity of metallic hydrogen remains an unsolved problem. Studies of the electrical conductivity at pressures above 400 GPa remain very challenging.

In 2004, N. Ashcroft suggested that the precompression effects from chemical bonding to other atoms may help to convert hydrogen to a metallic state. Fifteen years later, this idea was confirmed in the synthesis of many metallic and superconducting hydrides such as $H_3S$,[6,7] $LaH_{10}$,[8,9] $YH_6$[10,11] and $YH_9$,[11,12] $ThH_{10}$,[13] $CeH_9$,[14] $PrH_9$,[15] $NdH_9$,[16] and so forth. It is believed now that the superconducting properties of these compounds are due to the presence of a sublattice of metallic hydrogen, which is formed in pure hydrogen only at pressures of 500–700 GPa.

There must be an intermediate link between these two forms of hydrogen, metallic and superconducting, and a molecular dielectric or semiconducting phase. Moreover, this “bridge”, a molecular metal, has been found. In 2015–2017, it has been shown that lithium and sodium form molecular nonconducting hydrides $LiH_6$[17] and $NaH_7$.[18] Continuing these studies, we have recently synthesized a unique barium superhydride $BaH_{12}$[19] which is a molecular metal and a superconductor at moderate pressures of about 120–140 GPa. Pure hydrogen should demonstrate such properties at pressures above 350 GPa during the transition between the semiconducting and metallic modifications due to band overlap. Looking from a different perspective, $BaH_{12}$ is almost pure hydrogen doped by ~8 atom % of barium ($Ba_{0.08}H_{0.92}$). This kind of electron doping brings metallization closer and allows us (at very moderate pressures) to see what happens with pure hydrogen at more extreme conditions.

Lower strontium hydrides $SrH_{2-x}$ have been previously investigated.[20–22] Peterson and Colburn[20] have demonstrated that Sr can be dissolved in $SrH_2$ at high temperatures with formation of nonstoichiometric hydrides $SrH_{0.8-2}$. A detailed study of the $SrH_2$ behavior under pressure[21] has shown that strontium dihydride undergoes a series of phase transitions: *Pnma* (0–8 GPa)→ *P*6$_3$*/mmc* ($Ni_2In$ type, 8–115 GPa)→ *P6/mmm* ($AlB_2$-type, above 115 GPa). At higher pressures and excess hydrogen, the reaction of strontium with hydrogen proceeds further. Theoretical calculations[22] indicate stabilization of *I*4*/mmm*-$SrH_4$ (50–85 GPa) and *Cmcm*-$SrH_4$ (above 85 GPa), sequential formation of *C*2/*c*, *P*3 and $R\bar{3}m$ modifications of $SrH_6$, *Cmc*2$_1$-$SrH_8$ (50–75 GPa), and *P*2$_1$/*c*-$SrH_8$, and a series of decahydrides *P*2$_1$/*m*, *P*2/*c*, and $R\bar{3}m$-$SrH_{10}$ in the pressure range from 50 to 300 GPa. The research has demonstrated that a wide variety of polyhydrides can form in the Sr–H system at sufficiently low pressures, which is quite interesting and worth studying.

In this work, we investigated chemical reactions of strontium and strontium dihydride with hydrogen at high pressures (up to 181 GPa) and temperatures. In addition to a series of new molecular polyhydrides with compositions $Sr_3H_{13}$, $SrH_6$, and $SrH_{\sim 9}$, we discovered an amazing compound with a tetragonal strontium sublattice and the chemical formula $SrH_{22}$, which is hydrogen doped by 4 atom % of Sr. It is a yellow-colored semiconductor at 140 GPa. Together with $BaH_{12}$, this novel polyhydride, a successful model of hydrogen metallization via doping, does not require multi-megabar pressures and can be studied using a wide range of physical methods.

## II. Results

### *Structure prediction and synthesis of SrH$_{22}$*

In the first step of the research, we reexamined the Sr–H system at pressures of 50–300 GPa using the USPEX code.[23-25] From the chemical point of view, strontium has many of the characteristics of barium, which, as has been shown in our recent work,[19] reacts with hydrogen to form dodecahydride $BaH_{12}$. One of the goals of this research was to study the possibility of formation of similar strontium polyhydrides with the composition $SrH_{12}$ or higher.

An evolutionary search using USPEX for thermodynamically stable phases in the Sr–H system (Figure 1, Supporting Information Figures S1–S2) shows that this system is very rich in various compounds. At 150 GPa and 0 K (Figure 1f), there are several strontium hydrides on the convex hull: well-known *P*6/*mmm*-$SrH_2$; pseudocubic *P*1-$SrH_6$ (or $Sr_8H_{48}$) similar to recently discovered $Eu_8H_{46}$[26] and $Ba_8H_{46}$;[27] and superhydrides *C*2/*m*-$SrH_{10}$ and *P*1-$SrH_{17}$. At 100 GPa (Figure 1c), lower hydrides *Cmmm*-SrH and *Cmcm*-$SrH_4$ are stable, as well as $Sr_8H_{46}$, $Sr_8H_{48}$, and superhydride *P*1-$SrH_{22}$, whose structure is described below. Surprisingly, even at low pressures, there is wide variety of higher strontium polyhydrides (Supporting Information Figure S1), in agreement with previous studies.[28] At 50 GPa, $Sr_8H_{46}$ becomes unstable, whereas *C*2/*c*-$SrH_6$ and *C*2/*m*-$SrH_{12}$ appear on the convex hull. In addition, a large number of Sr polyhydrides are located at a small distance,

<30 meV, from the convex hull: $SrH_7$, $SrH_8$, $SrH_9$, $SrH_{10}$, $SrH_{16}$, $SrH_{20}$, and $SrH_{22}$. This indicates a high probability of discovering new compounds in the strontium–hydrogen system even at low pressures.

To synthesize the predicted strontium hydrides, we prepared several diamond anvil cells (DACs): Sr1–4, Sr50, Sr90, Sr165, and an electrical cell E1. We used diamond anvils with a 50–100 µm culet beveled to 250–300 µm at 8.5°. Electrical DACs were equipped with four Mo electrodes having a thickness of ~500 nm that were sputtered onto a diamond anvil. Gaskets consisting of a tungsten ring and a c-BN/epoxy mixture were used to isolate the electrical leads. Strontium (>99.9%) particles with a diameter of ~15–30 µm or $SrH_2$ powder (>99.9%) and sublimated ammonia borane (AB, >99.9%) were loaded into the gasket hole, with a thickness of 10–12 µm and a diameter of 35–60 µm, in an inert glove box. AB was used as a hydrogen source, following the technique that has shown excellent results in previous studies.[10–11,14–17] In DAC Sr3, the strontium particle was protected from reacting with AB by a thin sputtered layer of gold. Ammonia borane is a weak acid, therefore strontium may react with it upon contact during the DACs' loading. However, the salt $Sr(AB)_2$ forming on the surface of the Sr particle decomposes with emission of $H_2$ during laser heating and does not interfere with the synthesis of hydrides. A special test on DACs Sr4 and E1 confirmed a uniform pressure distribution in the sample area (Supporting Information Figure S30), with the accuracy of the pressure determination of ±6 GPa (4%). The heating of the samples above 1000 K at pressures of 70–170 GPa by a series of laser pulses with a duration of 0.1–0.5 s led to formation of strontium polyhydrides. The detailed description of the DACs and instrumental methods is presented in Supporting Information.

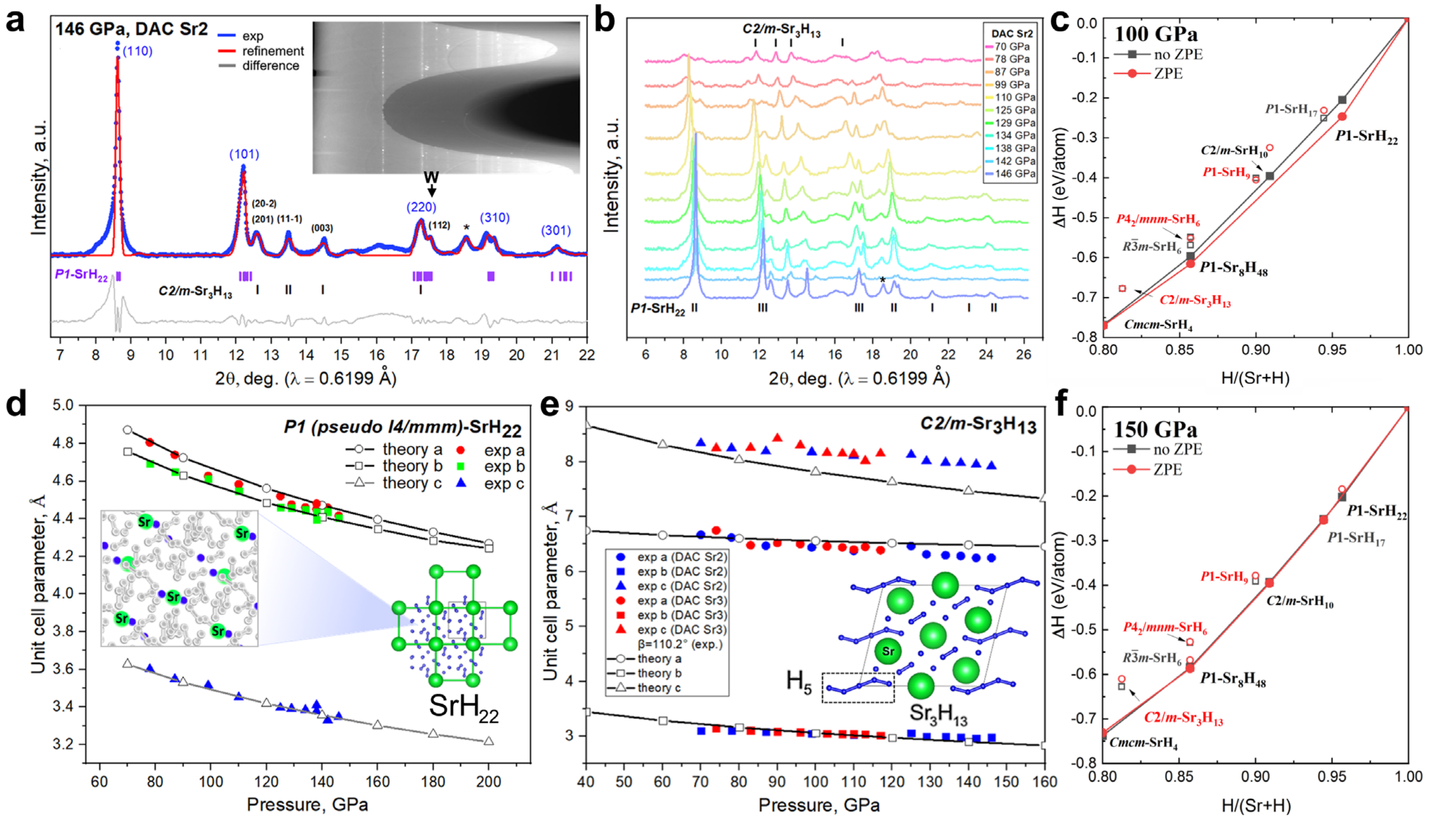

**Figure 1.** X-ray diffraction (XRD) study of strontium hydrides in DAC Sr2. (a) Experimental XRD pattern and the Le Bail refinements of the unit cell parameters of pseudotetragonal $P1$-$SrH_{22}$ and $C2/m$-$Sr_3H_{13}$ at 146 GPa. The experimental data, fit, and residues are shown in blue, red, and gray, respectively. Unidentified reflections are indicated by asterisks. The broadening of the (110) reflection is related to the proximity to the edge of WC seat. Inset shows the 2D diffraction image. (b) XRD patterns obtained during decompression of DAC Sr2 from 146 to 70 GPa. (c ,f) Convex hulls of the Sr–H system at 100 and 150 GPa calculated with and without the zero-point energy (ZPE) contribution. (d, e) Experimental and theoretical dependences of the unit cell parameters on the pressure for (d) $P1$-$SrH_{22}$ and (e) $C2/m$-$Sr_3H_{13}$. Insets: crystal structures of $SrH_{22}$ (blue circles indicate isolated H atoms) and $Sr_3H_{13}$.

We first investigated chemical reactions of strontium with hydrogen at pressures above 1 Mbar. The analysis of the diffraction patterns (Figure 1a,b) shows that a mixture of two strontium hydrides is formed after the laser heating of DAC Sr2 at 146 GPa, with a predominance of the phase having the *fcc* cubic set of reflections (110, 101, 220, 310, and 301, Figure 1a) and a very large unit cell volume. The comparison with USPEX calculations shows that the main phase can be immediately ascribed to pseudotetragonal *P*1-$SrH_{22}$, thermodynamically stable at 100–150 GPa, with a unit cell volume of ~65.1 $Å^3$/Sr at 100 GPa. The best candidate for the second phase, monoclinic *C*2/*m*-$Sr_3H_{13}$, with a unit cell volume of 25.26 $Å^3$/Sr at 100 GPa, was found 60 meV above the convex hull. For both phases, the predicted diffraction patterns, equation of state, and pressure dependence of the unit cell parameters are in very close agreement with the experimental data (Figure 1d,e). A similar situation was observed in DAC Sr1 loaded with $SrH_2$/AB, where an XRD pattern characteristic for $SrH_{22}$ was detected at 138 GPa with a minimum amount of impurities (Supporting Information Figure S24). For this reason, DAC Sr1 was used for subsequent optical and Raman measurements.

In *P*1-$SrH_{22}$, the H sublattice consists of $H_2$ molecules and $H^-$ anions. The minimum H–H bond length is 0.76 Å, Sr–Sr bond length is 3.49 Å at 100 GPa. Molecular polyhydrides are not something entirely new in the chemistry of hydrogen. In 2009, the formation of other molecular van der Waals polyhydride $Xe(H_2)_{7–8}$ was confirmed using single-crystal X-ray diffraction as well as IR and Raman spectroscopy in a Xe–$H_2$ mixture at about 5 GPa.[29] This discovery had been followed by an investigation of $(HI)(H_2)_{13}$ obtained as a small impurity in $HI(H_2)$ after laser heating of the $H_2 + I_2$ mixture above 25 GPa.[30] A significant difference of strontium hydrides from xenon and iodine (HI) polydydrides is a strong charge transfer and polarization of the Sr–H bonds. Indeed, Bader charge analysis performed in accordance with our previous experience[31,32] (Supporting Information Table S11) shows that the Sr atoms are a source of electrons for hydrogen. The charge of the strontium atoms in $SrH_{22}$ is $+1.23|e|$ at 120 GPa, whereas most of the $H_2$ molecules have a small negative charge ($\sim -0.1|e|$). About 10% of the H atoms are solitary anions $H^-$ with a charge of $-0.32|e|$ (Inset in Figure 1d, blue circles). Further theoretical study shows that anharmonic effects stabilize the structure of $SrH_{22}$ at 100 GPa (Supporting Information Figures S13). Calculations of the band structure point to a pronounced pseudogap of 1.5–1.9 eV in this material at 120 GPa. Increasing pressure to about 200 GPa leads to metallization of the molecular H sublattice and emergence of superconductivity $T_C$ about 21 K ($\mu^* = 0.1$, Supporting Information Figure S14).

Optical properties of *P*1-$SrH_{22}$ were studied in DAC Sr1 at 100–135 GPa. In transmitted light, this superhydride has a yellow or orange color at 100 GPa with the maximum transmission of ~630 nm wavelength. Increasing the pressure to 131 GPa leads to a significant darkening of the sample that corresponds to the gradual closure of the bandgap (Supporting Information Figures S33). The Raman spectra, measured using a 532 nm excitation laser, have the main peak at 4140 $cm^{-1}$ (123 GPa) whose pressure dependence $\nu(P)$ is very similar to the behavior of the molecular hydrogen vibron, and the intensity of this peak decreases with increasing pressure. The one-phonon resonant Raman calculations for *P*1-$SrH_{22}$ at 120 GPa (Supporting Information Figures S28) give the main signal at about 4154 $cm^{-1}$, in close agreement with experiment, whereas the nonresonant calculations predict several signals < 4100 $cm^{-1}$ which are not observed experimentally. Careful calculations show that all these Raman peaks have practically zero intensity.

Another strontium hydride, metastable $Sr_3H_{13}$, crystallizes in the monoclinic space group *C*2/*m*. A unique property of this compound is the presence of zigzag $H_5$ molecules with almost constant distance $d$(H–H) = 0.9–1.0 Å. Calculations of the band structure indicate that this material exhibits metallic properties and should have a superconducting transition temperature $T_C \sim 84$ K at 150 GPa ($\mu^* = 0.1$, Supporting Information Figure S9). A decrease in pressure in DAC Sr2 from 146 to 70 GPa demonstrates that monoclinic $Sr_3H_{13}$ remains stable in this pressure range, whereas $SrH_{22}$ superhydride decomposes below 100 GPa.

## *High-pressure synthesis of pseudocubic $SrH_6$*

At the next stage of the study, we reduced the pressure of synthesis and examined DAC Sr3 with 100 μm culet, loaded with ammonia borane (AB), gold foil (Au) and a strontium particle. Golden film was placed between AB and Sr to isolate them from each other and prevent the formation of $Sr(AB)_2$ salt. There was no evidence of the reaction between Sr and AB during the loading and compression. The sample was heated by a 1 μm IR fiber laser from the AB side at 123 GPa. After the reaction, the pressure decreased to 118 GPa. The reflectivity of the sample reduced without an obvious change in volume. After decompression to 48 GPa, the sample became transparent and demonstrated the Raman peaks at around 3635 and 735 $cm^{-1}$ (Supporting Information Figure S26).

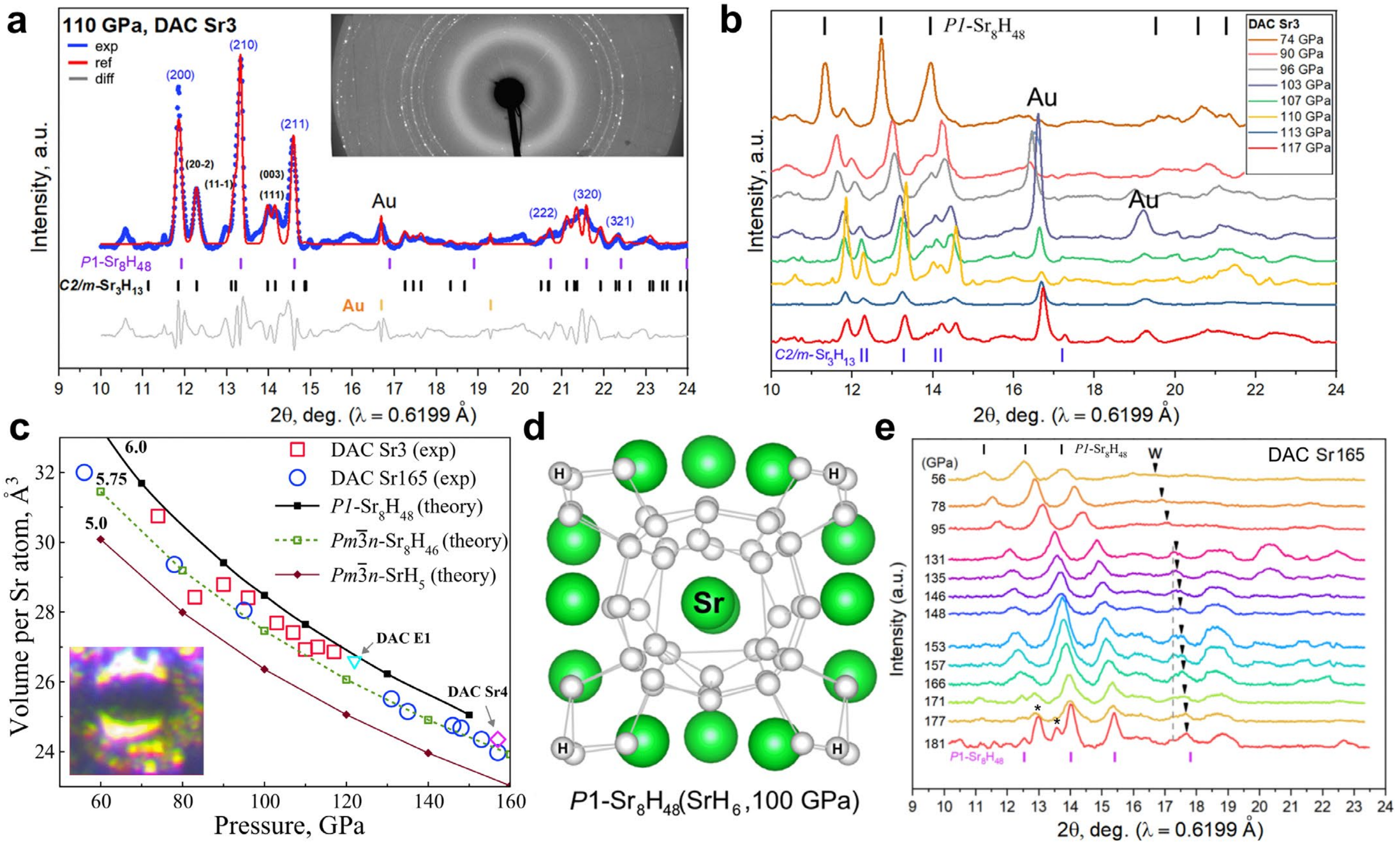

**Figure 2.** X-ray diffraction study of strontium hydrides in DACs Sr3 and Sr165. (a) Experimental XRD pattern and Le Bail refinements of the unit cell parameters of $Pm\bar{3}n$-like pseudocubic$P1$-$SrH_6$ and $C2/m$-$Sr_3H_{13}$ at 110 GPa. The experimental data, fit, and residues are shown in blue, red, and gray, respectively. Inset shows the 2D diffraction image. (b) XRD patterns measured during the decompression of DAC Sr3 from 117 to 74 GPa. During the XRD experiment, the sample was shifted, thus we can see various relative intensities of Au. (c) Experimental and theoretical pressure dependences of the unit cell volume for $P1$-$SrH_6$, $Pm\bar{3}n$-$Sr_8H_{46}$, and $SrH_5$. Inset: photo of the sample in DAC Sr165 at 160 GPa after the laser heating in transmitted and reflected light. (d) Crystal structure of $P1$-$SrH_6$ at 100 GPa. (e) XRD patterns obtained during the decompression of DAC Sr165 from 181 to 56 GPa. The signals at $2\theta > 18$ deg do not change with pressure and do not belong to the sample.

The observed diffraction pattern (Figure 2a,b) with main reflections (200), (210), and (211) corresponds to the $Pm\bar{3}n$ structure found earlier in the studies of europium ($Eu_8H_{46}$[26]) and barium ($Ba_8H_{46}$[27]) polyhydrides. However, both the harmonic and anharmonic calculations show dynamical instability at 300 K of structurally similar cubic strontium hydride $Sr_8H_{46}$, which distorts to $R\bar{3}c$ having a significantly different XRD pattern. For this reason, we performed an additional structural search for stable compounds with fixed Sr:H compositions of 2:12 and 8:48. As a result, we found thermodynamically and dynamically stable pseudocubic $P1$-$Sr_8H_{48}$, denoted as $P1$-$SrH_6$, whose Sr sublattice differs from that of $Pm\bar{3}n$-$Sr_8H_{46}$ by only a slight distortion and has an almost identical XRD spectrum. Slightly larger calculated cell volume of ~0.5 Å$^3$ (Figure 2c) can be explained by inaccuracy of the DFT methods or nonstoichiometric composition of the hydride (e.g., $Sr_8H_{47}$). The experiment shows that the obtained $Pm\bar{3}n$-like phase is stable to at least 74 GPa. In addition to $Sr_8H_{48}$, this sample probably also contains an admixture of the previously described $C2/m$-$Sr_3H_{13}$.

At 100 GPa, $Sr_8H_{48}$ crystallizes in the triclinic space group $P1$. The Sr–H bond length is 2.04–2.31 Å, the minimum H–H bond length is 0.84–0.87 Å. According to theoretical calculations, $P1$-$SrH_6$ is a narrow-bandgap semiconductor (Supporting Information Figures S16–S17) whose bandgap increases as pressure lowers. This may explain the fact that at pressures near or above 100 GPa the sample is opaque, whereas transparent regions and several Raman signals appear when pressure decreases (Supporting Information Figure S26). To study conductivity in $P1$-$SrH_6$, we made an electrical DAC E1 described further in the article.

The same pseudocubic $P1$-$SrH_6$ with a $Pm\bar{3}n$-like Sr sublattice was obtained in the experiment in high-pressure DAC Sr165 with a 50 µm culet, in which the loaded Sr/AB sample ($d$ ~ 15 µm) was heated by a laser to 1500–1800 K at 165 GPa. At this pressure, the synthesized compound is almost opaque, whereas below 56 GPa the sample became translucent (Supporting Information Figure S31). During the subsequent decompression from 181 to 56 GPa, a series of low-intensity XRD patterns was obtained (Figure 2e). The analysis of the XRD patterns points to pseudocubic $P1$-$SrH_6$ as the main component. The obtained hydride has a surprisingly high stability: the character of the XRD pattern does not change down to 56 GPa, in agreement with ab initio thermodynamic calculations (Supporting Information Figures S1–S2). In principle, this suggests that polyhydrides can maintain their structure during decompression, as has been recently shown for FeSe.[33]

## *Synthesis of $SrH_9$ below 1 Mbar*

To further investigate the stability of strontium superhydrides at low pressures, we loaded DAC Sr50 with Sr and ammonia borane and compressed it to 54 GPa. The opaque Sr particle showed metallic luster before being heated by a laser. After the laser heating at about 1000–1500 K, the pressure dropped to 50 GPa and the particle became heterogeneous and translucent, indicating an occurrence of a chemical reaction with generated hydrogen (Supporting Information Figure S31). The XRD study showed that the sample consists of two main components (Figure 3a,d). At about 100 GPa, the cubic phase is the main product, marked quite clearly, whereas the amount of impurity (second phase) is much smaller. Because Sr is a neighbor of Y, for simplicity this cubic set of reflections was preliminarily indexed using the $F\bar{4}3m$-$SrH_9$ structure common among metal polyhydrides (e.g., $PrH_9$[15] and $EuH_9$[26]). Despite $F\bar{4}3m$-$SrH_9$ ($a$ = 5.316 Å, $V$ = 37.55 Å$^3$ at 62 GPa) being dynamically unstable, its equation of state is in satisfactory agreement with the experimental data (Figure 3g). Molecular dynamics "annealing" of $F\bar{4}3m$-$SrH_9$ leads to a better candidate — $F\bar{4}3m$-like pseudocubic $P1$-$SrH_9$, or $Sr_4H_{36}$, with a molecular H sublattice which is almost stable within the harmonic approximation and lies near (25 meV/atom above) the convex hull at 100 GPa. The comparison of the calculated structure with experimental XRD patterns indicates that this solution is not ideal: the experimental structure is more symmetric and closer to $F\bar{4}3m$ than the predicted $P1$. Because of some uncertainty about the H content (Sr:H ratio is between 1:9 and 1:8, see also the results from DAC Sr90) and the structure of the compound, in this work the phase is denoted c-$SrH_{\sim 9}$.

The structure of the impurity phase cannot be determined unambiguously because of a superposition of reflections. One of the simplest possibilities is $Cmme$-$Sr_2H_3$ ($a$ = 5.62 Å, $b$ = 5.84 Å, $c$ = 5.03 Å, $V$ = 20.7 Å$^3$ at 62 GPa), which lies about 13 meV/atom above the convex hull at 100 GPa and is dynamically stable. During the compression from 62 to 93 GPa, the relative intensity of the XRD peaks of c-$SrH_{\sim 9}$ and $Cmme$-$Sr_2H_3$ changed greatly. This possibly indicates that c-$SrH_{\sim 9}$ is formed via a chemical reaction of the impurity phase $Sr_2H_3$ with excess hydrogen detected using the Raman spectroscopy (Supporting Information Figure S29). The sample, semitransparent at 62 GPa, darkens as pressure increases and becomes opaque above 100 GPa (Figure 3d). This indicates the semiconducting nature of the obtained molecular hydrides and their metallization in DAC Sr50 with increasing pressure. The experiment shows that hydrogen-rich polyhydrides such as c-$SrH_{\sim 9}$ can be synthesized at a relatively low pressure of about 50 GPa, which makes strontium a very attractive platform for the design of ternary hydride superconductors that are stable at pressures below 1 Mbar.

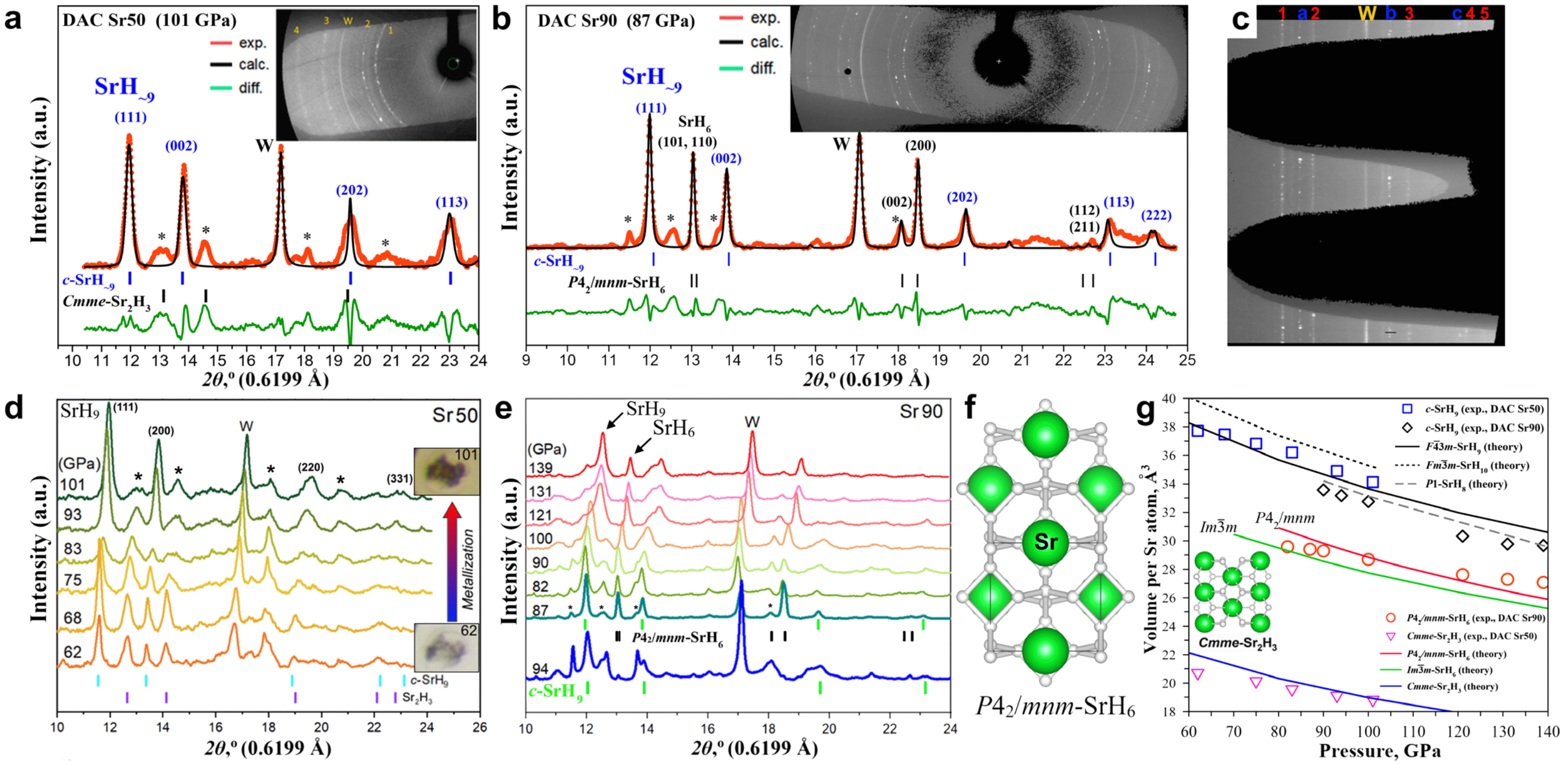

**Figure 3.** X-ray diffraction study of strontium hydrides in DACs Sr50 and Sr90. (a) Experimental XRD pattern and the Le Bail refinement of the unit cell parameters of cubic $SrH_9$ and *Cmme*-$Sr_2H_3$ at 101 GPa in DAC Sr50. Inset shows the 2D diffraction image. (b) Experimental XRD pattern and the Le Bail refinement of the unit cell parameters of cubic $SrH_9$ and $SrH_6$ at 87 GPa in DAC Sr90. The experimental data, fit, and residues are shown in red, black, and green, respectively. Unidentified reflections are indicated by asterisks. (c) Diffraction image ("cake") of the sample at 87 GPa, DAC Sr90. (d) XRD patterns obtained during the compression of DAC Sr50 from 63 to 101 GPa. (e) XRD patterns obtained during the compression of DAC Sr90 from 90–94 to 139 GPa. (f) Crystal structure of $P4_2/mnm$-$SrH_6$, which can also be considered as distorted $Im\bar{3}m$-$SrH_6$. (g) Experimental and theoretical pressure dependences of the unit cell volume of cubic $SrH_{\sim 9}$, $SrH_6$, and *Cmme*-$Sr_2H_3$.

In DAC Sr90, we discovered a compound similar to c-$SrH_{\sim 9}$ but with a lower H content. In this cell, strontium metal and ammonia borane were used for the synthesis at a slightly higher pressure of ~90 GPa. After laser heating, the pressure increased from 90 to 94 GPa and the sample obviously expanded (Supporting Information Figure S31). The obtained compound was almost opaque. There are no peaks from 500 $cm^{-1}$ to 4500 $cm^{-1}$ in the Raman spectrum of the sample at 94 GPa except for the molecular hydrogen vibron at 4200 $cm^{-1}$, which can correspond to both free and bound $H_2$ (Supporting Information Figure S29). The XRD data (Figure 3a,c) indicates the presence of two phases with cubic structures. The main cubic phase can be indexed using c-$SrH_9$ ($a$ = 5.141 Å, $V$ = 33.96 $Å^3$/Sr at 87 GPa) previously found in DAC Sr50, but with a 3% lower unit cell volume corresponding to the Sr:H ratio near 1:8.

To interpret the rest of the XRD reflections, we analyzed the diffraction rings at 87 GPa (Figure 3b,c). The reflections at 12.5° and 16° are diffuse, corresponding to a fine-crystalline phase of an unknown impurity (marked *), whereas cubic reflections "1–5" (c-$SrH_{\sim 9}$) and "a–c" (Figure 3c) correspond to coarse-crystalline phases and have a granular structure. The "a–c" series can be approximately indexed using the structure of recently discovered $Im\bar{3}m$-$CaH_6$[34] and $Im\bar{3}m$-$YH_6$[35] with the lattice parameter $a$ = 3.86 Å and $V$ = 28.73 $Å^3$/Sr at 87 GPa. The theoretical cell parameters for this phase at 90 GPa — $a$ = 3.851 Å and $V$ = 28.55 $Å^3$/Sr — are very close to the experimental data. Here the story with $SrH_9$ repeats: a highly symmetric $Im\bar{3}m$ structure with an atomic hydrogen sublattice is dynamically unstable in the case of $SrH_6$, whereas its distorted and molecular "isomer" turns out to be a much better candidate. Accurate DFT calculations show that $P4_2/mnm$-$SrH_6$ ($Sr_2H_{12}$), a tetragonally distorted $Im\bar{3}m$-$SrH_6$ with molecular hydrogen in the H sublattice, can also explain the experimental

XRD pattern. This tetragonal $SrH_6$ is dynamically stable and lies near (50 meV/atom above) the convex hull (Figure 1c).

Thus, both compounds found in DAC Sr90, c-$SrH_{\sim9}$ and $SrH_6$, can be considered molecular semiconducting "isomers" of known superhydrides such as $F\bar{4}3m$-$XH_9$ and $Im\bar{3}m$-$XH_6$, that is compounds with a similar composition, unit cell parameters, and X-ray diffraction patterns (metal sublattice) but significantly different structure of the hydrogen sublattice. This series of experiments also shows that although $SrH_{\sim9}$ can be synthesized at a relatively low pressure of about 62 GPa, the synthesis of hexahydride $SrH_6$ requires a higher pressure of ~90 GPa.

## *Electrical measurements of $SrH_6$*

Electrical measurements for strontium hydrides are hindered by their low conductivity. The active resistance of samples at low frequencies is about several MΩ; however, the use of high-frequency current up to 1 MHz makes it possible to obtain high-quality impedance patterns. We studied in detail the electrical DAC E1 with a pseudo-four-contact van der Pauw circuit, loaded with Sr/AB and heated at 126 GPa. According to the XRD study, the opaque sample consists of $Pm\bar{3}n$-like pseudocubic $P1$-$SrH_6$ (Supporting Information Figure S25).

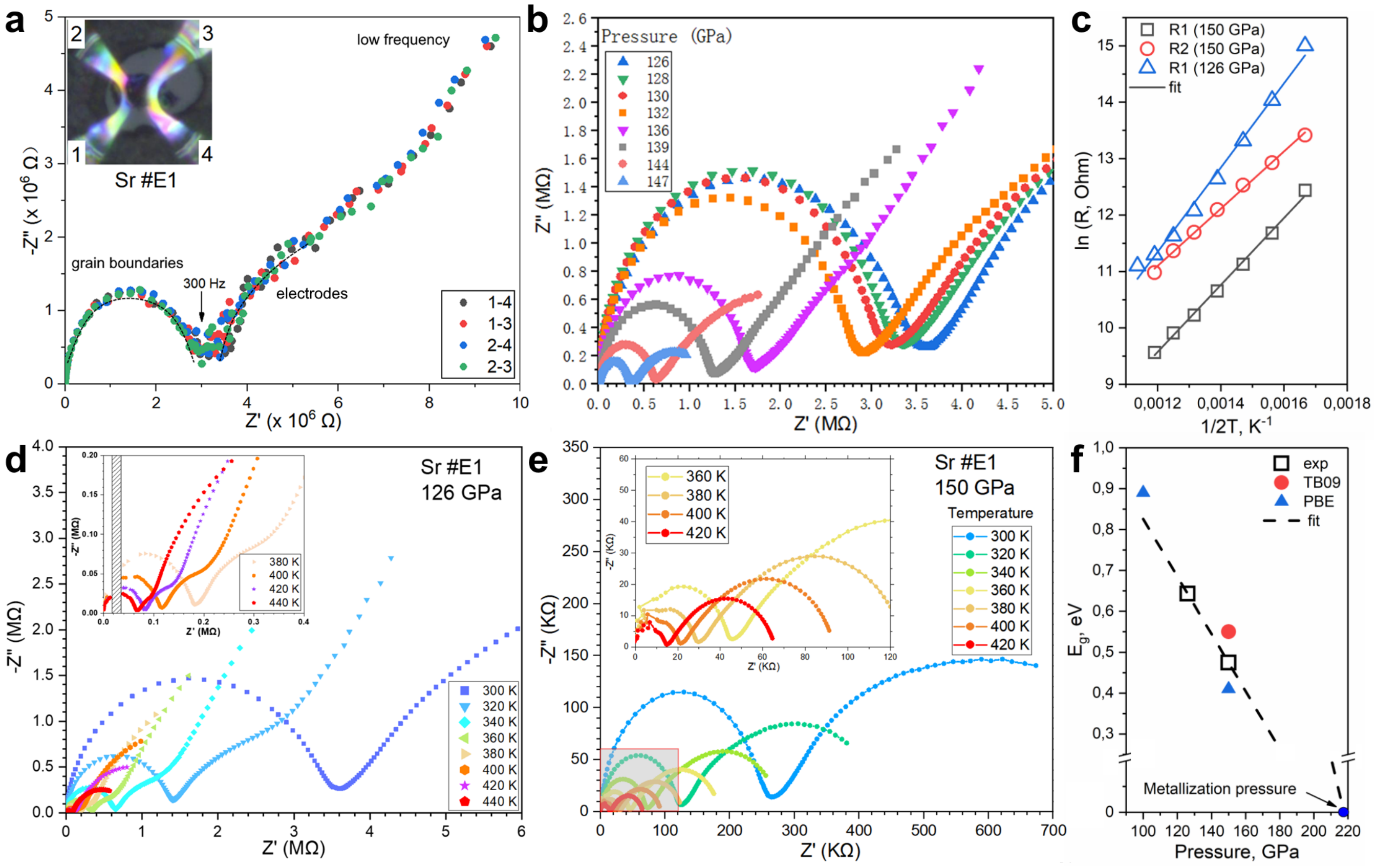

**Figure 4.** Impedance spectroscopy (Nyquist diagrams) of the pseudocubic $P1$-$SrH_6$ sample (DAC E1) in the frequency range from 0.1 to $10^7$ Hz. (a) The active resistance of the sample is about 2.5 MΩ, the capacitance of the circuit is ~0.5 µF with any combination of contacts. (b) Impedance at pressures of 126–147 GPa and 300 K. (c) Calculating the activation energy $E_g$ using the temperature dependence of the electrical resistance. (d) Impedance in the temperature interval of 300–440 K at 126 GPa, and (e) in the range of 300–420 K at 150 GPa. (f) Activation energy $E_g$ compared to the direct bandgap of $P1$-$SrH_6$ at 100–150 GPa. See Supporting Information for details.

Before the experiment, we checked that an impedance of the circuit is equal for any combination of contacts (Figure 4a). On a typical hodograph for the $SrH_6$ sample, there is a clear first semicircle, the radius of which has a pronounced dependence on the temperature and pressure (Figure 4b,d,e). In some cases, especially at high temperatures and pressures, a part of the second semicircle or, to be correct, half-ellipse, is visible, which continues with an oblique low-frequency tail. The dimensions of the second half-ellipse also significantly decrease as the temperature and pressure rise.

To interpret the obtained experimental data, we used a simplified $L(C, R)$(CPE, $R$) scheme of five elements, where $L$ reflects the inductance of the lead wires, ($C$, $R$) corresponds to ionic conductivity at the grain boundaries of the $SrH_6$ nanocrystals, and (CPE, $R$) describes the transport and hydrogen diffusion phenomena at the border of the Mo electrodes.[36–38]

Simple calculations show that in the formula $R(T) = R_0 \times \exp(-E_g/2k_BT)$, the activation energy $E_g$ approximately equals 0.44–0.51 eV at 150 GPa (the average value is 0.475 eV) and increases to 0.64 eV at 126 GPa. This agrees with the optical properties of the sample (its darkening) and the value of the direct bandgap for $P1$-$SrH_6$, predicted to be 0.55 eV at 150 GPa (Supporting Information Figure S17). Moreover, the pressure dependence allows us[39] to calculate $dE_g/dP$. The data from Figure 4b give $d\ln(R)/dP = -0.13$ GPa$^{-1}$ and $dE_g/dP = -0.0067$ eV/GPa at 300 K. Thus, the expected metallization pressure of $Pm\bar{3}n$-like pseudocubic $P1$-$SrH_6$ is about 220 GPa (Figure 4f).

Additional laser heating of the sample at 150 GPa (Supporting Information Figure S42) neutralizes the electrode part (the half-ellipse and low-frequency spike) of the hodograph. At the same time, the active resistance of the sample is reduced 20 times to 10 kΩ, which speaks in favor of the reordering of grain boundaries and the change in the nature of conductivity to predominantly electronic.

We calculated the diffusion coefficients of hydrogen for $SrH_6$ and $SrH_{22}$ at 150 GPa using machine learning interatomic potentials (MLIP, Supporting Information Table S2 and Figures S35–S36). Extrapolation of the data to 300 K gives $D_H = 0.18$ Å$^2$/ns for $SrH_6$ and a significantly larger $D_H = 17$ Å$^2$/ns for $SrH_{22}$. A better ordered structure of $P1$-$SrH_6$, which is close to the cubic prototypes $Pm\bar{3}n$-$Eu_8H_{46}$[26] and $Ba_8H_{46}$,[27] is more stable than the disordered molecular hydride $SrH_{22}$, where individual hydrogen atoms can migrate to a neighboring unit cell within 1 ns. This makes the concept of the hydrogen sublattice in this compound rather blurred.

## III. Conclusions

The importance of studying strontium polyhydrides stems from Sr being a promising element for the design of ternary and quaternary superconducting hydrides stable below 1 Mbar. Unlike Ca and Y, strontium forms binary superhydrides with a very high hydrogen content at relatively low pressures of ~50–90 GPa. These compounds exhibit semiconducting behavior below 120–150 GPa, whereas calcium and yttrium superhydrides are high-$T_C$ superconductors.

Complex and diverse behavior of the Sr–H system under pressure differs significantly from that of Ca–H, where $CaH_4$ and $CaH_6$ are present, and is closer to Ba–H in the chemical properties. In the Sr–H system, there is a series of molecular polyhydrides forming at pressures of 25–165 GPa: $Pm\bar{3}n$-like and $Im\bar{3}m$-like polymorphs of $SrH_6$, $F\bar{4}3m$-like $SrH_{\sim 9}$, and pseudotetragonal $P1$-$SrH_{22}$. Using the impedance spectroscopy, we estimated the direct bandgap in the $Pm\bar{3}n$-like polymorph $P1$-$SrH_6$ to be 0.44–0.51 eV at 150 GPa and the metallization pressure of this hydride to be 220 GPa.

The most amazing strontium hydride we observed is pseudotetragonal $P1$-$SrH_{22}$, the hydride of metal with the highest hydrogen content discovered so far, which can be considered a form of molecular hydrogen doped with 4 atom % of Sr. This compound is stable at a relatively low pressure of 100 GPa, and its metallization can be achieved during compression to about 200 GPa. Observing metallization in pure hydrogen is still a very difficult task associated with study of extremely small samples. Therefore, $SrH_{22}$ can be used as a helpful model of the hydrogen behavior above 300–350 GPa, realized at 1–2 Mbar. In a similar manner, barium hydride $BaH_{12}$ (or $Ba_{0.08}H_{0.92}$) that we have discovered earlier can be used as a model for the emergence of superconductivity in already metallic hydrogen.

## Author contributions

D.V.S., W.C., and X.H. contributed equally to this work. W.C. performed all experiments with diamond anvil cells. D.V.S. analyzed and interpreted the results of the XRD, impedance, and optical measurements, and wrote the manuscript. X.H., A.R.O. and T.C. directed the research and edited the manuscript. I.A.K. and A.B.M. performed the T-USPEX and anharmonic phonon density of states calculations and MLIP-LAMMPS diffusion simulations. M.G. wrote Python scripts for accelerated USPEX data processing and automatic interpretation of diffraction patterns. A.G.K., D.Z., and A.R.O. prepared the theoretical analysis and calculated the equation of states and electron and phonon band structures for various strontium hydrides. C.T. and X.G. performed the Bader analysis, ABINIT band structure calculations with TB09 and PBE, dielectric function calculations, and nonresonant (PEAD) and resonant Raman spectrum calculations. All the authors provided critical feedback and helped shape the research.

## Acknowledgments

In situ angle dispersive XRD experiments were performed at 4W2 HP-Station, Beijing Synchrotron Radiation Facility (BSRF), and 15U1 beamline, Shanghai Synchrotron Radiation Facility (SSRF). This work was supported by the National Key R&D Program of China (No. 2018YFA0305900), National Natural Science Foundation of China (Nos. 52072188, 51632002, 11804113, 51720105007), and Program for Changjiang Scholars, Innovative Research Team in University (No. IRT_15R23). A.R.O. thanks the Russian Science Foundation (grant 19-72-30043). D.V.S. thanks the Russian Foundation for Basic Research (project 20-32-90099). I.A.K. and A.B.M. thank the Russian Science Foundation (grant No. 21-73-10261) for the financial support of the anharmonic phonon density of states calculations and diffusion simulations. C.T. would like to thank the Siberian SuperComputer Center of the Siberian Branch of the Russian Academy of Sciences for computational resources and acknowledges funding from Russian state assignment to ISSCM SB RAS (project no. FWUS-2021-0005). X.G. acknowledges funding from the FRS-FNRS Belgium under grant number T.0103.19 - ALPS. S.V.L. acknowledges support by RSCF-INSF grant 20-53-56065. We also thank Igor Grishin (Skoltech) for proofreading the manuscript, Monan Ma for the laser heating simulations in COMSOL, and Song Hao (Jilin University) for helping to interpret the very first experiments with strontium hydrides in 2019.

# SUPPORTING INFORMATION

## Sr-Doped Molecular Hydrogen: Synthesis and Properties of $SrH_{22}$

Dmitrii V. Semenok,[1] Wuhao Chen,[2] Xiaoli Huang,[2,*] Di Zhou,[1] Ivan A. Kruglov,[3,4] Arslan B. Mazitov,[3,4] Michele Galasso,[1] Christian Tantardini,[1,7] Xavier Gonze,[1,5] Alexander G. Kvashnin,[1] Artem R. Oganov,[1,*] and Tian Cui[2,6,*]

[1] Skolkovo Institute of Science and Technology, Skolkovo Innovation Center, 30 Bolshoy Boulevard, Moscow 121205, Russia
[2] State Key Laboratory of Superhard Materials, College of Physics, Jilin University, Changchun 130012, China
[3] Dukhov Research Institute of Automatics (VNIIA), Moscow 127055, Russia
[4] Moscow Institute of Physics and Technology, 9 Institutsky Lane, Dolgoprudny 141700, Russia
[5] European Theoretical Spectroscopy Facility, Institute of Condensed Matter and Nanosciences, Université catholique de Louvain, Chemin des étoiles 8, bte L07.03.01, B-1348 Louvain-la-Neuve, Belgium
[6] School of Physical Science and Technology, Ningbo University, Ningbo 315211, China
[7] Institute of Solid State Chemistry and Mechanochemistry SB RAS, 630128, Russia

* Corresponding authors
E-mails: huangxiaoli@jlu.edu.cn (X.H.); a.oganov@skoltech.ru (A.R.O.); cuitian@jlu.edu.cn (T.C.)

## Contents

# Methods

## *Experiment*

The strontium metal samples with a purity of 99.9% were purchased from Alfa Aesar. All diamond anvil cells (with 100 µm and 50 µm culets) were loaded with metallic Sr or $SrH_2$ samples and sublimated ammonia borane (AB) in an argon glove box. Tungsten gaskets had a thickness of 10 ± 2 µm. Heating was carried out by several 0.1 s pulses of a Nd:YAG infrared laser (1.07 µm). The temperature was approximately determined using the decay of blackbody radiation within the Planck formula with an error of about ±200 K. The applied pressure was measured by the edge of diamond Raman signal[1] using Horiba LabRAM HR800 Ev spectrometer with an exposure time of 10 s at 532 nm. The X-ray diffraction (XRD) patterns from samples in diamond anvil cells (DACs) were recorded on BL15U1 synchrotron beamline at the Shanghai Synchrotron Research Facility (SSRF, China) using a focused (5 × 12 µm) monochromatic X-ray beam with a linear polarization (20 keV, 0.6199 Å). Mar165 CCD was used as a detector.

The experimental XRD images were analyzed and integrated using Dioptas software package (version 0.5).[2] The full profile analysis of the diffraction patterns and the calculations of the unit cell parameters were performed in Materials studio[3] and JANA2006[4] using the Le Bail method.[5]

To investigate the electrical resistivity of strontium polyhydrides, we performed measurements in Cu–Be DACs E1 using the pseudo-four-probe technique. The tungsten gasket with an initial thickness of 250 µm was precompressed to about 25 GPa. Then a hole with a diameter 20% larger than the culet diameter was drilled in the tungsten gasket using a pulse laser (λ = 532 nm). The cubic boron nitride (c-BN) powder mixed with epoxy was used as an insulating layer. We filled the chamber with MgO and compressed it to about 5 GPa. Then, in the obtained transparent MgO layer, a hole with a diameter of about 40 µm was drilled by a laser. Ultraviolet lithography was used to prepare four electrodes on the diamond culet. We deposited the 500 nm thick Mo layer by magnetron sputtering (with the field of 200 V at 300 K) and removed excess metal by acid etching. Four deposited Mo electrodes were extended by a platinum foil. The chamber was filled with sublimated ammonia borane (AB) and a small piece of Sr was placed on the culet of the upper diamond with four electrodes. All preparations were made in an argon glove box ($O_2$ < 0.1 ppm, $H_2O$ < 0.01 ppm). After that, the DAC was closed and compressed to the required pressure. We used the 1.07 µm infrared pulse laser (~0.1 s, ~1000 K) to heat the Sr/AB samples.

Impedance spectroscopy was performed using Solartron SI 1260 impedance analyzer equipped with Solartron 1296 dielectric interface. A 100 mV sine signal with a sweep frequency from 0.1 to $10^7$ Hz was irradiated on the sample. The symmetric BeCu cell was put on a thermostatic flat-plate heater for higher temperature, which was recorded using a T-type thermocouple inserted into the cell.

A summary of the stability parameters of all studied compounds and DACs is presented in Table S1.

**Table S1.** Studied strontium polyhydrides and regions of their experimental stability.

| Discovered phase | DAC | Lowest pressure of synthesis, GPa | Stability range, GPa |
|---|---|---|---|
| c-$SrH_{\sim 9}$ | Sr50, Sr90 | 62 | 62–139 |
| $P4_2/mnm$-$SrH_6$ | Sr90 | 90 | 87–139 |
| $Cmme$-$Sr_2H_3$ | Sr50 | 62 | 62–101 |
| $C2/m$-$Sr_3H_{13}$ | Sr3, Sr2 | 118 | 70–146 |
| $Pm\bar{3}n$-like $P1$-$Sr_8H_{48}$ | Sr3, Sr165, Sr4, E1 | 118 | 56–158 |
| $P1$-$SrH_{22}$ | Sr2, Sr1 | 146 | 99–146 |

## *Theory*

The non-self-consistent and self-consistent calculations were performed using the density functional theory (DFT)[6,7] within the Perdew–Burke–Ernzerhof functional (generalized gradient approximation)[8] as implemented in the Abinit code.[9,10] The computations of the total energy and optimization of the geometry of strontium hydrides were carried out using the optimized norm-conserving scalar-relativistic Vanderbilt pseudopotentials (ONCVPSP).[11] The kinetic energy cutoff for plane waves was found through the convergence tests for the total energy and the unit cell parameters performed in the interval from 5 to 80 Ha (ecut). The Brillouin zone was sampled using *k*-points meshes with a resolution found via the convergence tests for the total energy and the unit cell parameters performed in the interval from 1×1×1 to 8×8×8 (ngkpt). The band structure calculations were performed using the Hartwigsen–Goedecker–Hutter nonrelativistic local-density approximation (LDA) pseudopotentials[12] with the kinetic energy cutoff of 30 Ha and a 4×4×4 *k*-points mesh.

The non-self-consistent and self-consistent calculations of the equations of state and electron and phonon band structure of $Sr_3H_{13}$ and $SrH_{22}$ were performed using the density functional theory (DFT) within the Perdew–Burke–Ernzerhof functional (generalized gradient approximation) as implemented in the Abinit and VASP codes. For comparison of the total enthalpy and optimization of the geometry of $SrH_{22}$, calculations were also carried out using the optimized norm-conserving scalar-relativistic Vanderbilt pseudopotentials (ONCVPSP). Within the Abinit calculations, the kinetic energy cutoff of 40 Ha for plane waves was found through the convergence tests for the total energy and the unit cell parameters performed in the interval from 5 to 80 Ha. The Brillouin zone was sampled using *k*-points meshes with a resolution of 8×8×8 found via the convergence tests for the total energy and the unit cell parameters performed in the interval from 1×1×1 to 8×8×8 (ngkpt). The LDA band structure calculations were performed using the Hartwigsen–Goedecker–Hutter(HGH) nonrelativistic LDA pseudopotentials with the kinetic energy cutoff of 30 Ha and an 8×8×8 *k*-points mesh. The Bader analysis was performed using Critic2 software.[13,14]

The dynamic stability and phonon density of states of $SrH_{22}$ were studied using classical molecular dynamics and the interatomic potential based on machine learning. We used the Moment Tensor Potential (MTP)[15] whose applicability in calculations of the phonon properties of materials has been demonstrated previously. Moreover, within this approach we can explicitly take into account the anharmonicity of hydrogen vibrations. To train the potential, we first simulated Sr hydrides in quantum molecular dynamics in an *NPT*-ensemble at 100 GPa and 10, 100, and 300 K, with a duration of 5 ps using the VASP code.[16–18] We used the PAW PBE pseudopotentials for the H and Sr atoms, and $2\pi \times 0.06$ Å$^{-1}$ *k*-mesh with a cutoff energy of 400 eV. For training the MTP, sets of Sr–H structures were chosen using active learning. We checked the dynamical stability of the studied Sr hydrides with the obtained MTPs via several runs of molecular dynamics calculations at 300 K and 180 GPa. First, the *NPT* dynamics simulations were performed in a supercell with about 1000 atoms for 40 ps. During the last 20 ps, the cell parameters were averaged. In the second step, the coordinates of the atoms were averaged within the *NPT* dynamics with a duration of 20 ps and the final structures were symmetrized as implemented in T-USPEX method. Then, for the structures of Sr polyhydrides relaxed at 100 GPa and 10, 100, and 300 K, the phonon density of states (DOS) was calculated within the MTP using the velocity autocorrelator (VACF) separately for each type of atoms:

$$g(\theta) = 4\int_0^{\infty} \cos(2\pi\theta t)\frac{\langle \overline{V(0)V(t)} \rangle}{\langle V(0)^2 \rangle}\mathrm{d}t \quad \text{(S1)}$$

where $\theta$ is the frequency. The calculations were carried out in a 20×20×20 supercell. The velocity autocorrelator was calculated using molecular dynamics, then the phonon DOS was obtained.

All Raman investigations were performed using Abinit v.9.4.1. At 120 GPa, *P*1-$SrH_{22}$ is a metal with a band structure where the bands do not intersect with each other (see TB09 band structure). We optimized the crystal structure using the PBE DFT functional with the Fermi–Dirac distribution and a smearing temperature of 0.01 Ha. A Γ-centered *k*-point grid with a 6×6×6 mesh was used. The obtained initial band structure showed no intersection between the valence and conduction bands in the Brillouin zone. Therefore, we can treat $SrH_{22}$ as a semiconductor in the first approximation, fixing the number of the occupied bands, which allowed us to calculate the Raman spectra using the density functional perturbation theory for semiconductors.[19] In this case, we tested two different approaches using the same Γ-centered grid with 6×6×6 *k*-points:

(i) the crystal structure was optimized using the LDA DFT functional with norm-conserving pseudopotentials and its band structure generated using a non-self-consistent approach was compared with the band structure generated using the TB09 meta-GGA DFT functional;

(ii) the crystal structure was optimized using the PBE DFT functional with norm-conserving pseudopotentials and its band structure generated using a non-self-consistent approach was compared with the band structure generated using the TB09 meta-GGA DFT functional.

In the first and second approaches, $SrH_{22}$ was considered a semiconductor, maintaining the fixed band occupation. The subsequent one-phonon nonresonant Raman spectra were calculated using PEAD approach[19] and compared with the experimental one (Figure S34a). Both approaches give rather poor agreement with the experimental data, but the position of one of the groups of signals (about 4200 $cm^{-1}$) predicted in approach (ii) coincides with the experimental value.

Then, in more advanced one-phonon resonant Raman calculations, the Raman susceptibility was obtained from the derivative of the dielectric function for the incoming laser frequency (532 nm).[20,21] We considered the optimized crystal structure with the PBE and norm-conserving pseudopotentials for a metal with a Fermi–Dirac distribution and the temperature of smearing equal to 0.01 Ha to calculate the subsequent dynamical matrix(Δ) using the density functional perturbation theory with the LDA DFT functional and norm-conserving pseudopotentials. The right frequencies of metal *P*1-$SrH_{22}$ were obtained solving the secular equation

$$\Delta Q_\zeta = \omega_\zeta^2 Q_\zeta \tag{S2}$$

where $Q_\zeta = (e_{1\zeta}, \ldots, e_{N\zeta})$ is the eigenvector (i.e., displacement) of the phonon mode ζ with a frequency $\omega_\zeta$, which in general consists of 3*N* components, where *N* is the number of atoms per unit cell. The intensity *I* of the Raman spectra at each phonon frequency mode $\omega_\zeta$ for a photon of frequency $\omega_{\text{laser}}$ is defined as

$$I = (\omega_{\text{laser}} - \omega_\zeta)^4 \left| e_{\text{out}} \cdot \alpha^\zeta \cdot \boldsymbol{e}_{\text{in}} \right|^2 \frac{n_\zeta + 1}{2\omega_\zeta} \tag{S3}$$

where $n_\zeta = \frac{1}{e^{h\omega_\zeta/kT} - 1}$ is the phonon occupation factor that depends on the temperature *T.* Two $\omega_{\text{laser}}$ values were considered: 532 nm (green) and 650 nm (red), which can cover the energy of the bandgap seen in the electronic structure of *P*1-$SrH_{22}$ calculated using the PBE DFT functional. The fundamental bandgap is 0.08 eV, the direct bandgap is 1.39 eV (Figure S15).

The term $\alpha^\zeta$ in eq. S3 is defined as the Raman susceptibility:[19]

$$\alpha_{ij}^\zeta(\omega) = \sqrt{\Omega_0} \sum_{\tau\beta} \frac{\partial \chi_{ij}(\omega)}{\partial R_{\tau\beta}} u_{\tau\beta}^\zeta \tag{S4}$$

where $\Omega_0$ is the unit cell volume, $\chi_{ij}$ is the macroscopic dielectric susceptibility, and $u_{\tau\beta}^\zeta$ is the eigendisplacement of the phonon mode ζ of atom τ in the direction β. In the case of the Raman spectra of powders, the intensity of a peak at each frequency is the sum of the parallel intensity $I_{||}^{\text{powder}}$ and perpendicular intensity $I_{\perp}^{\text{powder}}$, which can be defined as

$$I_{||}^{\text{powder}} = C(10G_0 + 4G_2) \tag{S5}$$

$$I_{\perp}^{\text{powder}} = C(5G_1 + 3G_2) \tag{S6}$$

$$I_{\text{tot}}^{\text{powder}} = I_{||}^{\text{powder}} + I_{\perp}^{\text{powder}} \tag{S7}$$

where

$$G_0 = \frac{(\alpha_{xx}+\alpha_{yy}+\alpha_{zz})^2}{3} \tag{S8}$$

$$G_1 = \frac{(\alpha_{xy}-\alpha_{yz})^2+(\alpha_{yz}-\alpha_{zx})^2+(\alpha_{zx}-\alpha_{xy})^2}{2} \tag{S9}$$

$$G_2 = \frac{(\alpha_{xy}+\alpha_{yz})^2+(\alpha_{yz}+\alpha_{zx})^2+(\alpha_{zx}+\alpha_{xy})^2}{2} + \frac{(\alpha_{xx}+\alpha_{yy})^2+(\alpha_{yy}+\alpha_{zz})^2+(\alpha_{zz}-\alpha_{xx})^2}{2} \tag{S10}$$

$$C = (\omega_{\text{laser}} - \omega_{\zeta})^4 \frac{n_{\zeta}+1}{2\omega_{\zeta}} \tag{S11}$$

We calculated $I_{\text{tot}}^{\text{powder}}$ for the previously optimized structure of primitive triclinic $SrH_{22}$ at 120 GPa using the PBE DFT functional and norm-conserving pseudopotentials. The macroscopic dielectric susceptibility $\chi_{ij}$ was calculated using the LDA DFT functional and norm-conserving pseudopotentials with 80 bands for the derivative of energy with respect to *k*-points within the Brillouin zone, and the Fermi–Dirac distribution of occupation with a smearing temperature of 0.01 Ha. A Γ-centered grid with 16×16×16 *k*-points was used.

As a result, the Raman spectrum of $SrH_{22}$ was significantly simplified: only three signals, ~4000, 4150, and 4300 cm$^{-1}$, remained in the spectrum, the most intense of which corresponds well to the experimentally observed peak at 4140 cm$^{-1}$ (123 GPa, 100 K). Another peak, lower than 4000 cm$^{-1}$ and of much weaker intensity, is also seen in such one-phonon resonant Raman computations, but its intensity strongly depends on the match between the theoretical gap value and incoming laser frequency. Because of the absence of a pressure dependence in the cell, we assume that the signals at about 4500 cm$^{-1}$ and 4800 cm$^{-1}$ do not belong to the sample.

We estimated the diffusion rate in strontium hydrides using classical molecular dynamics with a machine learning interatomic potential implemented in the MLIP package.[15,22,23] The interatomic potential was actively trained on the ab initio molecular dynamics trajectories of $SrH_6$ and $SrH_{22}$ unit cells in the *NPT*-ensemble (*P* = 150 GPa, *T* = 1000 K) with the external electric field of $10^4$ V/m applied in the [100] direction. The reference ab initio data on energies, forces, and stresses were obtained on the DFT level using the VASP code.[16–18] We used the following VASP settings for training set preparation: the cutoff energy of the plane waves basis set was 450 eV, the first Brillouin zone was sampled by a Gamma-centered grid with 2π × 0.03 Å$^{-1}$ resolution, and the partial occupancies of electron states were set using a Gaussian method with 0.05 eV smearing width. The convergence criteria of the Self Consistent Field (SCF) cycle was $10^{-5}$ eV. The resulting actively selected training set had 2012 configurations, and the mean absolute errors (MAE) of the prediction of these configurations' energies and forces were 4.8 meV/atom and 0.27 eV/A (~25%), respectively. Molecular dynamics simulation in VASP for 30 ps shows discontinuous movement of H atoms; from this data, it is impossible to draw a clear conclusion about the diffusion parameters of the system.

The obtained interatomic potential was used to perform a large-scale molecular dynamics run on a 4×4×4 supercell of $SrH_{22}$ and 3×3×3 supercell of $Sr_8H_{48}$ at temperatures of 500, 550, and 600 K and a constant pressure of 150 GPa using LAMMPS package.[24] Each run lasted within 300 ps with a time step of 0.3 fs. Additionally, the external electric field of $10^6$ V/m was applied to each system. We found that applying this electric field has practically no effect on diffusion in the direction of the field in the MLIP calculations. Thus, the studied compounds $SrH_{22}$ and $SrH_6$ have no substantial ionic conductivity. However, estimates of ionic conductivity using the Nernst–Einstein relation give the following results: σ(300 K) ~ $10^{-3}$ S/cm and σ(500 K) ~ 3.5 × $10^{-2}$ S/cm for $SrH_6$ at 150 GPa. Higher ionic conductivity is expected in $SrH_{22}$: σ(500 K) ~ 0.2 S/cm at 150 GPa.

The initial atomic charges of the Sr and H atoms were obtained using the Bader charge analysis (Tables S11, S19, Figures S80, S81) of the DFT ground state charge distributions of $SrH_6$ and $SrH_{22}$ hydrides and kept fixed during the simulation. The Bader analysis of the simulations in VASP after 30 ps indicates that the charges on atoms change insignificantly (Figures S80, S81). Finally, the diffusion coefficients of hydrogen atoms in the [100] direction were calculated using the Einstein formula (projection on $x$): $\overline{x^2} = 2Dt$. An activation formula $D(T) = D_0 \times \exp(-E_a/k_BT)$ was used to extrapolate the temperature dependence of the diffusion coefficients (Table S2). In the order of magnitude, the obtained diffusion coefficients agree with the results of other works.[25]

**Table S2.** Calculated parameters of hydrogen diffusion in $SrH_6$ and $SrH_{22}$ at 150 GPa.

| Temperature, K | Diffusion coefficients, Å$^2$/ps | |
|---|---|---|
| | $Pm\bar{3}n$-like $P1$-$SrH_6$ | $P1$-$SrH_{22}$ |
| 500 | 0.0103 | 0.1259 |
| 550 | 0.0186 | 0.1592 |
| 600 | 0.0282 | 0.2076 |
| $D_0$, Å$^2$/ps | 4.505 | 2.471 |
| $E_a$, meV | 261.5 | 128.8 |

## Convex hulls

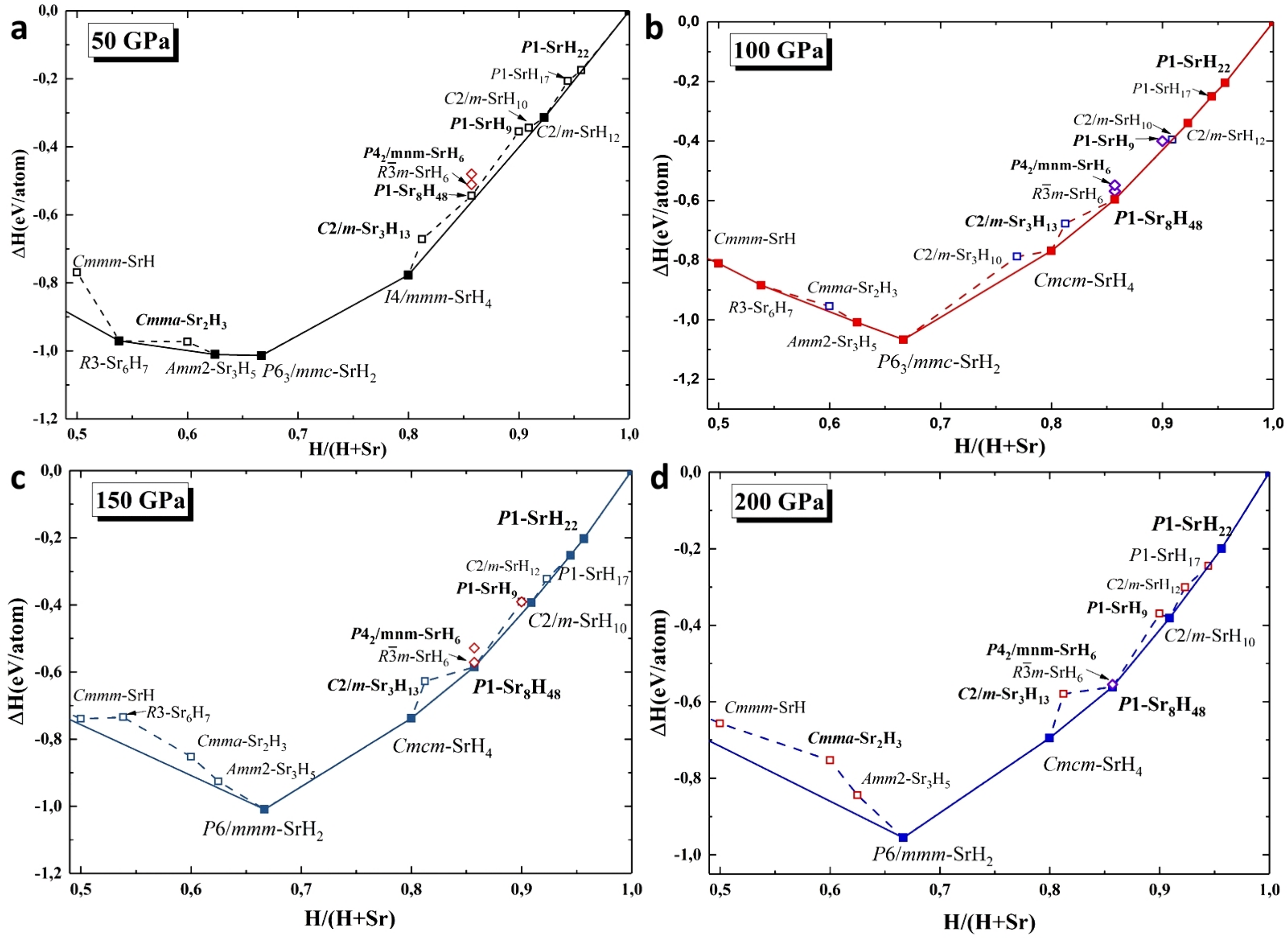

**Figure S1.** Convex hulls of the Sr–H system at (a) 50, (b) 100, (c) 150 and (d) 200 GPa at 0 K, calculated without the zero-point energy (ZPE) contribution. Stable and metastable phases are shown as filled and hollow squares, respectively. $P1$-$SrH_{22}$ and pseudocubic $P1$-$Sr_8H_{48}$ are located near the convex hull at 50 GPa. At the same time $C2/m$-$SrH_{12}$ is stabilized, whereas it is dynamically unstable at 50–150 GPa in the harmonic approximation and has a different XRD pattern than the experimentally obtained phases. At 100GPa, many strontium hydrides lie on the convex hull or in its immediate vicinity: $SrH_{22}$, $SrH_{17}$, $SrH_{12}$, $SrH_{10}$, $SrH_9$, $Sr_8H_{48}$, $Sr_2H_3$, and so forth. As the pressure rises to 200 GPa, the most important phases, $SrH_{22}$ and $Sr_8H_{48}$, retain their stability and position on the convex hull.

The Gibbs free energy was calculated as

$$G(T) = E_0 + PV + k_BT \int g(\omega) \ln [1 - \exp(-\hbar\omega/k_BT)]\, d\omega + 1/2 \int g(\omega)\hbar\omega d\omega, \tag{S12}$$

where $E_0 + PV$ is the total energy from the DFT calculations, $g(\omega)$ is the phonon density of states at a given pressure calculated using the finite displacements method as implemented in PHONOPY[26,27] with forces computed using VASP.[16–18]

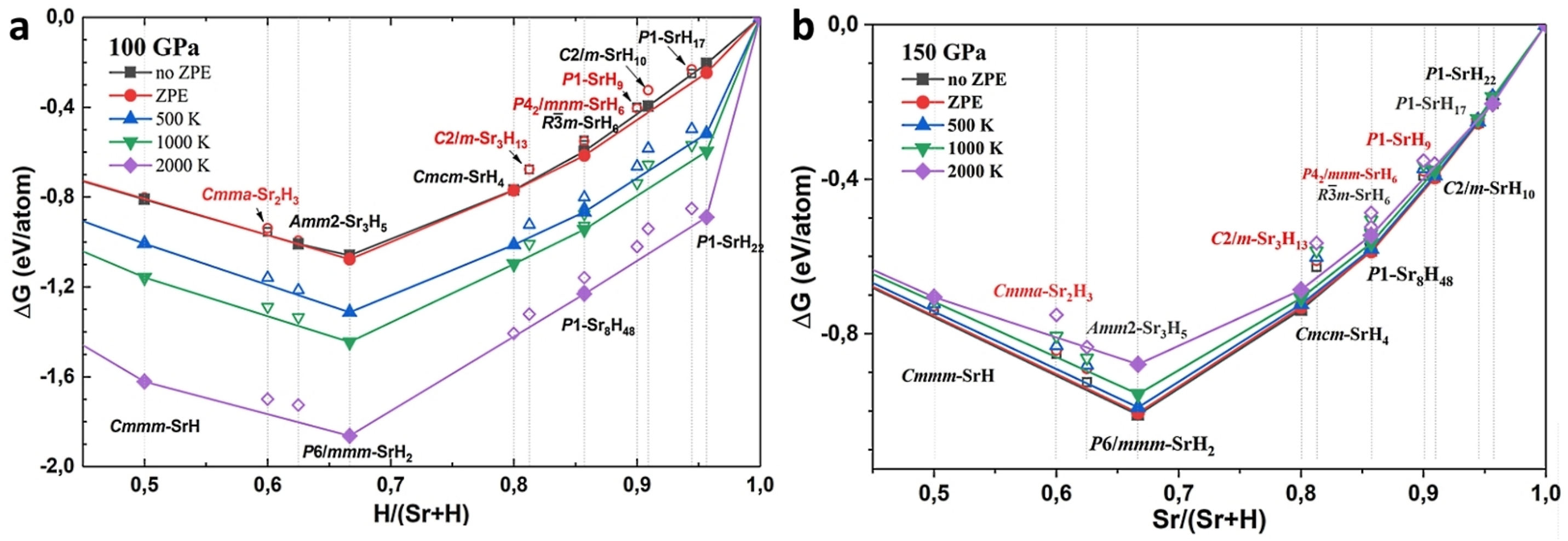

**Figure S2.** Convex hulls of the Sr–H system at (a) 100 GPa and (b) 150 GPa calculated with the zero-point energy (ZPE) and entropy (-TS) contributions at temperatures of 0, 500, 1000, and 2000 K. Two phase modifications of $SrH_2$ ($P6/mmm$ and $P6_3/mmc$) occupy almost the same position on the convex hull. Considering the ZPE makes $P6/mmm$ preferable at 100 GPa.

**Table S3**. Enthalpy, ZPE and Gibbs free energy of formation for various Sr–H phases at 100 GPa and 0, 500, 1000, and 2000 K.

| Compound | Enthalpy $H$, eV | $H$ per Sr, eV | $\Delta H_{form}$, eV/atom | ZPE, eV/cell | $\Delta(H+ZPE)_{form}$, eV/atom | $\Delta G_{form}$, 500K | $\Delta G_{form}$, 1000 K | $\Delta G_{form}$, 2000 K |
|---|---|---|---|---|---|---|---|---|
| $P6_3/mmc$-Sr | 300.50892 | 12.52121 | 0 | 0.76855 | 0 | 0 | 0 | 0 |
| $Cmmm$-$Sr_6H_6$ | 58.16262 | 9.69377 | -0.81045 | 1.86997 | -0.80109 | -1.00689 | -1.15781 | -1.62108 |
| $Cmma$-$Sr_8H_{12}$ | 66.60027 | 8.32503 | -0.95455 | 3.70136 | -0.93884 | -1.15878 | -1.28866 | -1.69947 |
| $Amm2$-$Sr_9H_{15}$ | 70.38155 | 7.82017 | -1.00881 | 4.56287 | -0.99377 | -1.21444 | -1.33519 | -1.72528 |
| $P6/mmm$-$SrH_2$ | 6.92977 | 6.92977 | -1.05946 | 0.50048 | -1.07725 | -1.31286 | -1.44516 | -1.86287 |
| $Cmcm$-$Sr_2H_8$ | 7.70518 | 3.85259 | -0.7685 | 2.13489 | -0.77015 | -1.01214 | -1.09795 | -1.40646 |
| $C2/m$-$Sr_6H_{26}$ | 22.08117 | 3.68019 | -0.67739 | 6.98828 | -0.677 | -0.92214 | -1.00813 | -1.32144 |
| $P1$-$SrH_6$ | 8.86737 | 1.10842 | -0.59623 | 11.70365 | -0.61545 | -0.86626 | -0.94386 | -1.23114 |
| $R$-$3m$-$SrH_6$ | 1.3041 | 1.3041 | -0.56827 | 1.37014 | -0.60076 | -0.84985 | -0.92716 | -1.21982 |
| $P4_2/mnm$-$Sr_2H_{12}$ | 2.88729 | 1.44365 | -0.54834 | 3.15383 | -0.55128 | -0.80062 | -0.87522 | -1.15834 |
| $P1$-$SrH_9$ | -9.36689 | -2.34172 | -0.40042 | 9.32768 | -0.40525 | -0.66336 | -0.73784 | -1.02039 |
| $C2/m$-$Sr_2H_{20}$ | -7.79019 | -3.89509 | -0.39555 | 6.85252 | -0.32417 | -0.58334 | -0.65497 | -0.94037 |
| $P1$-$Sr_2H_{34}$ | -24.99765 | -12.49882 | -0.2505 | 9.62845 | -0.23124 | -0.49652 | -0.56883 | -0.85088 |
| $P1$-$SrH_{22}$ | -18.73969 | -18.73969 | -0.2051 | 4.81316 | -0.24679 | -0.51748 | -0.59651 | -0.88911 |
| $C2/c$-H | -28.95672 | -1.20653 | 0 | 6.26195 | 0 | 0 | 0 | 0 |

**Table S4**. Enthalpy, ZPE and Gibbs free energy of formation for various Sr–H phases at 150 GPa and 0, 500, 1000, and 2000 K.

| Compound | Enthalpy $H$, eV | $H$ per Sr, eV | $\Delta H_{form}$, eV/atom | ZPE, eV/cell | $\Delta(H+ZPE)_{form}$, eV/atom | $\Delta G_{form}$, 500K | $\Delta G_{form}$, 1000 K | $\Delta G_{form}$, 2000 K |
|---|---|---|---|---|---|---|---|---|
| $P6_3/mmc$-Sr | 403.77031 | 16.82376 | 0 | 0.94392 | 0 | 0 | 0 | 0 |
| $Cmmm$-$Sr_6H_6$ | 88.75235 | 14.79206 | -0.73927 | 1.98652 | -0.7272 | -0.72253 | -0.71337 | -0.70515 |
| $Cmma$-$Sr_8H_{12}$ | 110.90891 | 13.86361 | -0.85217 | 3.72533 | -0.8422 | -0.83132 | -0.80578 | -0.75075 |
| $Amm2$-$Sr_9H_{15}$ | 120.9014 | 13.43349 | -0.92563 | 5.23784 | -0.88939 | -0.88118 | -0.86231 | -0.83391 |
| $P6/mmm$-$SrH_2$ | 12.6907 | 12.6907 | -1.00892 | 0.58811 | -1.0044 | -0.9902 | -0.95592 | -0.87885 |
| $Cmcm$-$Sr_2H_8$ | 21.84666 | 10.92333 | -0.73756 | 2.28638 | -0.73088 | -0.72308 | -0.70739 | -0.68601 |
| $C2/m$-$Sr_6H_{26}$ | 66.48349 | 11.08058 | -0.62741 | 7.74212 | -0.61028 | -0.60217 | -0.5851 | -0.56494 |
| $P1$-$SrH_6$ | 75.30214 | 9.41277 | -0.58458 | 13.01941 | -0.58709 | -0.58009 | -0.5649 | -0.5446 |
| $R$-$3m$-$SrH_6$ | 9.51012 | 9.51012 | -0.57067 | 1.66278 | -0.56813 | -0.55817 | -0.54044 | -0.52378 |
| $P4_2/mnm$-$Sr_2H_{12}$ | 19.61206 | 9.80603 | -0.5284 | 3.30839 | -0.52709 | -0.52036 | -0.50525 | -0.48666 |
| $P1$-$SrH_9$ | 31.75144 | 7.93786 | -0.39075 | 10.27908 | -0.37856 | -0.37315 | -0.36203 | -0.35171 |
| $C2/m$-$Sr_2H_{20}$ | 13.93573 | 6.96787 | -0.39312 | 5.38182 | -0.39535 | -0.39026 | -0.3778 | -0.35905 |
| $P1$-$Sr_2H_{34}$ | 5.77364 | 2.88682 | -0.25185 | 9.10114 | -0.25397 | -0.25037 | -0.2446 | -0.24348 |
| $P1$-$SrH_{22}$ | -0.00129 | -0.00129 | -0.20242 | 6.34131 | 150 | -0.18503 | -0.18798 | -0.20438 |
| $C2/c$-H | -13.27573 | -0.55316 | 0 | 6.42274 | 0 | 0 | 0 | 0 |

To determine the crystal structure of three phases with expected stoichiometries $SrH_5$, $SrH_6$, and $SrH_9$, we ran the USPEX code with a multiobjective optimization of stability and agreement with a given experimental XRD pattern. In these calculations, USPEX reads an XRD spectrum file that contains the minimum and maximum diffraction angles 2θ of the pattern, the wavelength of the X-ray radiation in Å, the parameter *match_tol* (the maximum distance in degrees at which the calculated and experimental peaks are considered matching), and all major peaks of the pattern in the form of "angle–intensity" doublets on separate lines (the peak intensities are automatically normalized to 100). The disagreement (fitness) $F$ between the given diffraction pattern and the calculated pattern of each structure is expressed as

$$F = \sum_{i,j}^{match} \left( \frac{h_i^{exp} - h_j^{th}}{100} \right)^2 \left( \frac{h_i^{exp}}{100} \right)^2 + \sum_{i}^{exp\ rest} \left( \frac{h_i^{exp}}{100} \right)^2 + \sum_{i}^{th\ rest} \left( \frac{h_i^{th}}{100} \right)^2 \tag{S13}$$

where $h^{exp}$ and $h^{th}$ are the intensities of the peaks in the experimental and calculated XRD patterns, respectively. The first sum in eq. S13 runs over the matching peaks, the second sum runs over the unmatched experimental peaks, and the third sum runs over the unmatched calculated peaks. In our calculations, we used *match_tol* = 0.3° because the experimental peaks are quite broad. As a rule of thumb, this parameter should be equal to the half of the full width at half maximum (FWHM/2) of the broadest peak in the experimental XRD pattern.

For $SrH_5$, we did a single USPEX calculation with the stoichiometry $Sr_4H_{20}$ at 150 GPa. In this and the following USPEX runs, the following parameters were used: a population size of 60 structures, a maximum number of 250 generations, and a stopping criterion of the evolutionary algorithm, which was always set equal to the total number of atoms in the unit cell. For $Sr_4H_{20}$, we got a single good match with the experimental spectrum (Figure S3) that lies 0.05 eV/atom above the best structure found during this run.

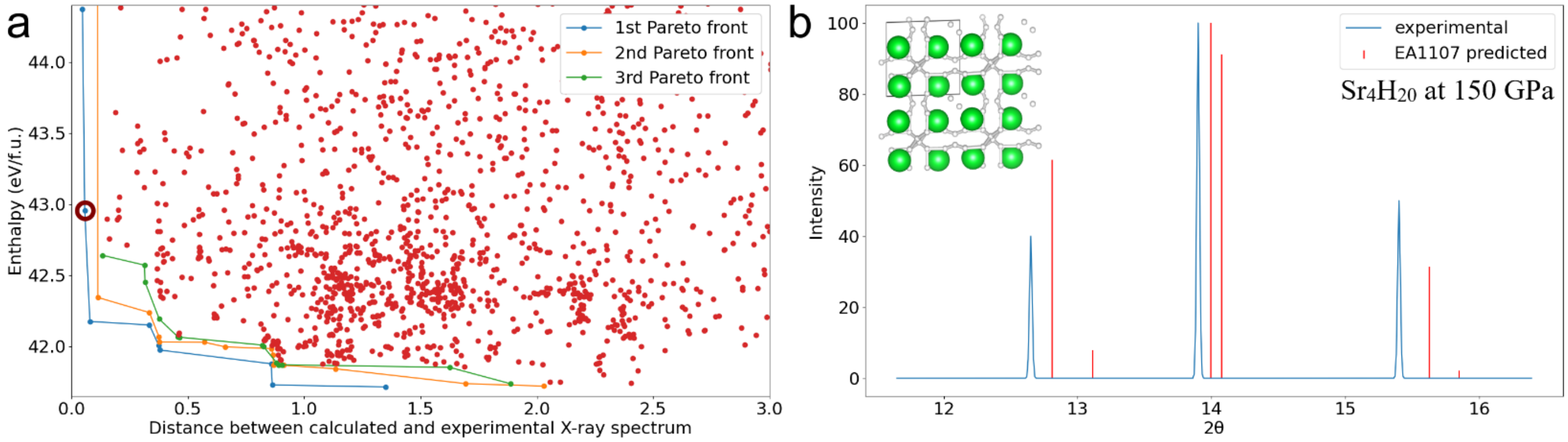

**Figure S3.** Crystal structure search of $Sr_4H_{20}$ at 150 GPa with a multiobjective optimization. (a) Enthalpy–fitness Pareto diagram with three first Pareto fronts. The circle marks one of the good matches with the experimental spectrum. (b) Comparison of the experimental (153 GPa, λ = 0.62 Å, see main text Figure 2e) and theoretical XRD patterns of the $Sr_4H_{20}$ structure circled in panel (a). The structure, which is lower in energy (42.2 eV/f.u.) than the one highlighted by the circle, has a XRD pattern without reflection at 15.5°.

For $SrH_6$, we did three USPEX calculations with the stoichiometries $Sr_2H_{12}$, $Sr_4H_{23}$ (or $SrH_{5.75}$), and $Sr_4H_{24}$ at 100 GPa. During the first run, three good matches were found, all lying less than 0.05 eV/atom above the most stable structure found with USPEX (Figure S4). The second run did not bring any good matches, and the enthalpy–fitness plot shows that the structures tend to group around two different minima of the potential energy surface. The third run resulted in several good matches (Figure S5).

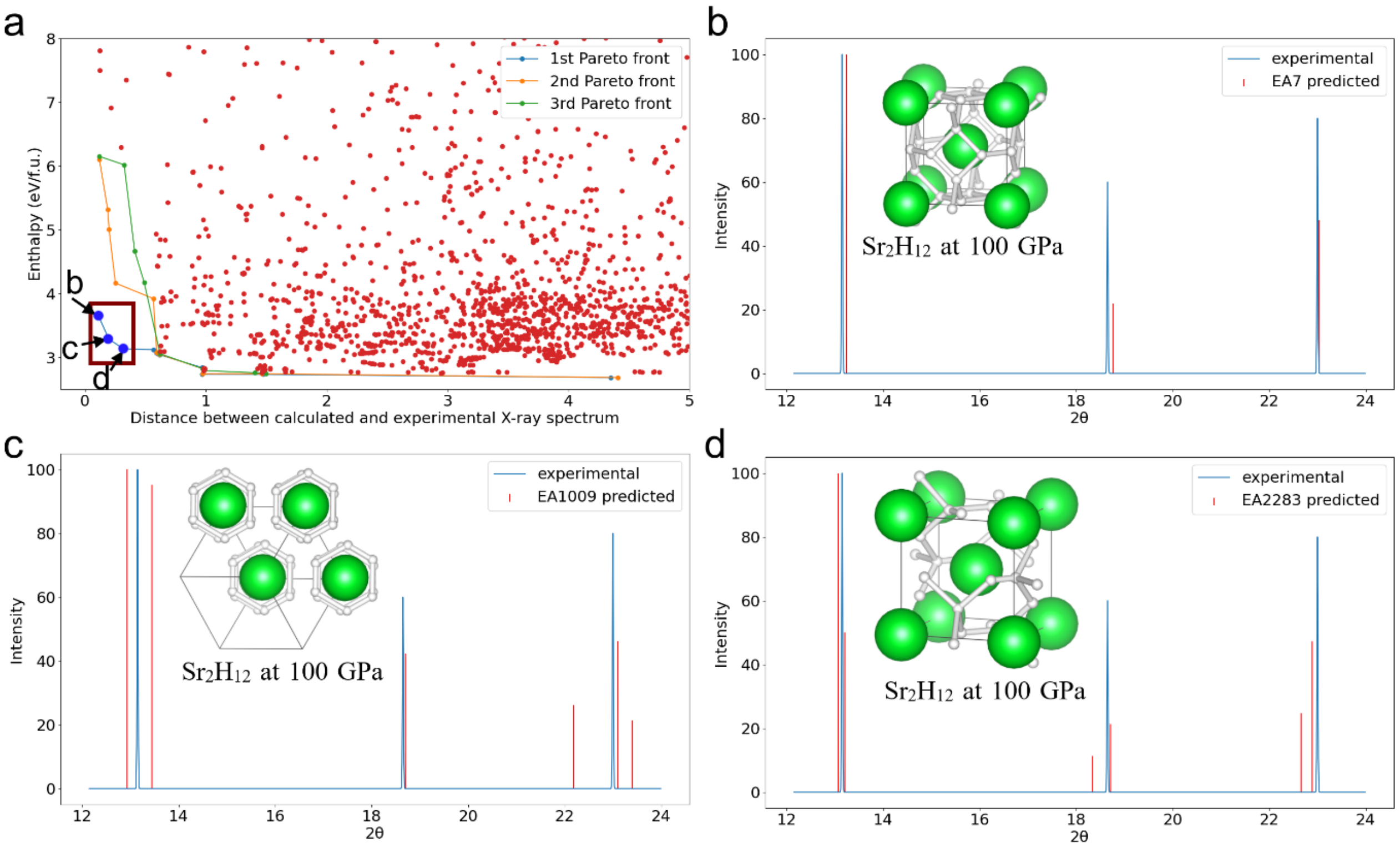

**Figure S4.** Crystal structure search of $Sr_2H_{12}$ at 100 GPa with a multiobjective optimization. (a) Enthalpy–fitness Pareto diagram with three first Pareto fronts. Good structures are marked with a rectangle. (b, c, d) Comparison of the experimental (100 GPa, λ = 0.62 Å, see main text Figure 3e) and theoretical XRD patterns of three best structures of $Sr_2H_{12}$ found in the search. Structure (b) is almost ideal $Im\bar{3}m$-$SrH_6$, whereas structure (d) can be symmetrized to $P4_2/mnm$-$SrH_6$.

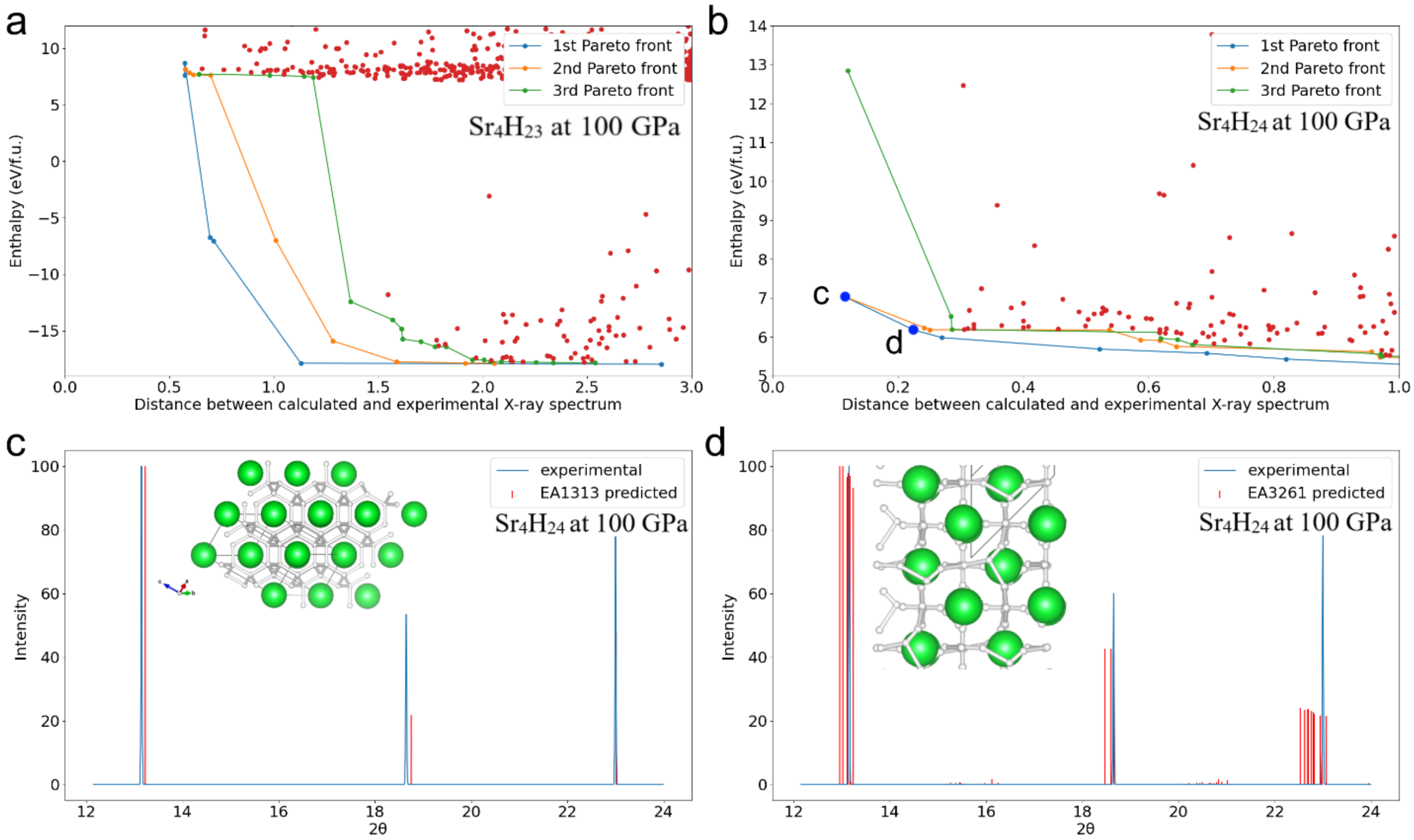

**Figure S5.** Crystal structure search of $Sr_4H_{23}$ and $Sr_4H_{24}$ at 100 GPa with a multiobjective optimization. (a) Enthalpy–fitness Pareto diagram with three first Pareto fronts for $Sr_4H_{23}$. No good candidates were found. (b) Enthalpy–fitness Pareto diagram with three first Pareto fronts for $Sr_4H_{24}$. Two good structures are indicated by blue circles. (c, d) Comparison of the experimental (100 GPa, λ = 0.62 Å, see main text Figure 3e) and theoretical XRD patterns for two best structures of $Sr_4H_{24}$ found in the search. Structure (c) is almost ideal $Im\bar{3}m$-$SrH_6$.

For $SrH_9$, we did two USPEX runs with the stoichiometries $Sr_4H_{36}$ at 100 GPa and $Sr_4H_{35}$ (or $SrH_{8.75}$). In both cases, several good matches were found (Figure S6).

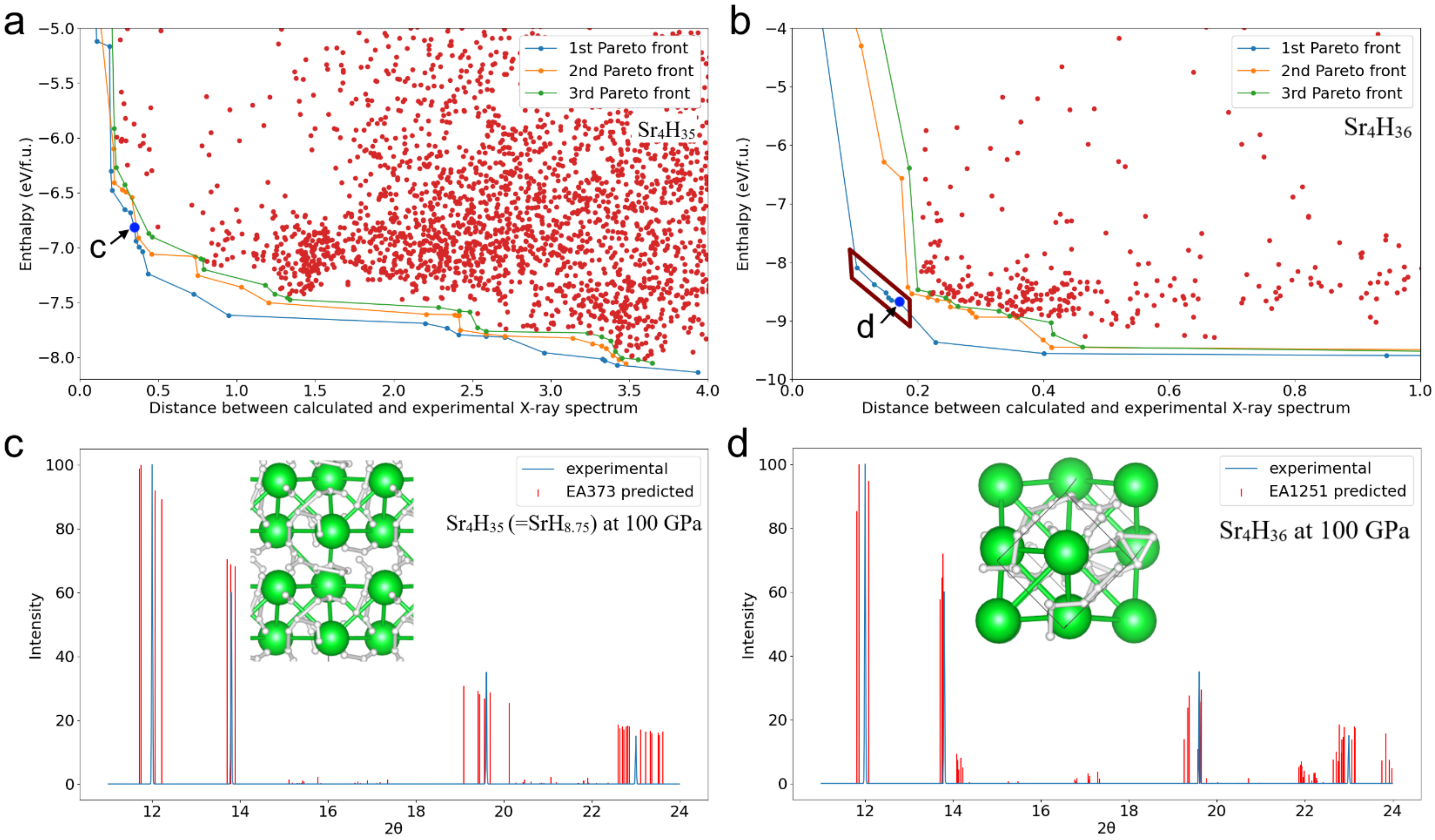

**Figure S6.** Crystal structure search of $Sr_4H_{35}$ and $Sr_4H_{36}$ at 100 GPa with a multiobjective optimization. Enthalpy–fitness Pareto diagram with three first Pareto fronts for (a) $Sr_4H_{35}$ and (b) $Sr_4H_{36}$. Good structures are indicated by blue circles. (c, d) Comparison of the experimental (101 GPa, $\lambda = 0.62$ Å, see main text Figure 3a) and theoretical XRD patterns for two best structures of $Sr_4H_{35}$ and $Sr_4H_{36}$ found in the search. Structure (d) is $F\bar{4}3m$-like $P1$-$SrH_9$ described in the main text.

## Description of individual Sr–H phases

## 1. $SrH_2$

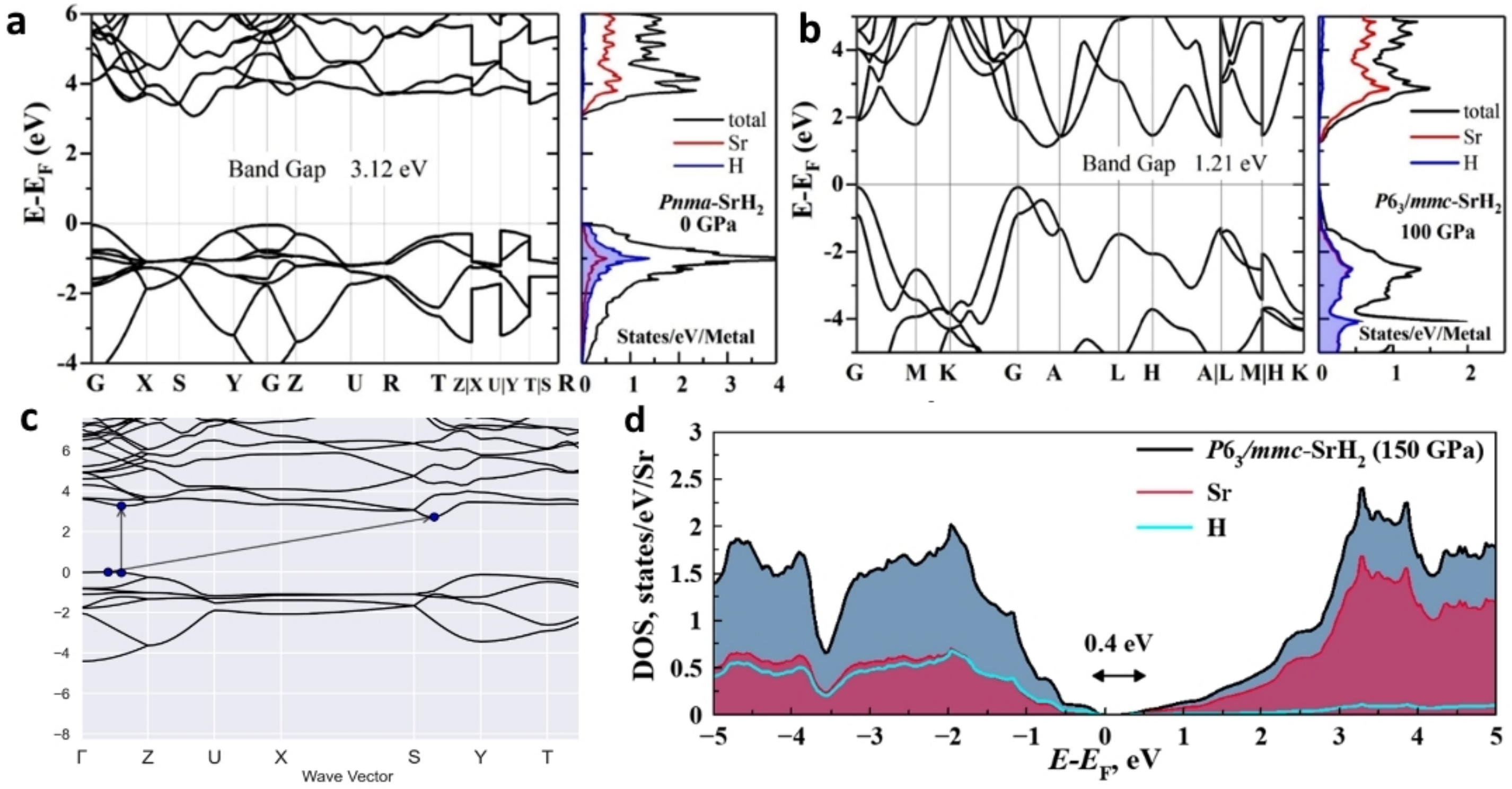

**Figure S7.** Band structure and the density of electron states for $SrH_2$. (a) Band structure and the density of electron states of *Pnma*-$SrH_2$ at 0 GPa calculated using the GGA–PBE exchange–correlation functional (VASP code). The bandgap is 3.12 eV (indirect). (b) Band structure and the density of electron states of *P*6$_3$/*mmc*-$SrH_2$ at 100 GPa calculated using the GGA–PBE functional (VASP code). The bandgap is 1.21 eV (indirect). (c) Band structure of *Pnma*-$SrH_2$ at 0 GPa calculated using the TB09 exchange–correlation functional (Abinit code). The bandgap is 2.72 eV (indirect). (d) Density of electron states of *P*6$_3$/*mmc*-$SrH_2$ at 150 GPa calculated using the GGA–PBE functional (VASP code). The bandgap decreases to ~0.4 eV.

## 2. *C*2/*m*-$Sr_3H_{13}$

**Table S7.** Experimental and calculated unit cell parameters of *C*2/*m*-$Sr_3H_{13}$ synthesized in DACs Sr2 and Sr3.

| Pressure, GPa | *a*, Å | *b*, Å | *c*, Å | *V*, Å$^3$ |
|---|---|---|---|---|
| DAC Sr2 | | | | |
| 146 | 6.25 | 2.98 | 7.92 | 23.07 |
| 142 | 6.25 | 2.96 | 7.955 | 23.02 |
| 138 | 6.284 | 2.9819 | 7.9797 | 23.38 |
| 134 | 6.2982 | 2.9886 | 7.9881 | 23.52 |
| 129 | 6.3134 | 2.9953 | 8.0136 | 23.70 |
| 125 | 6.465 | 3.0548 | 8.128 | 24.20 |
| 110 | 6.379 | 3.03 | 8.11 | 24.52 |
| 99 | 6.4407 | 3.05 | 8.18 | 25.26 |
| 87 | 6.47 | 3.08 | 8.19 | 25.53 |
| 78 | 6.62 | 3.098 | 8.25 | 26.47 |
| 70 | 6.67 | 3.1 | 8.34 | 26.96 |
| DAC Sr3 | | | | |
| 117 | 6.39 | 3.015 | 8.15 | 24.57 |
| 113 | 6.45 | 3.035 | 8.02 | 24.55 |
| 110 | 6.40146 | 3.0435 | 8.14436 | 24.82 |
| 107 | 6.44 | 3.042 | 8.15 | 24.97 |
| 103 | 6.46 | 3.044 | 8.165 | 25.12 |
| 96 | 6.5 | 3.075 | 8.3 | 25.95 |
| 90 | 6.52 | 3.085 | 8.4237 | 26.50 |
| 83 | 6.48 | 3.1048 | 8.256 | 26.10 |
| 74 | 6.75 | 3.15 | 8.25 | 27.40 |
| Theory | | | | |
| 160 | 6.456 | 2.833 | 7.331 | 21.74 |
| 140 | 6.487 | 2.897 | 7.465 | 22.76 |
| 120 | 6.518 | 2.974 | 7.63 | 23.98 |
| 100 | 6.559 | 3.06 | 7.814 | 25.44 |
| 80 | 6.604 | 3.162 | 8.035 | 27.26 |
| 60 | 6.663 | 3.286 | 8.308 | 29.67 |
| 40 | 6.747 | 3.445 | 8.665 | 33.07 |

**Table S8.** Crystal structure of the discovered *C*2/*m*-$Sr_3H_{13}$ phase at 100 GPa.

| Phase | Pressure, GPa | Lattice parameters | Coordinates | | | |
|---|---|---|---|---|---|---|
| *C*2/*m*-$Sr_3H_{13}$ | 100 | *a* = 6.559 Å | Sr1(4i) | -0.34408 | 0.0 | 0.16684 |
| | | *b* = 3.060 Å | Sr2(2c) | 0.0 | 0.0 | 0.5 |
| | | *c* = 7.814 Å | H1(4i) | 0.02193 | 0.0 | -0.2372 |
| | | α = 90° | H2(4i) | 0.05862 | 0.0 | -0.11121 |
| | | β = 103.3° | H3(4i) | -0.3624 | 0.0 | -0.29667 |
| | | γ = 90° | H4(4i) | -0.33268 | 0.0 | -0.10272 |
| | | | H5(4i) | -0.32295 | 0.0 | 0.43516 |
| | | | H6(4i) | -0.28826 | 0.0 | -0.38035 |
| | | | H7(2a) | 0.0 | 0.0 | 0.0 |

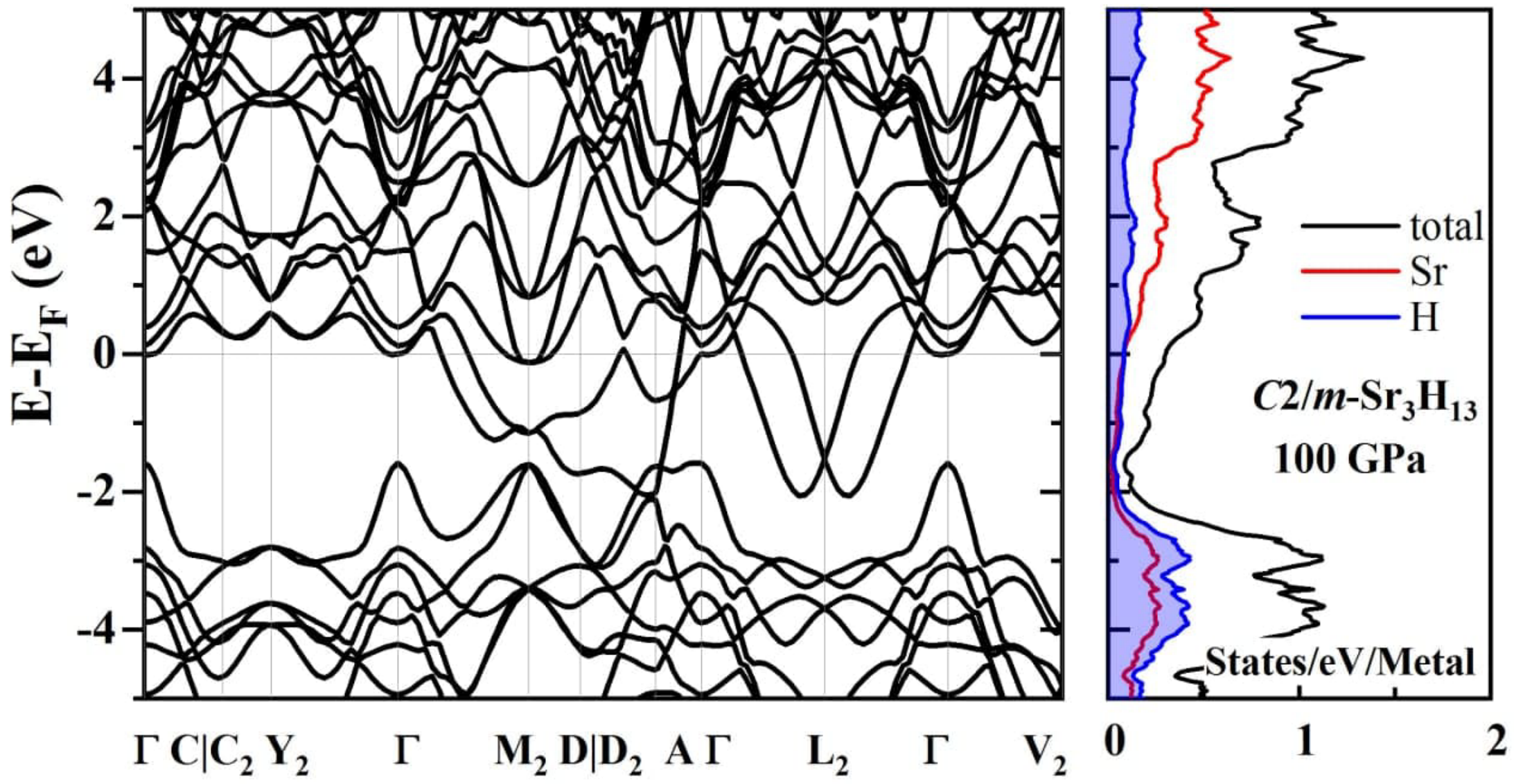

**Figure S8.** Band structure and the density of electron states of $C2/m$-$Sr_3H_{13}$ at 100 GPa calculated using the PBE–GGA exchange–correlation functional (VASP code).

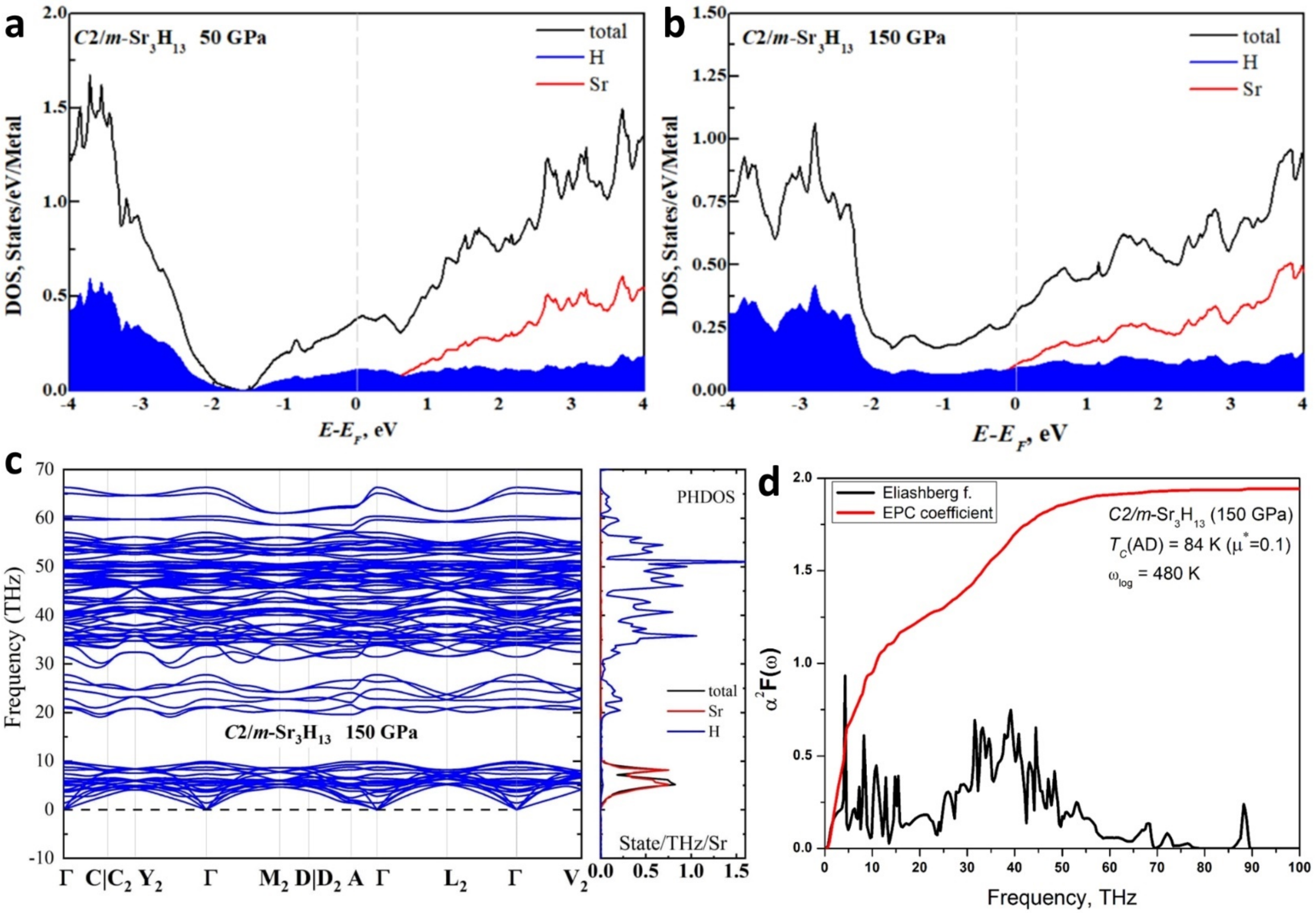

**Figure S9.** Density of electron states of $C2/m$-$Sr_3H_{13}$ at (a) 50 and (b) 150 GPa calculated using the PBE GGA exchange–correlation functional (VASP code). The contributions from hydrogen and strontium are approximately equal at the Fermi level. (c) Phonon band structure and the density of states of $C2/m$-$Sr_3H_{13}$ at 150 GPa calculated within the harmonic approximation. (d) Ab initio calculated harmonic Eliashberg function $\alpha^2F(\omega)$ and the electron–phonon coupling (EPC) parameter at 150 GPa for $C2/m$-$Sr_3H_{13}$.

## 3. $P1$-$SrH_{22}$

**Table S9.** Experimental and calculated unit cell parameters of pseudocubic $P1$-$SrH_{22}$ synthesized in DACs Sr1 and Sr2.

| Pressure, GPa | *a*, Å | *b*, Å | *c*, Å | *V*, Å$^3$ |
|---|---|---|---|---|
| DAC Sr2 | | | | |
| 146 | 4.4157 | 4.4045 | 3.348 | 56.6 |
| 142 | 4.4585 | 4.4371 | 3.3308 | 57.3 |
| 138 | 4.451 | 4.4337 | 3.3784 | 58.0 |
| 134 | 4.4603 | 4.4459 | 3.3858 | 58.4 |
| 129 | 4.4756 | 4.4576 | 3.3923 | 58.9 |
| 125 | 4.52 | 4.46 | 3.3965 | 59.4 |
| 110 | 4.583 | 4.5462 | 3.4525 | 62.5 |
| 99 | 4.627 | 4.611 | 3.5156 | 65.1 |
| 87 | 4.7393 | 4.6496 | 3.5498 | 67.7 |
| 78 | 4.8051 | 4.6919 | 3.6017 | 70.2 |
| 70 | 4.872 | 4.758 | 3.627 | 72.8 |
| DAC Sr1 | | | | |
| 138 | 4.457 | 4.437 | 3.369 | 58.15 |
| Theory | | | | |
| 200 | 4.269 | 4.243 | 3.215 | 50.438 |
| 180 | 4.33 | 4.285 | 3.256 | 52.4921 |
| 160 | 4.396 | 4.345 | 3.302 | 54.8482 |
| 140 | 4.472 | 4.409 | 3.357 | 57.5752 |
| 120 | 4.561 | 4.485 | 3.419 | 60.7747 |
| 90 | 4.724 | 4.631 | 3.531 | 67.0454 |
| 70 | 4.872 | 4.758 | 3.627 | 72.8096 |

**Table S10.** Crystal structure of the discovered pseudocubic $P1$-$SrH_{22}$ at 100 GPa.

| Phase | Pressure, GPa | Lattice parameters | Coordinates | | | |
|---|---|---|---|---|---|---|
| $P1$-$SrH_{22}$ | 100 | *a* = 4.663Å<br>*b* = 4.578 Å<br>*c* = 3.490 Å<br>α = 68.38°<br>β = 69.11°<br>γ = 82.29° | Sr1(4a) | -0.33906 | -0.16093 | 0.43474 |
| | | | H1(4a) | 0.42825 | -0.03638 | -0.01041 |
| | | | H2(4a) | 0.4416 | 0.13656 | -0.10975 |
| | | | H3(4a) | 0.01787 | 0.19755 | 0.21107 |
| | | | H4(4a) | 0.20074 | -0.23729 | 0.41284 |
| | | | H5(4a) | -0.26366 | 0.27274 | -0.12205 |
| | | | H6(4a) | 0.21803 | -0.24499 | -0.01764 |
| | | | H7(4a) | 0.03253 | 0.19286 | -0.00745 |
| | | | H8(4a) | -0.38882 | 0.38617 | 0.38085 |
| | | | H9(4a) | 0.25424 | 0.15117 | -0.4687 |
| | | | H10(4a) | 0.03201 | 0.49453 | 0.31907 |
| | | | H11(4a) | 0.34898 | 0.26444 | 0.3225 |
| | | | H12(4a) | 0.12802 | -0.08907 | 0.44684 |
| | | | H13(4a) | -0.27421 | 0.28439 | 0.48285 |
| | | | H14(4a) | 0.31581 | 0.47923 | -0.35829 |
| | | | H15(4a) | 0.00551 | 0.46212 | -0.44802 |
| | | | H16(4a) | 0.12502 | -0.1001 | -0.10473 |
| | | | H17(4a) | -0.12189 | 0.05326 | -0.27026 |
| | | | H18(4a) | -0.38648 | 0.38415 | -0.02934 |
| | | | H19(4a) | -0.12661 | -0.45177 | -0.02902 |
| | | | H20(4a) | -0.09512 | -0.27865 | -0.12742 |
| | | | H21(4a) | 0.45099 | -0.38208 | 0.1379 |
| | | | H22(4a) | 0.30301 | 0.47229 | -0.13433 |

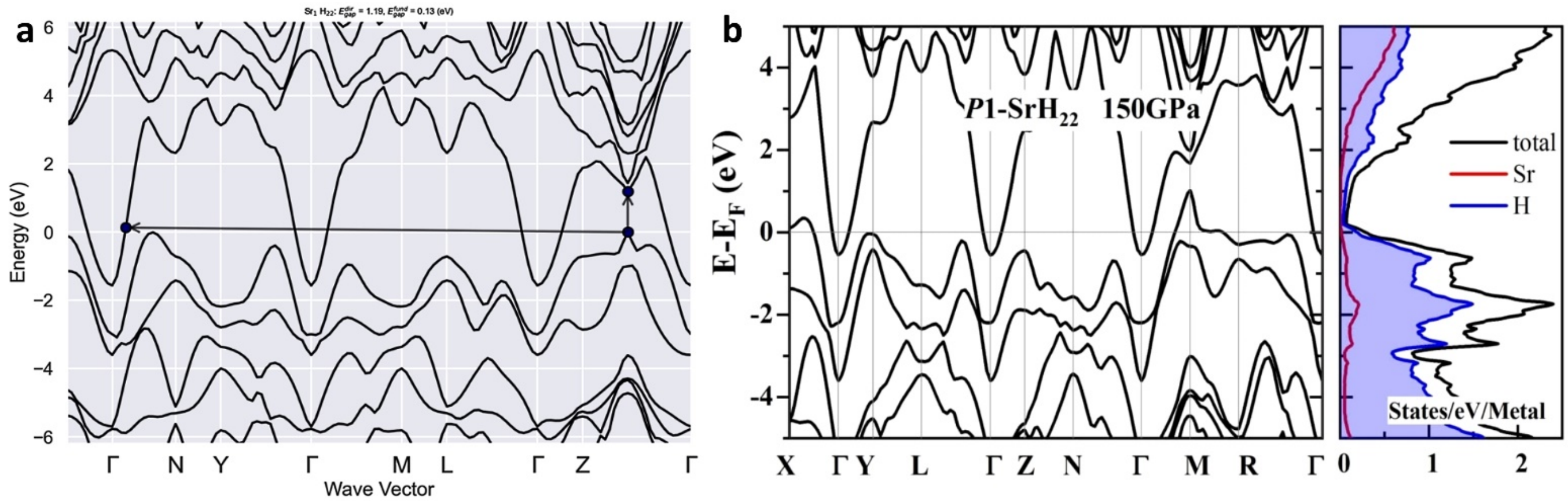

**Figure S10.** Band structure and the density of electron states of $P1$-$SrH_{22}$: (a) at 120 GPa, calculated using the LDA norm conserving (NC) exchange–correlation functional, Abinit code; (b) at 150 GPa, calculated using the PBE-GGA, VASP code.

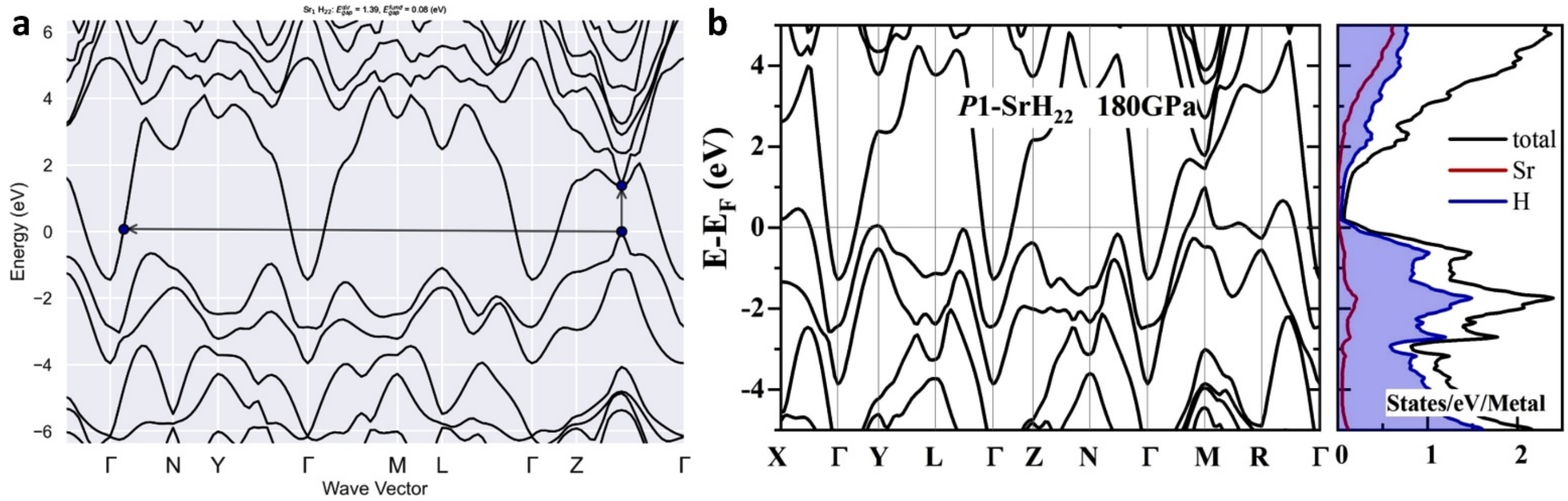

**Figure S11.** Band structure and the density of electron states of $P1$-$SrH_{22}$: (a) at 120 GPa, calculated using the PBE–GGA exchange–correlation functional, Abinit code; (b) at 180 GPa, calculated using the PBE-GGA, VASP code.

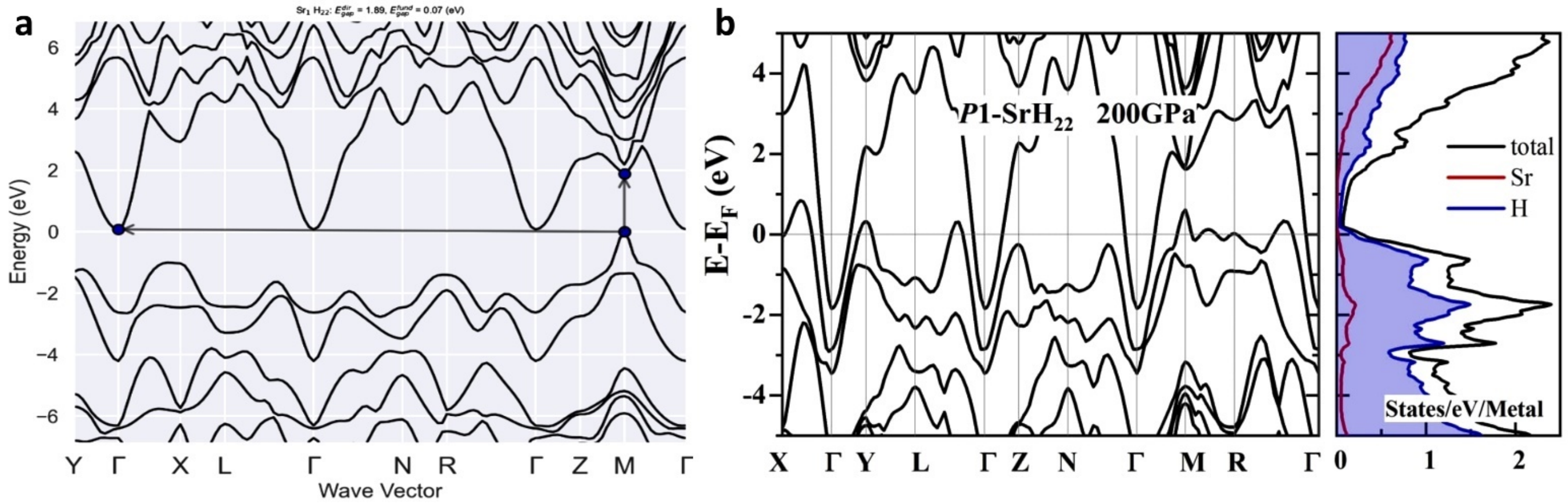

**Figure S12.** Band structure and the density of electron states of $P1$-$SrH_{22}$: (a) at 120 GPa, calculated using the TB09-HGH exchange–correlation functional, Abinit code; (b) at 200 GPa, calculated using the PBE-GGA, VASP code.

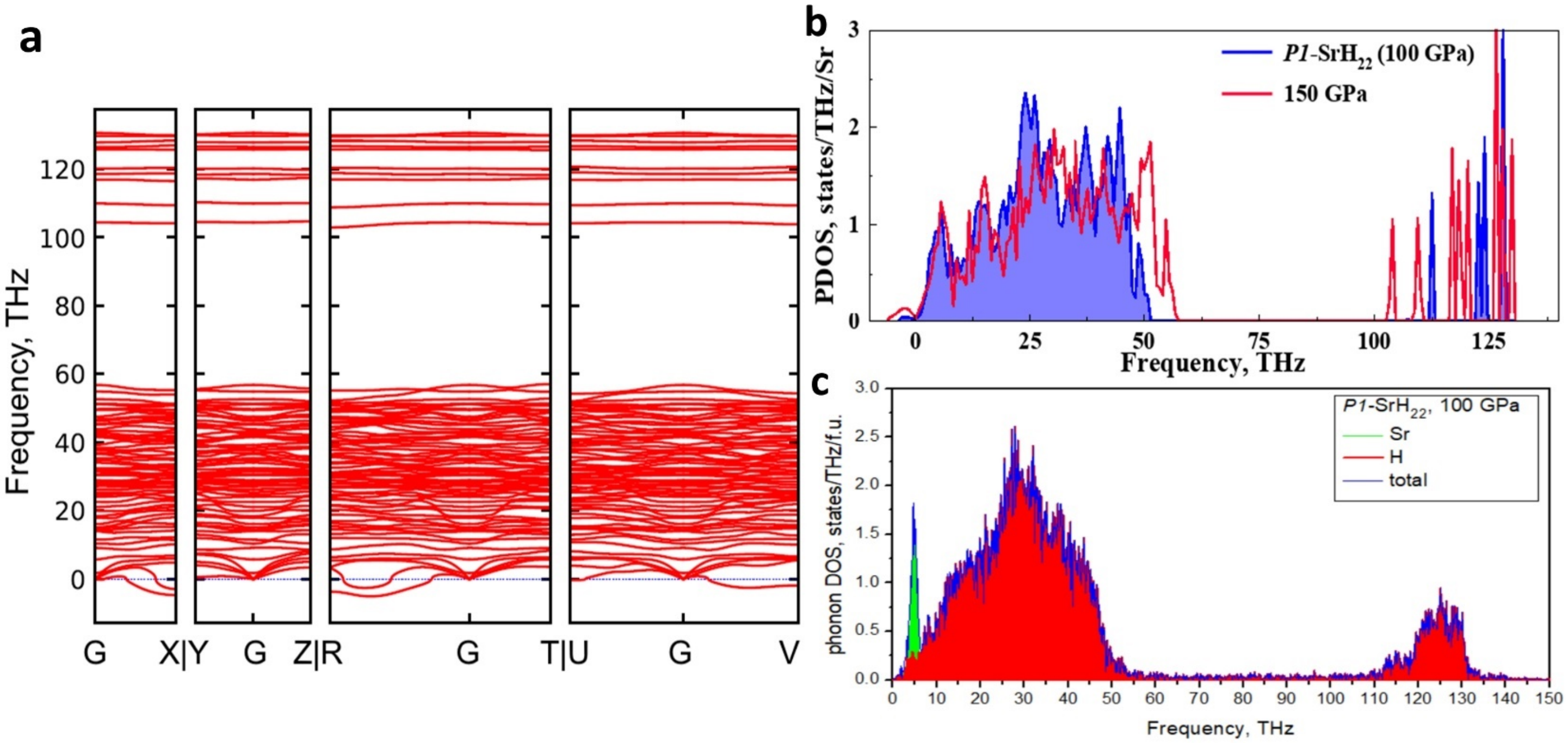

**Figure S13.** Phonon spectra of $P1$-$SrH_{22}$ at different pressures. (a) Harmonic phonon band structure of $P1$-$SrH_{22}$ at 150 GPa (VASP code). The phonon branches show very limited dispersion. (b) Phonon density of states of $P1$-$SrH_{22}$ calculated using the DFPT PBE at 100 and 150 GPa within the harmonic approximation (VASP). There are several imaginary acoustic modes. (c) Anharmonic phonon density of states of $P1$-$SrH_{22}$ calculated at 100 GPa using molecular dynamics with the MTP and MLIP at 300 K.

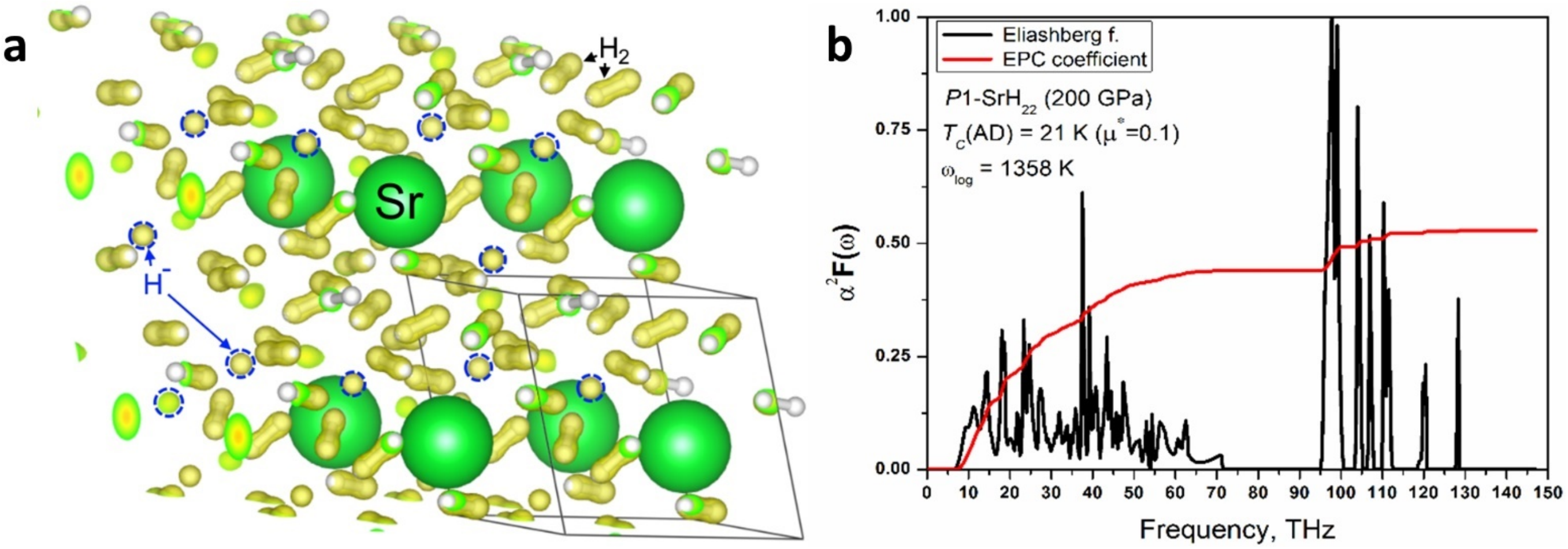

**Figure S14**. (a) Electron density at the isovalue of 0.01 of $P1$-$SrH_{22}$ at 120 GPa. The electron density was calculated using the LDA functional. Hydrogen mostly forms slightly charged (~ –0.1|$e$|) $H_2$ molecules with a bond length $d$(H–H) = 0.8 Å, and the charge density ρ at the bond critical point is 2.8|$e$|/bohr$^3$. The hydrogen molecules form stronger bonds with the Sr atoms with ρ = 4.7|$e$|/bohr$^3$. (b) Ab initio-calculated harmonic Eliashberg functions $\alpha^2F(\omega)$ and the electron–phonon coupling (EPC) parameter at 200 GPa for $P1$-$SrH_{22}$.

**Table S11.** Bader charge analysis of $P1$-$SrH_{22}$ at 120 GPa.

| Element | Bader charge | Coordinate X | Coordinate Y | Coordinate Z |
|---|---|---|---|---|
| Sr | 1.2383 | 0.5766593 | 0.1630336 | 0.3344873 |
| H | -0.0342 | 0.0143542 | 0.0478963 | 0.5734584 |
| H | -0.0581 | 0.0120759 | 0.7964527 | 0.9605628 |
| H | -0.0311 | 0.0147627 | 0.2304206 | 0.8058419 |
| H | -0.0905 | 0.0305188 | 0.6335676 | 0.373611 |
| H | -0.0174 | 0.034599 | 0.455148 | 0.1262341 |
| H | -0.0274 | 0.1208795 | 0.0742438 | 0.8817379 |
| H | -0.0178 | 0.1184602 | 0.8709192 | 0.5427404 |
| H | 0.0117 | 0.1491269 | 0.7362817 | 0.2457107 |
| H | -0.0344 | 0.138188 | 0.2781709 | 0.0955161 |
| H | -0.0080 | 0.0791917 | 0.5138391 | 0.6712897 |
| H | **-0.3276** | 0.2959499 | 0.9461376 | 0.1192135 |
| H | -0.0569 | 0.2715705 | 0.5296145 | 0.7084117 |
| H | -0.0150 | 0.467839 | 0.5280785 | 0.9656285 |
| H | -0.0155 | 0.4640251 | 0.7979888 | 0.703346 |
| H | -0.0888 | 0.5401766 | 0.6924997 | 0.2953635 |
| H | -0.0322 | 0.5535044 | 0.0956467 | 0.8631326 |
| H | -0.0243 | 0.5793404 | 0.2518234 | 0.7872366 |
| H | 0.0119 | 0.6335984 | 0.5897855 | 0.4232638 |
| H | -0.0247 | 0.677696 | 0.5213226 | 0.9842153 |
| H | -0.0238 | 0.685713 | 0.8047447 | 0.6847592 |
| H | -0.0057 | 0.7667996 | 0.8122281 | 0.9976848 |
| H | **-0.3285** | 0.8637802 | 0.3799296 | 0.549761 |
| Total | 0.0000 | | | |

## 4. Possible formation of barium polyhydrides $BaH_{21–23}$

Studying the Ba–H system,[28] we have found that $BaH_{12}$ is not the highest hydride lying on the convex hull: superhydride *C*2/*m*-$BaH_{23}$ with a pseudotetragonal (*I*4/*mmm*) structure is also thermodynamically stable at 150 GPa. In that research, we have also predicted the stability of $BaH_{21}$ and $BaH_{22}$ at lower pressures, though it was difficult to imagine compounds with such a high hydrogen content. The retrospective analysis shows that the previously unidentified reflections at 8.5–9° and 12.5–12.8° (Figure S15) observed in the diffraction patterns of $BaH_{12}$ at 119–160 GPa may be explained by the presence of an impurity of thermodynamically stable molecular superhydride *C*2/*m*-$BaH_{21–23}$ (Table S12) similar to *P*1-$SrH_{22}$ we discovered in the Sr–H system. Therefore, there is a high probability that superhydrides with extremely high hydrogen content, $XH_{21–23}$, exist in both Ba–H and Sr–H systems at 100–150 GPa.

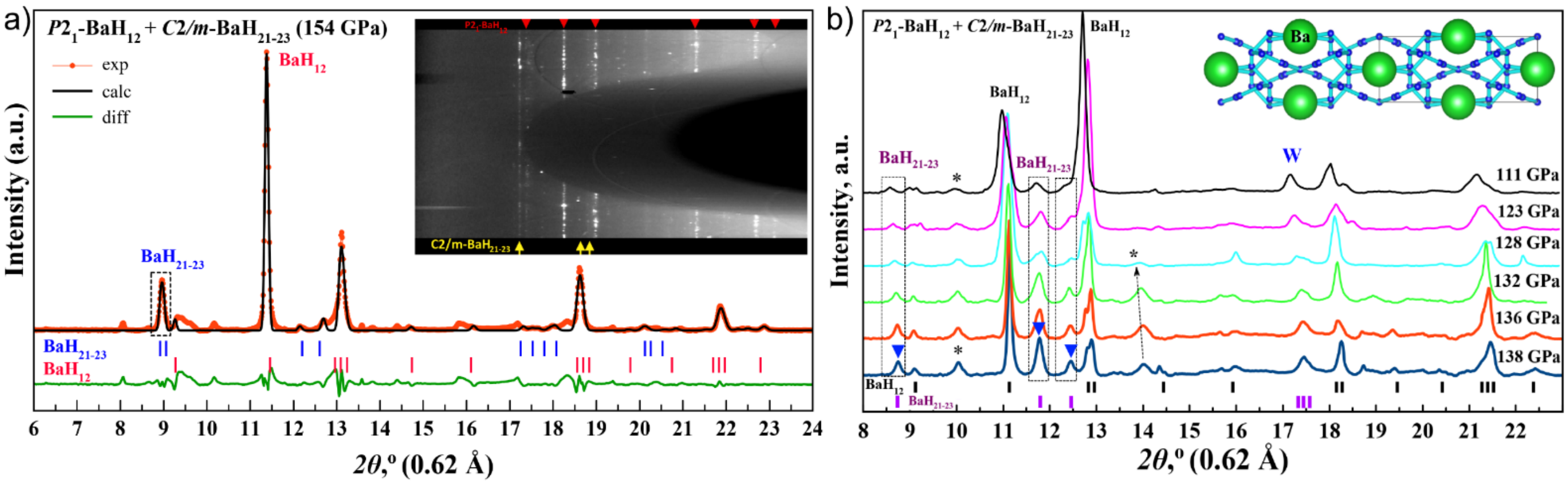

**Figure S15.** Experimental X-ray diffraction patterns from (a) DAC B1 and (b) DAC B2[28] and the Le Bail refinement of the pseudotetragonal phase *C*2/*m*-$BaH_{21–23}$. Low content of this phase in the sample does not permit to determine the exact Ba:H ratio. Unidentified reflections are indicated by asterisks. Insets show the diffraction image ("cake") and projection of the *C*2/*m*-$BaH_{23}$ structure to the (*ab*) plane. The hydrogen network is shown by light blue lines.

**Table S12.** Experimental parameters of the unit cell of the previously unknown barium superhydride *C*2/*m*-$BaH_{21–23}$.

| Crystal structure | Pressure, GPa | *a*, Å | *b*, Å | *c*, Å | β, ° | Cell volume, $Å^3$/Ba atom |
|---|---|---|---|---|---|---|
| 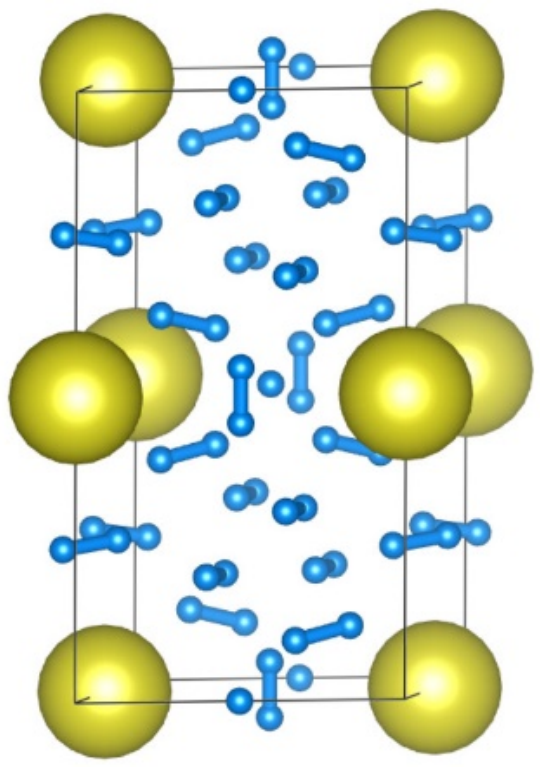 | 126 | 3.583 | 8.060 | 4.399 | 67.74 | 58.80 |
| | 132 | 3.502 | 8.053 | 4.390 | 67.84 | 57.35 |
| | 135 | 3.522 | 7.996 | 4.367 | 67.57 | 56.85 |
| | 145 | 3.538 | 7.953 | 4.338 | 67.70 | 56.49 |
| | 154 | 3.471 | 7.904 | 4.304 | 67.60 | 54.59 |
| | 160 | 3.473 | 7.886 | 4.294 | 67.60 | 54.38 |

## 5. $P1$-$SrH_6$ ($Sr_8H_{48}$) and proposed $Pm\bar{3}n$-$Sr_8H_{46}$

**Table S13.** Experimental and calculated unit cell parameters of $Pm\bar{3}n$-like pseudocubic $P1$-$SrH_6$ synthesized in DACs Sr3, Sr165, E1, and Sr4.

| Pressure, GPa | $a$, Å (as $Pm\bar{3}n$) | $V$, Å$^3$ per Sr atom (as $Pm\bar{3}n$) |
|---|---|---|
| DAC Sr3 | | |
| 117 | 5.989 | 26.86 |
| 113 | 5.999 | 27.0 |
| 110 | 5.994 | 26.925 |
| 107 | 6.030 | 27.4125 |
| 103 | 6.050 | 27.6875 |
| 96 | 6.102 | 28.41 |
| 90 | 6.129 | 28.7875 |
| 83 | 6.103 | 28.425 |
| 74 | 6.265 | 30.75 |
| DAC Sr165 | | |
| 181.0 | 5.677 | 22.875 |
| 177.0 | 5.690 | 22.9125 |
| 171.0 | 5.688 | 23.0125 |
| 166.0 | 5.735 | 23.5875 |
| 157.0 | 5.766 | 23.975 |
| 153.0 | 5.707 | 24.35 |
| 148.0 | 5.823 | 24.675 |
| 146.0 | 5.828 | 24.75 |
| 135.0 | 5.860 | 25.15 |
| 131.0 | 5.888 | 25.5125 |
| 95.0 | 6.076 | 28.05 |
| 78.0 | 6.170 | 29.3625 |
| 56.0 | 6.350 | 32.0 |
| DAC E1 | | |
| 122 | 5.970 | 26.6 |
| DAC Sr4 | | |
| 157 | 5.796 | 24.35 |
| Theory (as $Pm\bar{3}n$) | | |
| 50 | 6.533 | 34.860 |
| 70 | 6.328 | 31.687 |
| 90 | 6.173 | 29.412 |
| 100 | 6.107 | 28.480 |
| 110 | 6.047 | 27.647 |
| 130 | 5.942 | 26.226 |
| 150 | 5.852 | 25.052 |

**Table S14.** Pressure dependence of the calculated unit cell parameters of $Pm\bar{3}n$-like pseudocubic $P1$-$SrH_6$ ($Sr_8H_{48}$, $Z = 8$).

| Pressure, GPa | $a$, Å | $b$, Å | $c$, Å | α, ° | β, ° | γ, ° | $V$, Å$^3$ |
|---|---|---|---|---|---|---|---|
| 50 | 6.529 | 6.514 | 6.558 | 90.85 | 89.96 | 90.25 | 278.88 |
| 70 | 6.312 | 6.307 | 6.368 | 90.73 | 90.14 | 90.31 | 253.5 |
| 90 | 6.152 | 6.148 | 6.222 | 90.57 | 90.19 | 90.31 | 235.3 |
| 100 | 6.084 | 6.080 | 6.160 | 90.48 | 90.21 | 90.30 | 227.84 |
| 110 | 6.024 | 6.016 | 6.103 | 90.40 | 90.23 | 90.25 | 221.18 |
| 130 | 5.912 | 5.912 | 6.003 | 90.25 | 90.25 | 90.52 | 209.81 |
| 150 | 5.821 | 5.821 | 5.915 | 90.13 | 90.14 | 90.28 | 200.42 |

**Table S15.** Crystal structure of discovered $Pm\bar{3}n$-like pseudocubic $P1$-$SrH_6$ at 100 GPa.

| Phase | Pressure, GPa | Lattice | Coordinates | | | |
|---|---|---|---|---|---|---|
| $P1$-$SrH_6$ 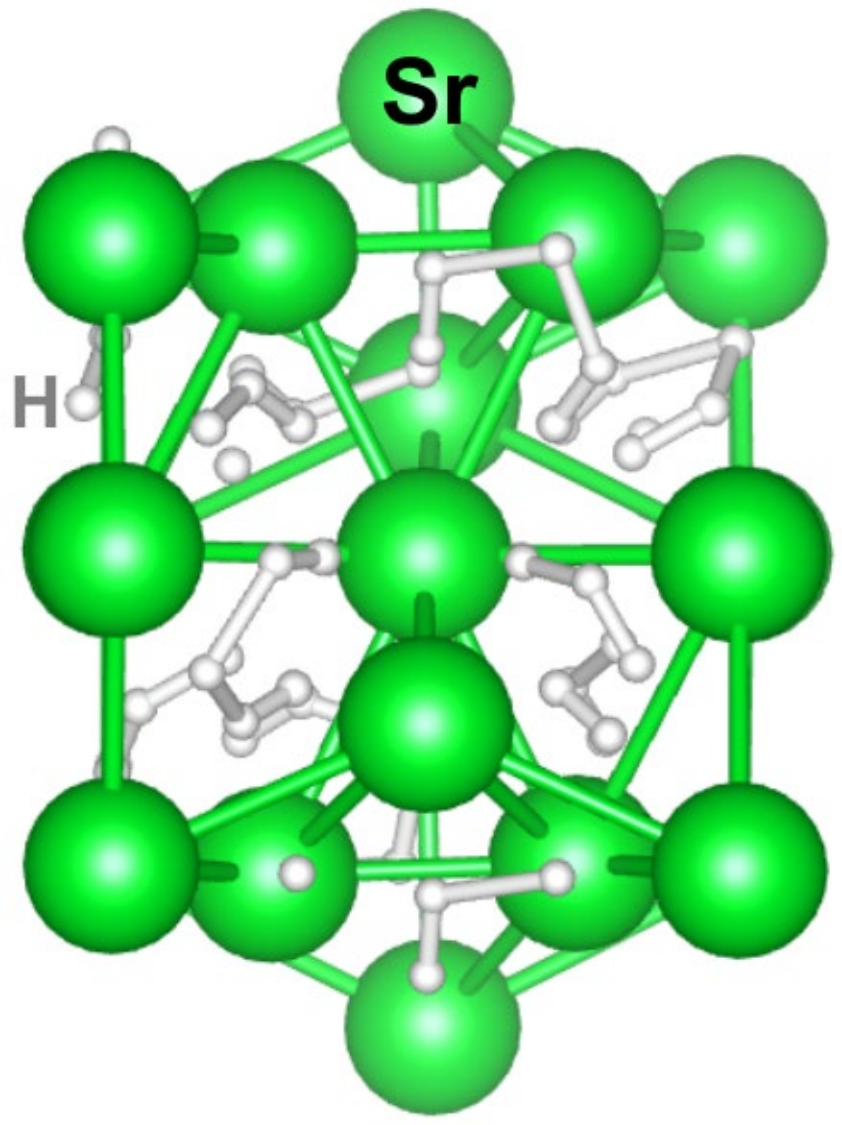  | 100 | $a$ = 6.084 Å<br>$b$ = 6.080 Å<br>$c$ = 6.160 Å<br>α = 90.48°<br>β = 90.21°<br>γ = 90.30° | Sr1(1a) | 0.12251 | 0.62481 | 0.87321 |
| | | | Sr2(1a) | 0.86195 | 0.13111 | 0.62262 |
| | | | Sr3(1a) | 0.63559 | 0.86933 | 0.1373 |
| | | | Sr4(1a) | 0.37682 | 0.11185 | 0.63943 |
| | | | Sr5(1a) | 0.62243 | 0.35237 | 0.12526 |
| | | | Sr6(1a) | 0.13328 | 0.62641 | 0.37244 |
| | | | Sr7(1a) | 0.62007 | 0.6192 | 0.6537 |
| | | | Sr8(1a) | 0.12384 | 0.11843 | 0.1535 |
| | | | H1(1a) | 0.83108 | 0.10887 | 0.97125 |
| | | | H2(1a) | 0.9734 | 0.81083 | 0.12742 |
| | | | H3(1a) | 0.12569 | 0.96391 | 0.83565 |
| | | | H4(1a) | 0.62632 | 0.60953 | 0.33588 |
| | | | H5(1a) | 0.30726 | 0.45652 | 0.62636 |
| | | | H6(1a) | 0.61939 | 0.32533 | 0.47245 |
| | | | H7(1a) | 0.26967 | 0.43008 | 0.12216 |
| | | | H8(1a) | 0.12572 | 0.28312 | 0.46802 |
| | | | H9(1a) | 0.45582 | 0.11496 | 0.30449 |
| | | | H10(1a) | 0.62604 | 0.95054 | 0.80334 |
| | | | H11(1a) | 0.77602 | 0.61062 | 0.95804 |
| | | | H12(1a) | 0.95377 | 0.78727 | 0.62779 |
| | | | H13(1a) | 0.42108 | 0.11143 | 0.98695 |
| | | | H14(1a) | 0.96016 | 0.45059 | 0.12861 |
| | | | H15(1a) | 0.11818 | 0.96836 | 0.45703 |
| | | | H16(1a) | 0.4607 | 0.61632 | 0.96987 |
| | | | H17(1a) | 0.93158 | 0.47973 | 0.62245 |
| | | | H18(1a) | 0.61829 | 0.91833 | 0.48649 |
| | | | H19(1a) | 0.33135 | 0.35301 | 0.94544 |
| | | | H20(1a) | 0.91214 | 0.32993 | 0.34409 |
| | | | H21(1a) | 0.34042 | 0.9089 | 0.34199 |
| | | | H22(1a) | 0.41952 | 0.8247 | 0.85143 |
| | | | H23(1a) | 0.83306 | 0.41253 | 0.85688 |
| | | | H24(1a) | 0.83811 | 0.83144 | 0.43647 |
| | | | H25(1a) | 0.41825 | 0.8435 | 0.43171 |
| | | | H26(1a) | 0.41229 | 0.4035 | 0.84926 |
| | | | H27(1a) | 0.85631 | 0.41406 | 0.43398 |
| | | | H28(1a) | 0.91205 | 0.33316 | 0.93143 |
| | | | H29(1a) | 0.91975 | 0.90726 | 0.36027 |
| | | | H30(1a) | 0.34651 | 0.89223 | 0.95227 |
| | | | H31(1a) | 0.77755 | 0.61264 | 0.33374 |
| | | | H32(1a) | 0.31471 | 0.76685 | 0.6236 |
| | | | H33(1a) | 0.62302 | 0.30606 | 0.78698 |
| | | | H34(1a) | 0.81068 | 0.11671 | 0.28645 |
| | | | H35(1a) | 0.28222 | 0.81498 | 0.11848 |
| | | | H36(1a) | 0.12372 | 0.2724 | 0.82894 |
| | | | H37(1a) | 0.63129 | 0.07832 | 0.86913 |
| | | | H38(1a) | 0.88945 | 0.57795 | 0.11076 |
| | | | H39(1a) | 0.07988 | 0.85877 | 0.61495 |
| | | | H40(1a) | 0.38572 | 0.61711 | 0.09097 |
| | | | H41(1a) | 0.15492 | 0.37096 | 0.58984 |
| | | | H42(1a) | 0.58367 | 0.12309 | 0.37057 |
| | | | H43(1a) | 0.32822 | 0.32415 | 0.35251 |
| | | | H44(1a) | 0.41893 | 0.38965 | 0.43476 |
| | | | H45(1a) | 0.4712 | 0.6145 | 0.33036 |
| | | | H46(1a) | 0.12 | 0.12949 | 0.83601 |
| | | | H47(1a) | 0.83019 | 0.82987 | 0.84235 |
| | | | H48(1a) | 0.8978 | 0.92451 | 0.92017 |

**Table S16.** Crystal structure of proposed $Pm\bar{3}n$-$Sr_8H_{46}$ at 120 GPa.

| Phase | Pressure, GPa | Lattice parameters | Coordinates | | | |
|---|---|---|---|---|---|---|
| $Pm\bar{3}n$-$Sr_8H_{46}$ | 120 | $a$ = 5.930 Å | Sr1 | 0 | 0 | 0 |
| | | | Sr2 | 0.25 | 0 | 0.5 |
| | | | H1 | 0 | 0.15833 | 0.31308 |
| | | | H2 | 0.25 | 0.5 | 0 |
| | | | H3 | 0.20572 | 0.20572 | 0.20572 |

**Table S17.** Calculated unit cell parameters of cubic $Pm\bar{3}n$-$SrH_5$ ($Sr_8H_{40}$). Isostructural $Pm\bar{3}n$-$EuH_5$ has been previously proposed to explain the formation of new phases in the Eu–H system.[29] However, ab initio calculations show that the ideal $Pm\bar{3}n$-$SrH_5$ structure is completely dynamically unstable and therefore cannot be used to interpret the experimental data.

| Pressure, GPa | $a$, Å (as $Pm\bar{3}n$, $Z$ = 8) | $V$, Å$^3$ per Sr atom (as $Pm\bar{3}n$) |
|---|---|---|
| 40 | 6.413 | 32.977 |
| 60 | 6.220 | 30.081 |
| 80 | 6.072 | 27.99 |
| 100 | 5.952 | 26.36 |
| 120 | 5.852 | 25.059 |
| 140 | 5.765 | 23.96 |
| 160 | 5.689 | 23.026 |

**Table S18.** Bader charge analysis of $Pm\bar{3}n$-like pseudocubic $P1$-$SrH_6$ at 100 GPa.

| Element | Bader charge | Coordinate X | Coordinate Y | Coordinate Z |
|---|---|---|---|---|
| Sr | 1,124 | 0,870808 | 0,6227229 | 0,8749857 |
| Sr | 1,148 | 0,6251439 | 0,1269839 | 0,1393409 |
| Sr | 1,151 | 0,1387032 | 0,869925 | 0,3666065 |
| Sr | 1,148 | 0,6400509 | 0,1157757 | 0,6227973 |
| Sr | 1,147 | 0,127004 | 0,3532591 | 0,3766161 |
| Sr | 1,122 | 0,3700049 | 0,6252475 | 0,8674192 |
| Sr | 1,065 | 0,6535769 | 0,6199615 | 0,3768713 |
| Sr | 1,065 | 0,153609 | 0,1199978 | 0,8761696 |
| H | -0,349 | 0,973023 | 0,1099248 | 0,1701267 |
| H | -0,414 | 0,1248227 | 0,8119182 | 0,0253724 |
| H | -0,291 | 0,8344744 | 0,9633547 | 0,8752897 |
| H | 0,031 | 0,3366999 | 0,6092284 | 0,3712572 |
| H | -0,338 | 0,6212006 | 0,4625102 | 0,6830264 |
| H | -0,376 | 0,4735865 | 0,32395 | 0,3804787 |
| H | -0,350 | 0,1200279 | 0,4307898 | 0,7278987 |
| H | -0,212 | 0,4687275 | 0,2824226 | 0,875339 |
| H | -0,217 | 0,3041117 | 0,1157773 | 0,5445022 |
| H | -0,218 | 0,8040751 | 0,9499099 | 0,3754673 |
| H | -0,324 | 0,9620443 | 0,6124613 | 0,2207319 |
| H | -0,238 | 0,6259821 | 0,7844539 | 0,0455015 |
| H | -0,409 | 0,9869781 | 0,1124019 | 0,5776545 |
| H | -0,256 | 0,1265185 | 0,4475407 | 0,0375389 |
| H | -0,337 | 0,460331 | 0,9686129 | 0,8808238 |
| H | -0,211 | 0,9682716 | 0,6166244 | 0,5363793 |
| H | -0,370 | 0,6197945 | 0,4756087 | 0,0664667 |
| H | -0,390 | 0,4859749 | 0,9187875 | 0,382016 |
| H | -0,040 | 0,9459562 | 0,3520889 | 0,6678609 |
| H | -0,149 | 0,3462164 | 0,3289077 | 0,0862806 |
| H | -0,159 | 0,3415474 | 0,90997 | 0,661004 |
| H | -0,182 | 0,8485356 | 0,8260172 | 0,5821452 |
| H | -0,085 | 0,8535336 | 0,4110209 | 0,1677138 |
| H | -0,093 | 0,4372868 | 0,8302066 | 0,1584418 |
| H | -0,004 | 0,4373558 | 0,8408169 | 0,5887506 |
| H | -0,124 | 0,8474416 | 0,4049217 | 0,5865531 |
| H | -0,011 | 0,4418416 | 0,4140313 | 0,1395418 |
| H | -0,076 | 0,9323183 | 0,3350721 | 0,0874485 |
| H | -0,068 | 0,3562624 | 0,9084872 | 0,0821558 |
| H | 0,008 | 0,9529067 | 0,8918327 | 0,6557067 |
| H | -0,246 | 0,3320068 | 0,6123312 | 0,2207249 |
| H | -0,396 | 0,6212498 | 0,7700113 | 0,6831097 |
| H | -0,409 | 0,7884712 | 0,3056371 | 0,3775097 |
| H | -0,416 | 0,2879164 | 0,1171889 | 0,1894833 |
| H | -0,332 | 0,1173372 | 0,8153535 | 0,7183523 |
| H | -0,240 | 0,8290418 | 0,2720012 | 0,876988 |
| H | -0,018 | 0,8717853 | 0,0799804 | 0,3701721 |
| H | 0,000 | 0,1037027 | 0,5828141 | 0,115619 |
| H | -0,007 | 0,6088659 | 0,8619859 | 0,9132206 |
| H | -0,031 | 0,0915592 | 0,6161131 | 0,6135648 |
| H | -0,026 | 0,5895225 | 0,3641036 | 0,8663512 |
| H | -0,014 | 0,3724567 | 0,1237747 | 0,4152746 |
| H | -0,176 | 0,3477018 | 0,3248401 | 0,669414 |
| H | 0,009 | 0,4422296 | 0,3968609 | 0,5928269 |
| H | -0,287 | 0,3308854 | 0,6147967 | 0,5284857 |
| H | 0,032 | 0,8364467 | 0,1246422 | 0,880615 |
| H | -0,179 | 0,8431626 | 0,8285789 | 0,1688118 |
| H | 0,019 | 0,9264109 | 0,9175624 | 0,099606 |
| Total | 0,000 | | | |

**Table S19.** Bader charge analysis of $Pm\overline{3}n$-like pseudocubic $P1$-$SrH_6$ at 150 GPa.

| Element | Bader charge | Coordinate X | Coordinate Y | Coordinate Z |
|---|---|---|---|---|
| Sr | 1,038 | 0,1216972 | 6,23E-01 | 0,8643163 |
| Sr | 1,074 | 0,8609518 | 1,22E-01 | 0,6303453 |
| Sr | 1,076 | 0,622266 | 8,77E-01 | 0,1382668 |
| Sr | 1,076 | 0,3768387 | 1,22E-01 | 0,6382668 |
| Sr | 1,074 | 0,6223045 | 3,61E-01 | 0,1303453 |
| Sr | 1,038 | 0,1233579 | 6,22E-01 | 0,3643163 |
| Sr | 0,977 | 0,6225154 | 6,22E-01 | 0,6567168 |
| Sr | 0,977 | 0,1220142 | 1,23E-01 | 0,1567168 |
| H | -0,333 | 0,8249147 | 1,19E-01 | 0,9746721 |
| H | -0,311 | 0,9682427 | 8,14E-01 | 0,1168925 |
| H | -0,255 | 0,1201653 | 9,64E-01 | 0,8306556 |
| H | 0,029 | 0,6227485 | 6,18E-01 | 0,3406195 |
| H | -0,311 | 0,3143441 | 4,68E-01 | 0,6168925 |
| H | -0,333 | 0,6186704 | 3,25E-01 | 0,4746721 |
| H | -0,291 | 0,2759871 | 4,33E-01 | 0,1159938 |
| H | -0,229 | 0,1205407 | 2,82E-01 | 0,4639115 |
| H | -0,244 | 0,4491025 | 1,21E-01 | 0,3052344 |
| H | -0,244 | 0,6210622 | 9,49E-01 | 0,8052344 |
| H | -0,229 | 0,7824073 | 6,21E-01 | 0,9639115 |
| H | -0,291 | 0,9330509 | 7,76E-01 | 0,6159938 |
| H | -0,359 | 0,4182123 | 1,19E-01 | 0,9804885 |
| H | -0,297 | 0,9673734 | 4,33E-01 | 0,1171524 |
| H | -0,224 | 0,1205212 | 9,61E-01 | 0,4631023 |
| H | -0,224 | 0,4606402 | 6,21E-01 | 0,9631023 |
| H | -0,297 | 0,9326445 | 4,67E-01 | 0,6171524 |
| H | -0,359 | 0,6186795 | 9,18E-01 | 0,4804885 |
| H | -0,028 | 0,3282164 | 3,47E-01 | 0,9472817 |
| H | -0,136 | 0,9095175 | 3,27E-01 | 0,3451277 |
| H | -0,167 | 0,3314733 | 9,16E-01 | 0,3448325 |
| H | -0,167 | 0,4162157 | 8,31E-01 | 0,8448325 |
| H | -0,136 | 0,8265974 | 4,10E-01 | 0,8451277 |
| H | -0,028 | 0,8469436 | 8,28E-01 | 0,4472817 |
| H | 0,000 | 0,4003662 | 8,33E-01 | 0,4498928 |
| H | -0,135 | 0,4154069 | 4,11E-01 | 0,8458325 |
| H | -0,023 | 0,8455518 | 4,14E-01 | 0,4458395 |
| H | -0,023 | 0,9141696 | 3,46E-01 | 0,9458395 |
| H | -0,135 | 0,9114189 | 9,15E-01 | 0,3458325 |
| H | 0,000 | 0,3327845 | 9,00E-01 | 0,9498928 |
| H | -0,242 | 0,7789434 | 6,20E-01 | 0,3293645 |
| H | -0,304 | 0,3149984 | 7,75E-01 | 0,6146242 |
| H | -0,339 | 0,6211853 | 3,04E-01 | 0,7942044 |
| H | -0,339 | 0,8038242 | 1,21E-01 | 0,2942044 |
| H | -0,304 | 0,2746191 | 8,15E-01 | 0,1146242 |
| H | -0,242 | 0,1196015 | 2,79E-01 | 0,8293645 |
| H | -0,028 | 0,6270762 | 1,01E-01 | 0,8786761 |
| H | -0,039 | 0,8692059 | 6,19E-01 | 0,0968528 |
| H | -0,046 | 0,1174326 | 8,76E-01 | 0,5969566 |
| H | -0,046 | 0,375621 | 6,17E-01 | 0,0969566 |
| H | -0,039 | 0,1191597 | 3,69E-01 | 0,5968528 |
| H | -0,028 | 0,6005894 | 1,27E-01 | 0,3786761 |
| H | -0,164 | 0,3323698 | 3,26E-01 | 0,3447297 |
| H | 0,002 | 0,4047915 | 4,09E-01 | 0,4461622 |
| H | -0,255 | 0,4641598 | 6,20E-01 | 0,3306556 |
| H | 0,029 | 0,1175383 | 1,23E-01 | 0,8406195 |
| H | -0,164 | 0,8261942 | 8,32E-01 | 0,8447297 |
| H | 0,002 | 0,9086104 | 9,05E-01 | 0,9461622 |
| Total | 0,000 | | | |

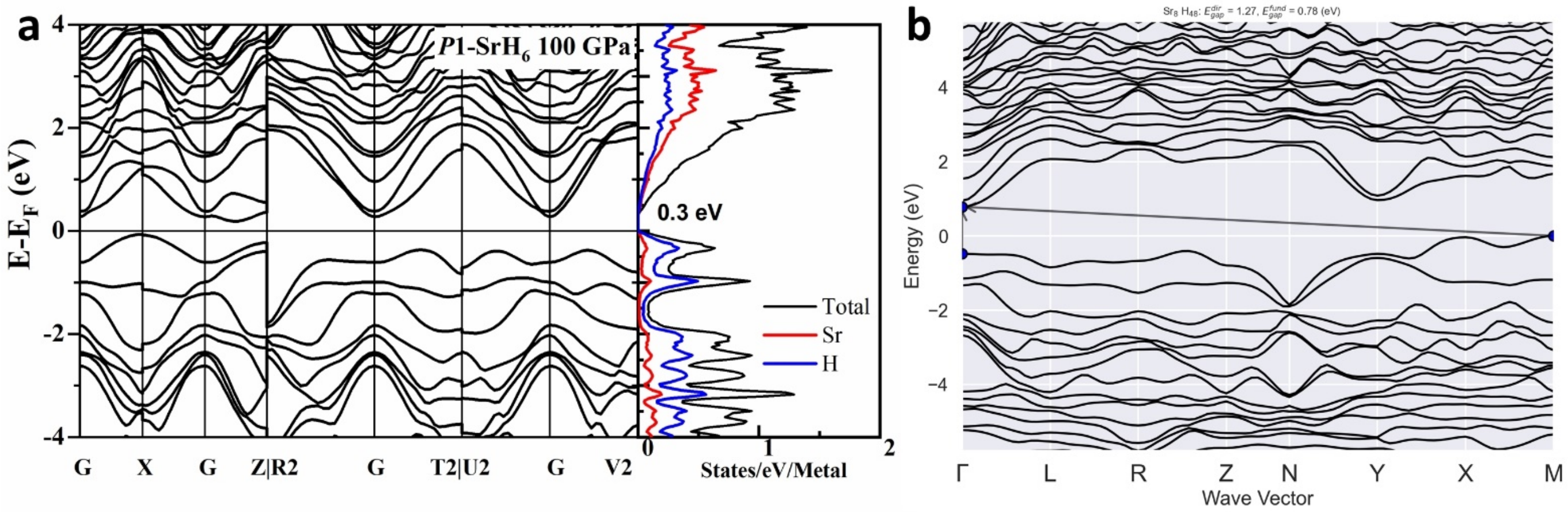

**Figure S16.** Band structure of *P*1-$SrH_6$ ($Sr_8H_{48}$, $Pm\bar{3}n$-like) calculated using different methods at 100 GPa. Band structure and the density of electron states of *P*1-$SrH_6$ per Sr atom calculated using the PBE–GGA exchange–correlation functional (VASP code). (b) Band structure of *P*1-$SrH_6$ at 100 GPa calculated using the TB09-HGH functional (Abinit code).

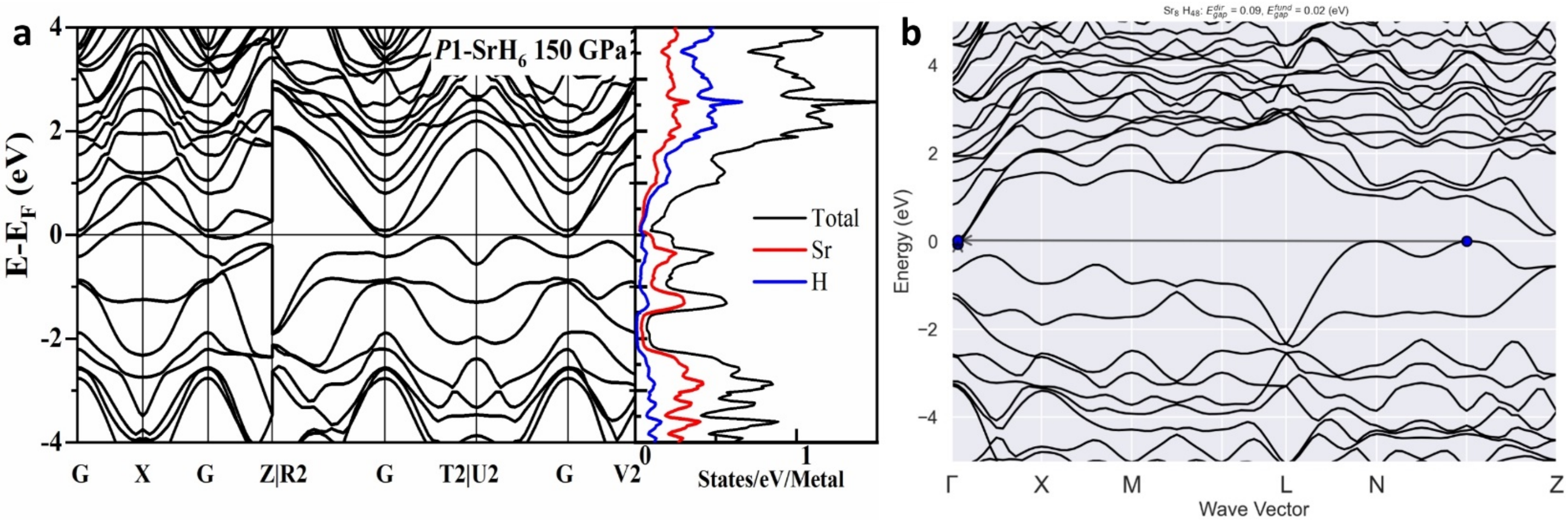

**Figure S17.** Band structure of $P1$-$SrH_6$ ($Sr_8H_{48}$, $Pm\bar{3}n$-like) calculated using different methods at 150 GPa. (a) Band structure and the density of electron states of $P1$-$SrH_6$ per Sr atom at 150 GPa calculated using the PBE–GGA exchange–correlation functional (VASP code). (b) Band structure of $P1$-$SrH_6$ at 150 GPa calculated using the TB09-HGH functional (Abinit code). An accurate estimation of the direct bandgap gives 0.55 eV.

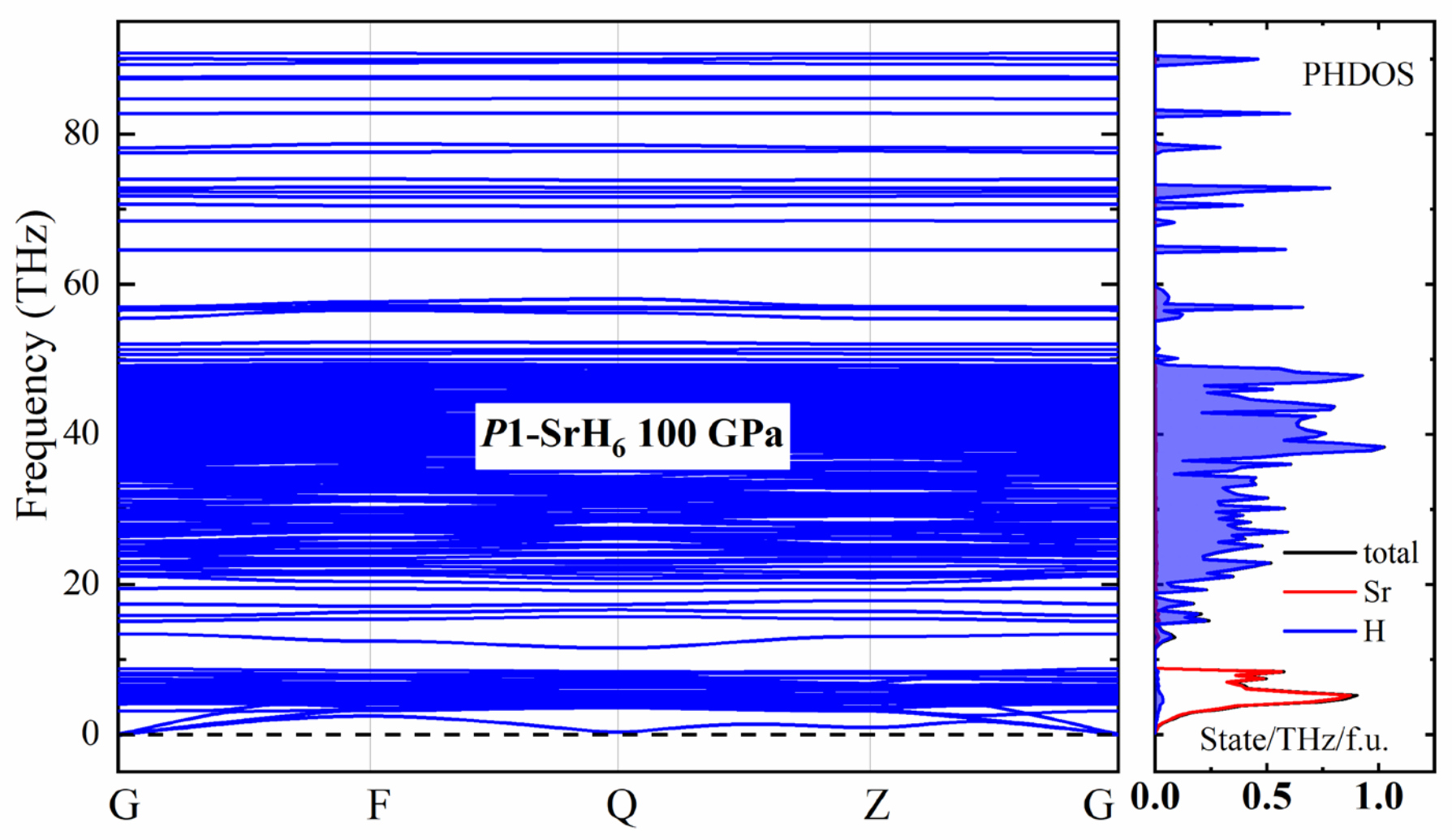

**Figure S18**. Phonon band structure and the density of states of $P1$-$SrH_6$ at 100 GPa calculated within the harmonic approximation.

## 5. $F\bar{4}3m$-like pseudocubic $P1$-$SrH_9$ and $P4_2/mnm$-$SrH_6$

**Table S20.** Crystal structure of $P4_2/mnm$-$SrH_6$ ($Sr_2H_{12}$) at 150 GPa. This structure can be considered tetragonally distorted $Im\bar{3}m$-$SrH_6$ similar to $Im\bar{3}m$-$YH_6$.

| Phase | Pressure, GPa | Lattice parameters | Coordinates | | | |
|---|---|---|---|---|---|---|
| $P4_2/mnm$-$SrH_6$ | 150 | $a$ = 3.672 Å<br>$c$ = 3.747 Å | Sr1 | 0 | 0 | 0.5 |
| | | | H1 | -0.22659 | 0.05509 | 0.0000 |
| | | | H2 | 0.0000 | 0.5 | 0.25 |

**Table S21.** Experimental and calculated unit cell parameters of $P4_2/mnm$-$SrH_6$ ($Sr_2H_{12}$), proposed to explain the results of the synthesis in DAC Sr90.

| Pressure, GPa | $a$, Å | $c$, Å | $V$, Å$^3$ per Sr atom |
|---|---|---|---|
| DAC Sr90 | | | |
| 82.0 | 3.962 | 3.863 | 29.57 |
| 87.0 | 3.946 | 3.860 | 29.40 |
| 90.0 | 3.942 | 3.854 | 29.28 |
| 100.0 | 3.923 | 3.824 | 28.68 |
| 121.0 | 3.882 | 3.771 | 27.61 |
| 131.0 | 3.866 | 3.757 | 27.29 |
| 139.0 | 3.862 | 3.743 | 27.06 |
| Theory (PBE GGA) | | | |
| 80.0 | 3.911 | 4.040 | 30.89 |
| 90.0 | 3.868 | 3.986 | 29.82 |
| 100.0 | 3.829 | 3.935 | 28.84 |
| 110.0 | 3.793 | 3.891 | 27.99 |
| 130.0 | 3.729 | 3.813 | 26.51 |
| 150.0 | 3.670 | 3.747 | 25.26 |

**Table S22.** Experimental and calculated unit cell parameters of $Im\bar{3}m$-$SrH_6$ ($Sr_2H_{12}$), isostructural to $YH_6$, which can also be used to explain the results of the synthesis in DAC Sr90.

| Pressure, GPa | $a$, Å | $V$, Å$^3$ per Sr atom |
|---|---|---|
| DAC Sr90 | | |
| 82 | 3.8659 | 28.88 |
| 87 | 3.8599 | 28.75 |
| 90 | 3.8539 | 28.62 |
| 94 | 3.8619 | 28.79 |
| 100 | 3.8219 | 27.91 |
| 112 | 3.7939 | 27.30 |
| 121 | 3.7759 | 26.91 |
| 131 | 3.7589 | 26.55 |
| 139 | 3.7439 | 26.23 |
| Theory (PBE GGA) | | |
| 70 | 3.934 | 30.44 |
| 90 | 3.851 | 28.55 |
| 100 | 3.814 | 27.74 |
| 120 | 3.751 | 26.38 |
| 140 | 3.696 | 25.24 |
| 150 | 3.670 | 24.71 |

**Table S23.** Crystal structure of discovered $F\bar{4}3m$-like pseudocubic $P1$-$SrH_9$ at 100 GPa (PBE GGA, VASP).

| Phase | Pressure, GPa | Lattice parameters | Coordinates | | | |
|---|---|---|---|---|---|---|
| $P1$-$SrH_9$ 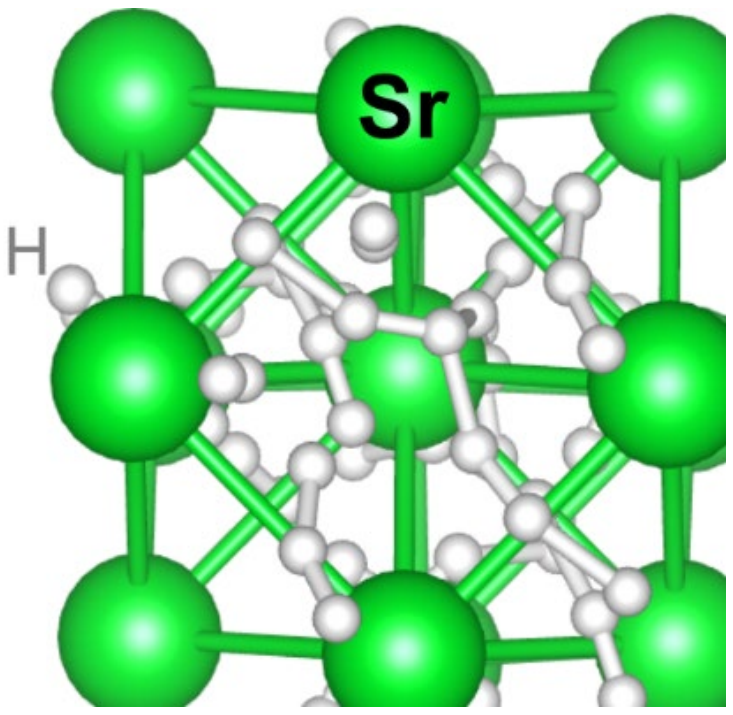 | 100 | $a$ = 10.487 Å, $b$ = 3.626 Å, $c$ = 3.713 Å, α = 89.74°, β = 89.74°, γ = 91.13° | Sr1(1a) | 0.47972 | -0.48646 | -0.43961 |
| | | | Sr2(1a) | 0.24665 | 0.02859 | 0.08032 |
| | | | Sr3(1a) | -0.25475 | 0.04565 | -0.03816 |
| | | | Sr4(1a) | -0.01671 | -0.44344 | 0.44443 |
| | | | H1(1a) | 0.05993 | -0.19215 | -0.09202 |
| | | | H2(1a) | 0.29038 | -0.29703 | -0.41027 |
| | | | H3(1a) | -0.19594 | -0.48626 | -0.35321 |
| | | | H4(1a) | 0.3632 | -0.48927 | 0.09005 |
| | | | H5(1a) | -0.1844 | -0.45981 | 0.18107 |
| | | | H6(1a) | 0.16356 | -0.15118 | -0.39663 |
| | | | H7(1a) | -0.24083 | -0.20211 | 0.45121 |
| | | | H8(1a) | 0.16103 | 0.48637 | -0.20572 |
| | | | H9(1a) | 0.05091 | 0.0739 | 0.29906 |
| | | | H10(1a) | -0.20722 | 0.33221 | 0.45729 |
| | | | H11(1a) | 0.15763 | -0.46454 | 0.11255 |
| | | | H12(1a) | -0.01912 | 0.27171 | -0.04864 |
| | | | H13(1a) | -0.45594 | 0.33218 | 0.06266 |
| | | | H14(1a) | 0.05479 | 0.29748 | -0.08112 |
| | | | H15(1a) | 0.43134 | -0.00624 | -0.16695 |
| | | | H16(1a) | -0.43302 | 0.00504 | -0.303 |
| | | | H17(1a) | -0.43392 | -0.22674 | 0.10296 |
| | | | H18(1a) | 0.10317 | 0.01614 | 0.47694 |
| | | | H19(1a) | -0.39081 | -2.5E-4 | -0.4786 |
| | | | H20(1a) | 0.29142 | 0.27035 | -0.43781 |
| | | | H21(1a) | -0.32467 | -0.39039 | 0.34932 |
| | | | H22(1a) | -0.01353 | -0.15813 | -0.04492 |
| | | | H23(1a) | -0.36331 | -0.42517 | -0.07634 |
| | | | H24(1a) | 0.42874 | -0.00554 | 0.3255 |
| | | | H25(1a) | -0.11955 | 0.04779 | -0.48823 |
| | | | H26(1a) | -0.31863 | -0.46483 | -0.24704 |
| | | | H27(1a) | 0.18898 | -0.48774 | 0.33226 |
| | | | H28(1a) | -0.08282 | 0.04282 | -0.29448 |
| | | | H29(1a) | 0.43856 | -0.06203 | 0.07688 |
| | | | H30(1a) | -0.14928 | -0.44826 | -0.04547 |
| | | | H31(1a) | 0.17071 | 0.35013 | -0.3979 |
| | | | H32(1a) | -0.44385 | 0.13201 | 0.13254 |
| | | | H33(1a) | 0.34135 | 0.11014 | -0.42028 |
| | | | H34(1a) | -0.07801 | 0.04967 | 0.18258 |
| | | | H35(1a) | 0.30736 | -0.39937 | -0.2113 |
| | | | H36(1a) | -0.33151 | 0.37592 | 0.37801 |

**Table S24.** Experimental and calculated unit cell parameters of cubic ($F\bar{4}3m$) $SrH_9$, proposed to explain the results of the synthesis in DACs Sr50 and Sr90.

| Pressure, GPa | $V$, Å$^3$ per Sr atom | Pressure, GPa | $V$, Å$^3$ per Sr atom | Pressure, GPa | $V$, Å$^3$ per Sr atom |
|---|---|---|---|---|---|
| DAC Sr50 | | DAC Sr90 | | Theory ($F\bar{4}3m$) | |
| 101 | 34.12 | 139 | 29.67 | 60 | 38.28 |
| 93 | 34.90 | 131 | 29.77 | 80 | 35.68 |
| 83 | 36.20 | 121 | 30.32 | 100 | 33.65 |
| 75 | 36.82 | 100 | 32.75 | 120 | 31.98 |
| 68 | 37.47 | 94 | 33.17 | 140 | 30.59 |
| 62 | 37.72 | 90 | 33.57 | | |

**Table S25.** Experimental and calculated unit cell parameters of *Cmme*-$Sr_2H_3$, proposed to explain the results of the synthesis in DAC Sr50.

| Pressure, GPa | *a*, Å | *b*, Å | *c*, Å | *V*, Å$^3$ per Sr atom |
|---|---|---|---|---|
| DAC Sr50 | | | | |
| 62 | 5.62 | 5.83 | 5.04 | 20.7 |
| 75 | 5.57 | 5.74 | 5.02 | 20.13 |
| 83 | 5.53 | 5.70 | 4.95 | 19.56 |
| 93 | 5.48 | 5.63 | 4.95 | 19.11 |
| 101 | 5.45 | 5.63 | 4.89 | 18.82 |
| Theory | | | | |
| 60 | 5.485 | 6.111 | 5.281 | 22.13 |
| 80 | 5.401 | 5.907 | 5.096 | 20.32 |
| 100 | 5.337 | 5.747 | 4.949 | 18.97 |
| 120 | 5.282 | 5.620 | 4.830 | 17.92 |

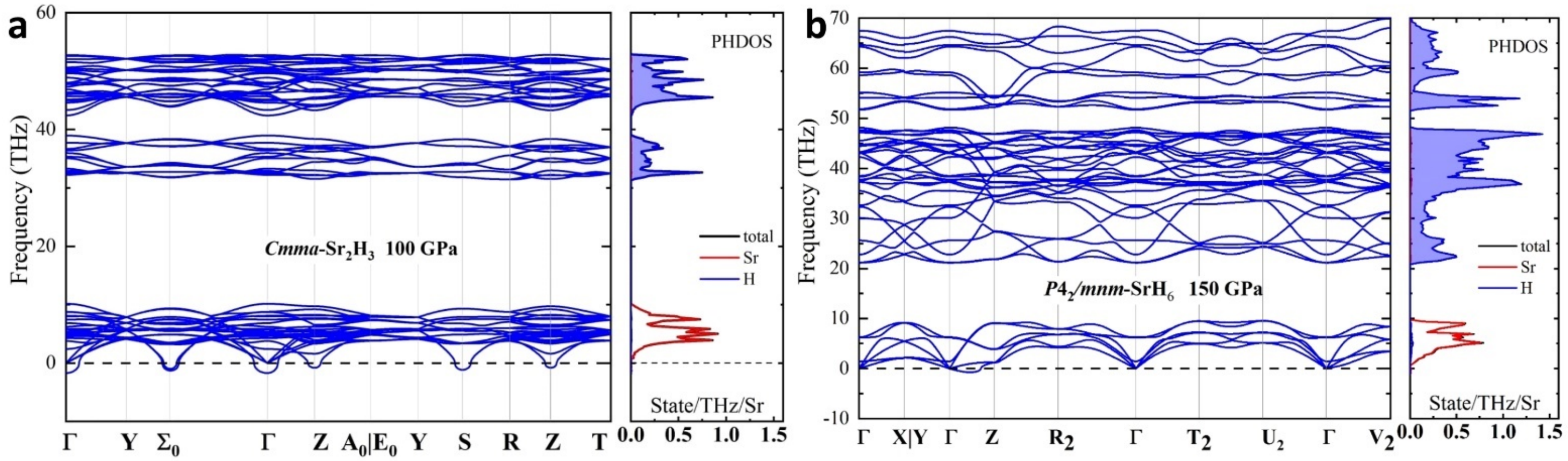

**Figure S19**. Phonon band structure and the density of states calculated within the harmonic approximation: (a) *Cmma*-$Sr_2H_3$ at 100 GPa. (b) *P*$4_2$/*mnm*-$SrH_6$ ($Sr_2H_{12}$) at 150 GPa.

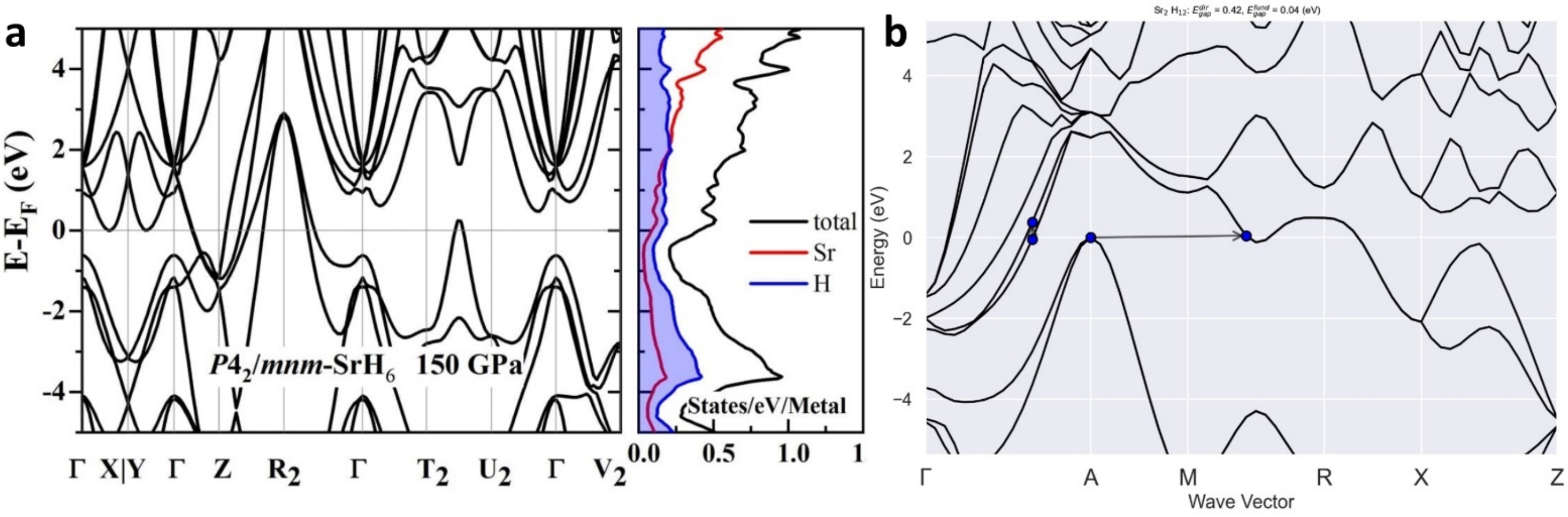

**Figure S20.** (a) Band structure and the density of electron states of *P*$4_2$/*mnm*-$SrH_6$ ($Sr_2H_{12}$) at 150 GPa per Sr atom calculated using the PBE–GGA exchange–correlation functional (VASP code). (b) Band structure of *P*$4_2$/*mnm*-$SrH_6$ ($Sr_2H_{12}$) at 150 GPa calculated using the TB09-HGH functional (Abinit code). The compound should demonstrate metallic properties at this pressure.

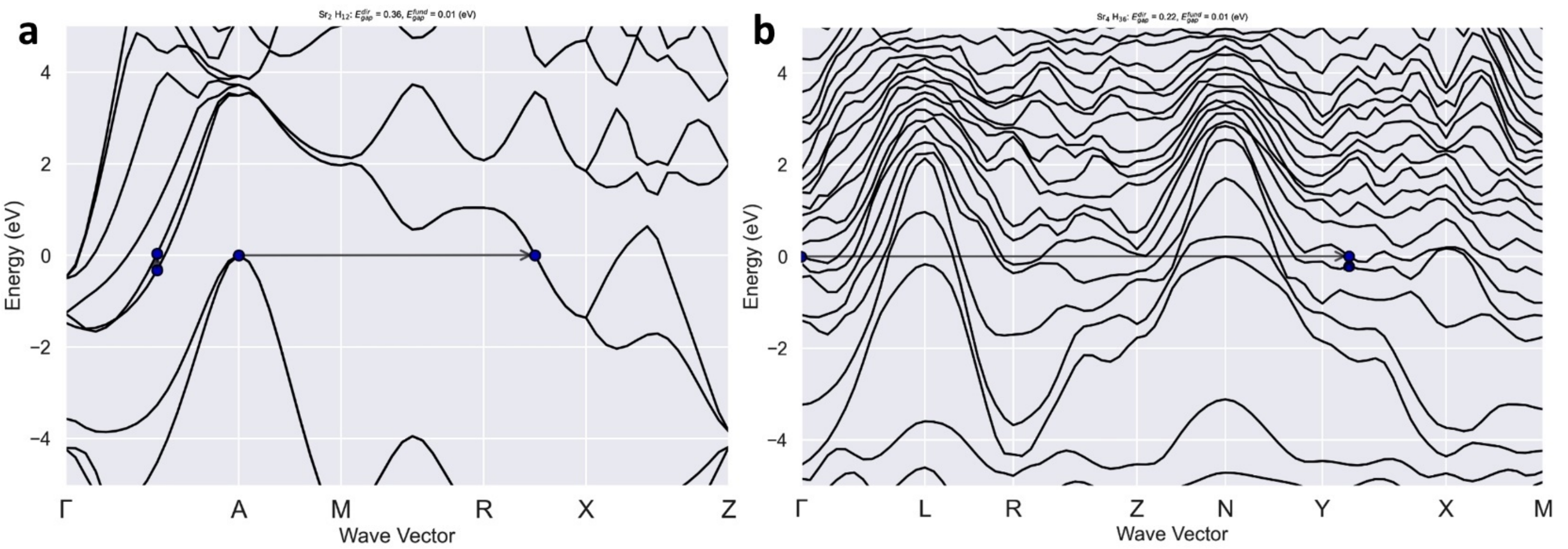

**Figure S21.** (a) Band structure $P4_2/mnm$-$SrH_6$ ($Sr_2H_{12}$) at 150 GPa calculated using the PBE–GGA exchange–correlation functional (Abinit code). This compound should demonstrate metallic properties at this pressure. (b) Band structure of $P1$-$SrH_9$ (c-$SrH_{\sim 9}$) at 100 GPa calculated using the PBE–GGA exchange–correlation functional (Abinit code).

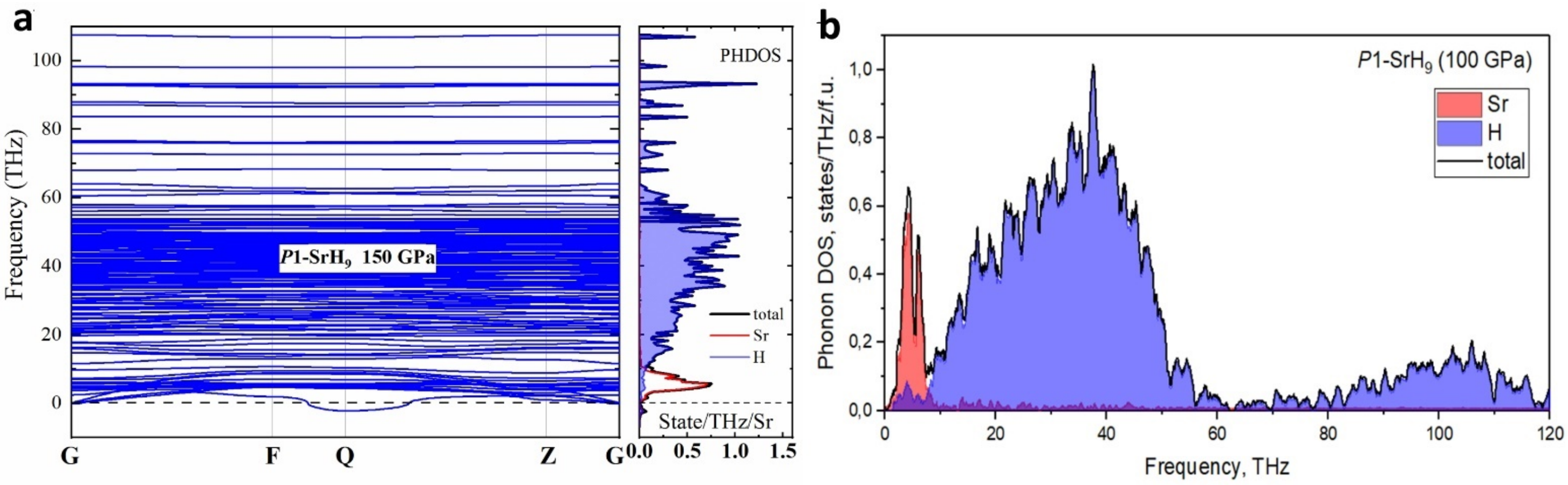

**Figure S22**. (a) Phonon band structure and the density of states of $F\bar{4}3m$-like pseudocubic $P1$-$SrH_9$ ($Sr_4H_{36}$) at 150 GPa calculated within the harmonic approximation. (b) Anharmonic phonon density of states of $P1$-$SrH_9$ calculated at 100 GPa using molecular dynamics with the MTP and MLIP at 10 K. In the harmonic approximation, this structure is unstable at 100 GPa.

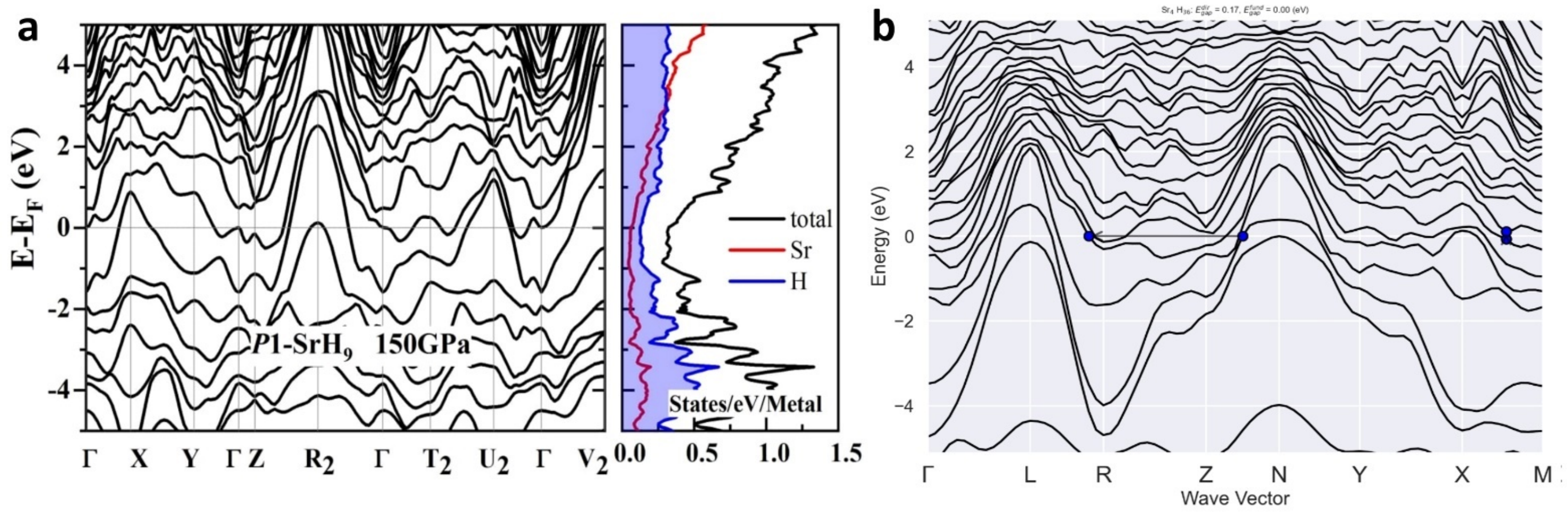

**Figure S23.** (a) Band structure and the density of electron states of $F\bar{4}3m$-like pseudocubic $P1$-$SrH_9$ ($Sr_4H_{36}$) per Sr atom at 150 GPa calculated using the PBE–GGA exchange–correlation functional (VASP code). (b) Band structure of $P1$-$SrH_9$ (c-$SrH_{\sim 9}$) at 100 GPa calculated using the TB09 HGH functional (Abinit code).

# Experiment

## *Additional X-ray diffraction data*

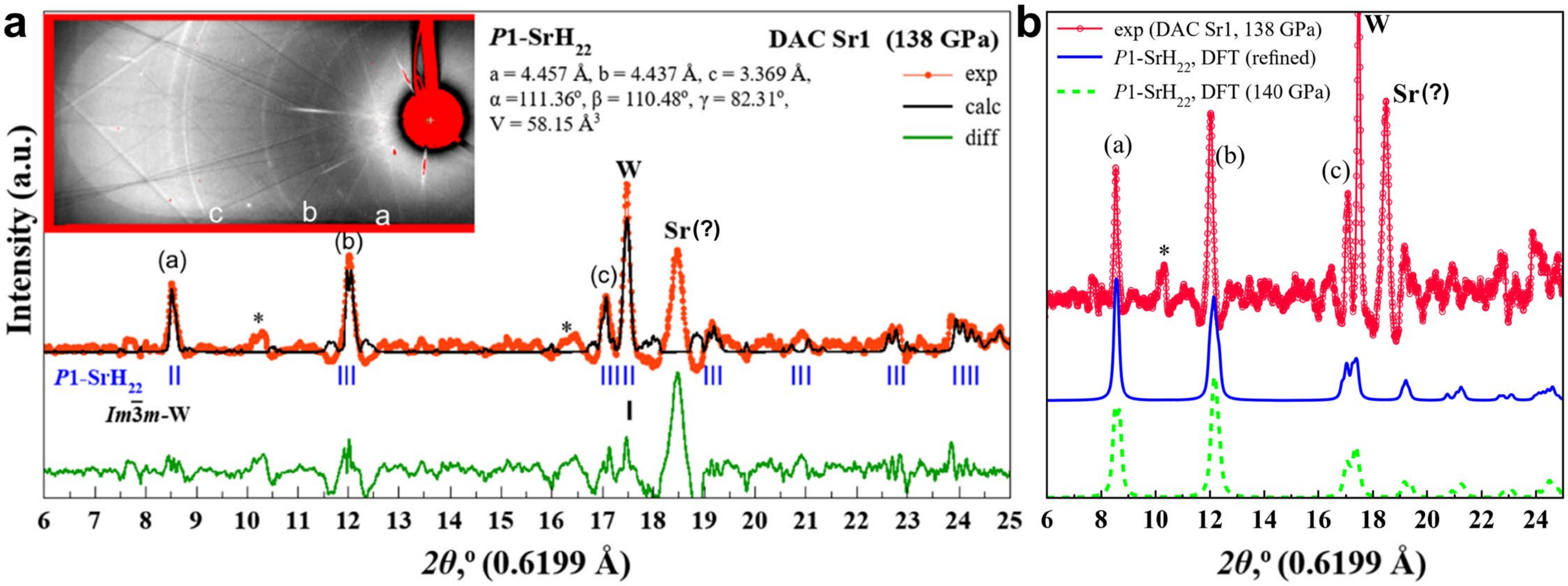

**Figure S24.** X-ray diffraction study of strontium hydrides in DAC Sr1. (a) Experimental diffraction pattern and the Le Bail refinement of the unit cell parameters of pseudotetragonal $P1$-$SrH_{22}$ at 138 GPa. The experimental data, fit, and residues are shown in red, black, and green, respectively. Unidentified reflections are indicated by asterisks. (b) Comparison of the experimental and predicted XRD patterns of DFT relaxed (PBE GGA) $P1$-$SrH_{22}$.

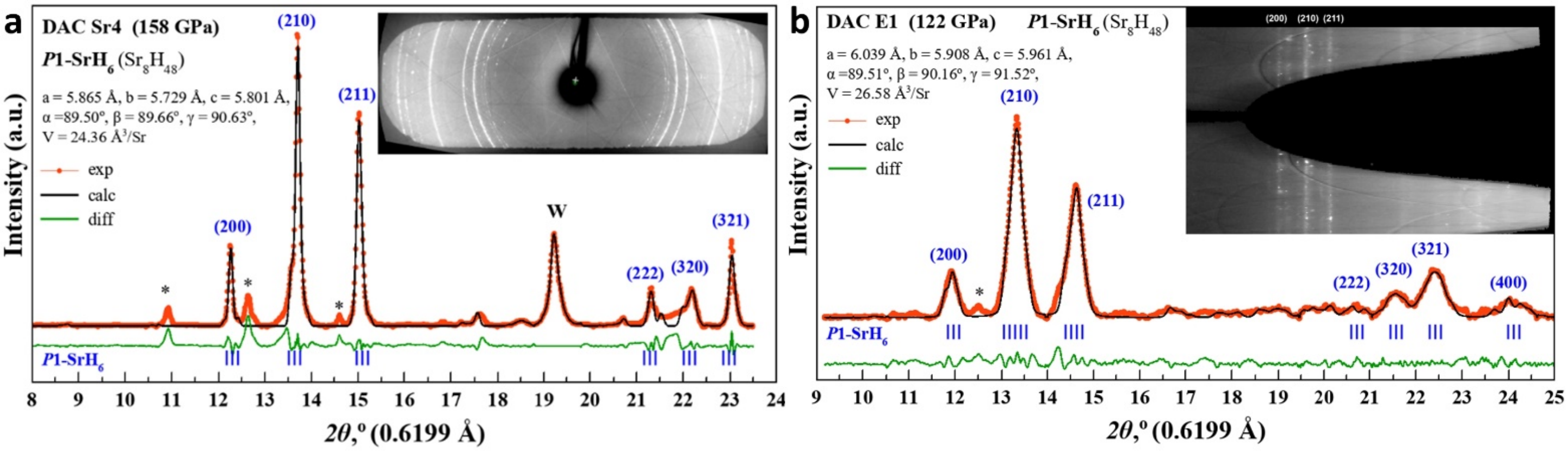

**Figure S25.** X-ray diffraction study of strontium hydrides in (a) DAC Sr4 and (b) the electrical DAC E1. Experimental diffraction pattern and the Le Bail refinement of the unit cell parameters of $Pm\bar{3}n$-like pseudocubic $P1$-$SrH_6$ ($Sr_8H_{48}$) at (a) 158 GPa and (b) 122 GPa. The experimental data, fit, and residues are shown in red, black, and green, respectively.

## *Raman measurements*

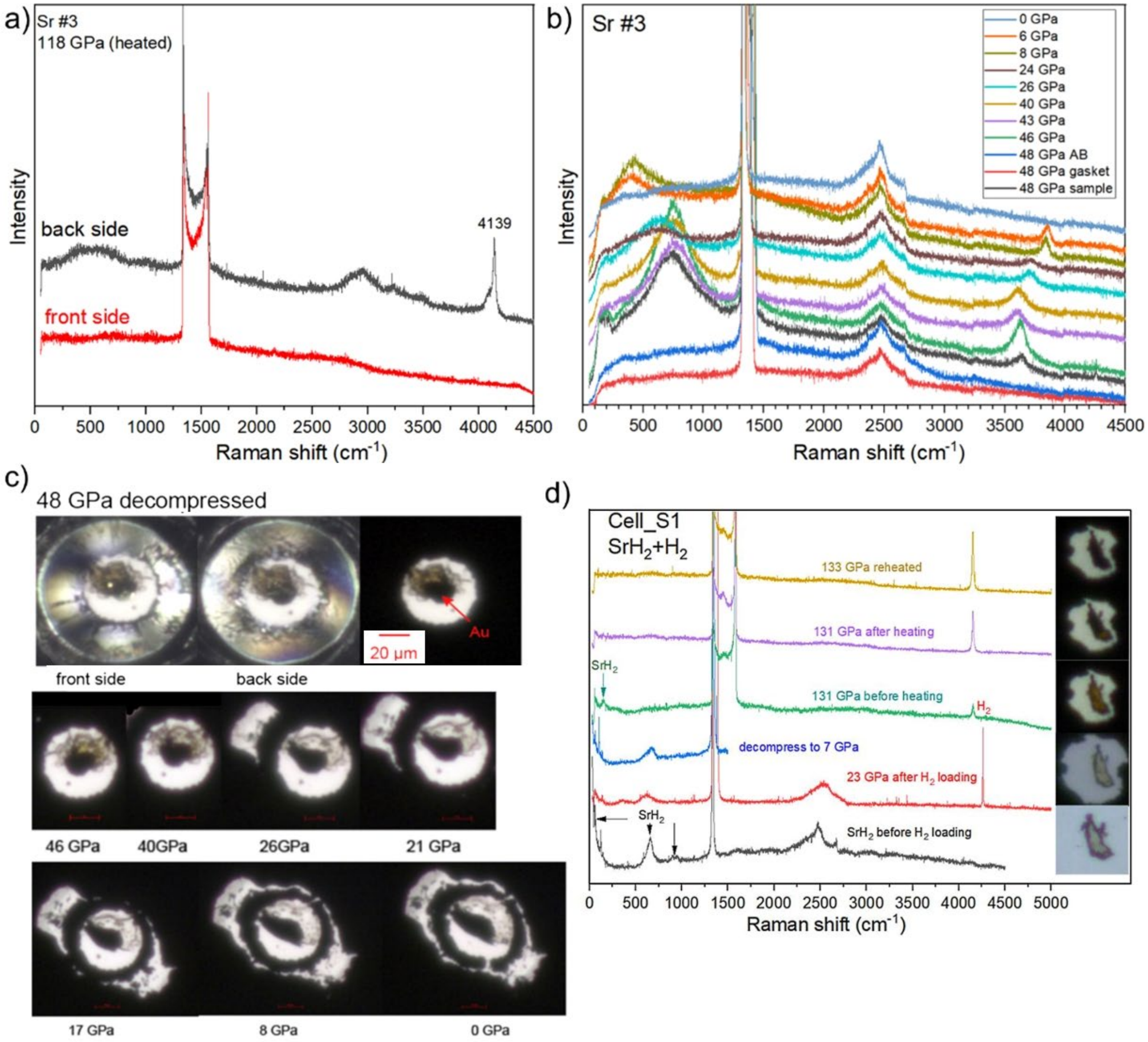

**Figure S26.** (a) Raman spectra from the back and front sides of DAC Sr3 after the laser heating (the laser wavelength was 532 nm). There is a very strong signal corresponding to the $H_2$ vibron. (b) Raman spectra of DAC Sr3 during decompression from 48 to 0 GPa. Comparison of the Raman spectra from the sample, AB, and gasket shows that the peaks at 3635 and 735 $cm^{-1}$ belong to lower Sr hydrides — the products of decomposition. Hydrogen was expected to be one of the decomposition products as the pressure decreased; however, no Raman signals of hydrogen were detected. (c) The sample after a pressure drop to 48 GPa (upper panel). Expansion of the loaded material during decompression (middle and bottom panels). (d) Raman spectra of strontium hydrides synthesized from $SrH_2$ loaded with $H_2$ in DAC S1 before and after the laser heating (the laser wavelength was 532 nm). Spectra at 0–131 GPa recorded before the laser heating show the presence of the $SrH_2$ and $H_2$ peaks, in accordance with the previous study.[30] After the double laser heating at 131–133 GPa, the sample became opaque without an obvious volume change, the $SrH_2$ peaks disappeared, and no additional Raman signals were detected except a signal at ~4140–4150 $cm^{-1}$. This peak can be due to both excess hydrogen not reacted with Sr and to the resulting molecular strontium superhydride.

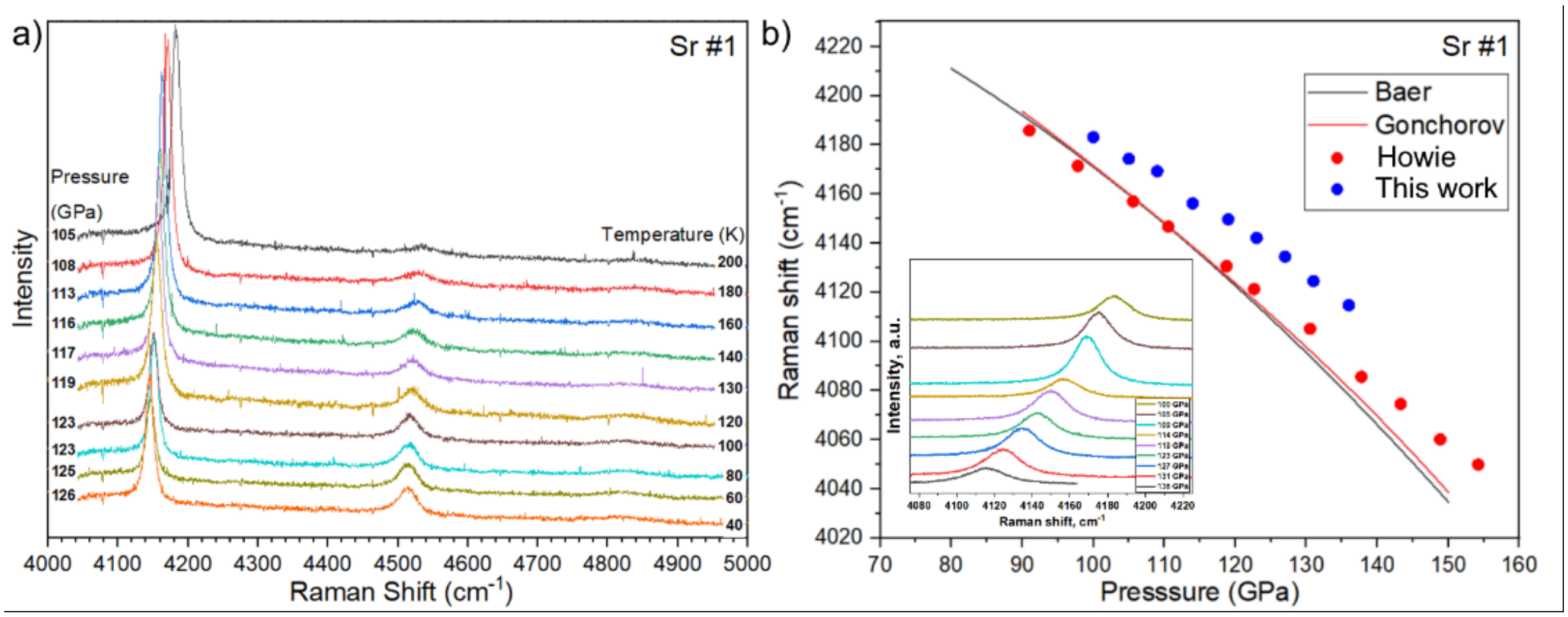

**Figure S27.** (a) Raman spectra of strontium hydrides synthesized using $SrH_2$/AB in DAC Sr1 at different temperatures (40–200 K) and pressures (105–126 GPa, the laser wavelength was 532 nm). The XRD pattern (Figure S31) shows *P*1-$SrH_{22}$ as the main component of the sample. (b) Pressure dependence of the hydrogen signal of the sample compared with the literature data.[31]

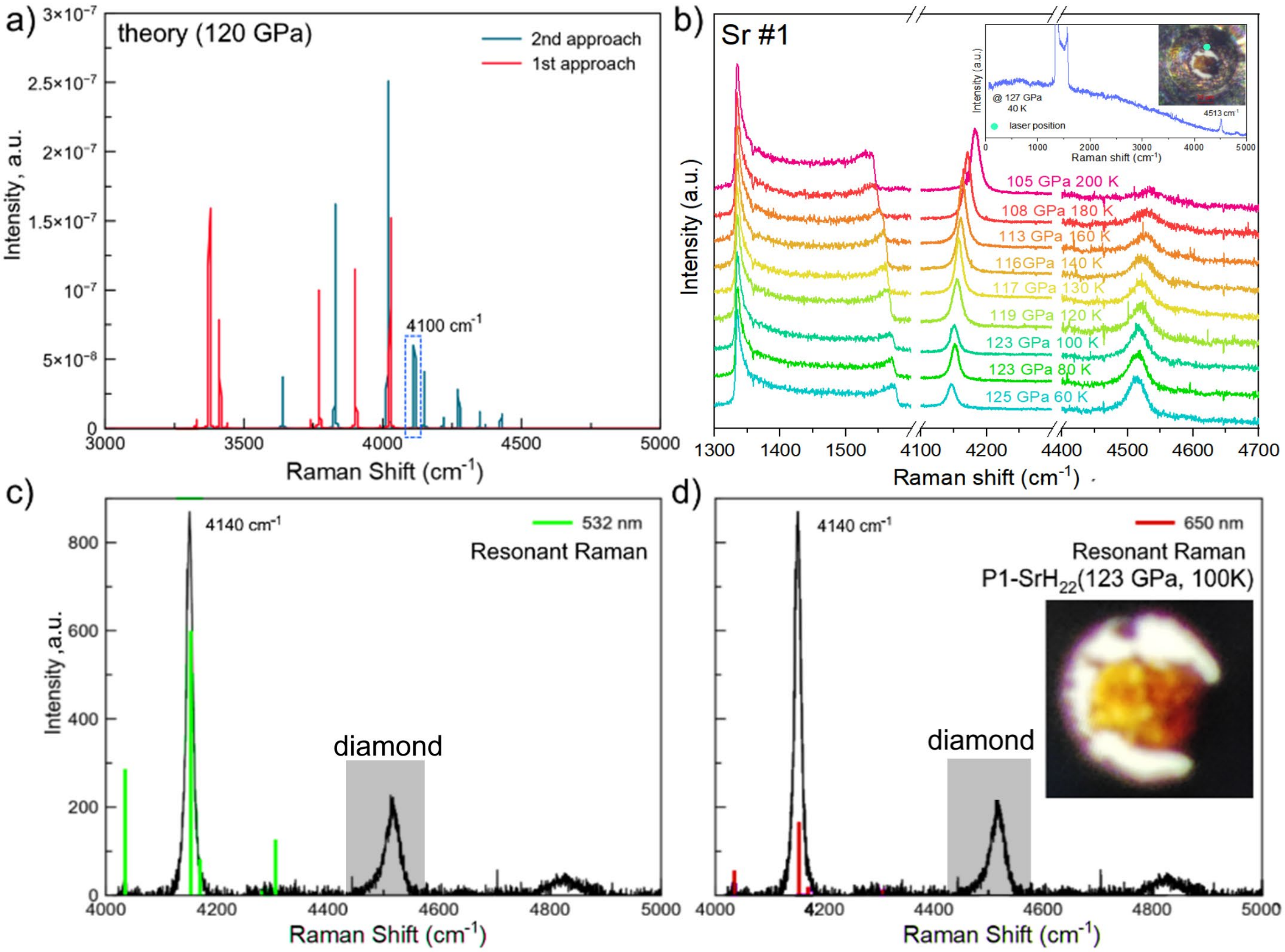

**Figure S28.** Calculated and experimental Raman spectra of *P*1-$SrH_{22}$ (DAC Sr1) at 120 and 123 GPa. (a) Nonresonant Raman spectra of $SrH_{22}$ calculated within the LDA NC (1st approach) and PBE NC (2nd approach) functionals. (b) Experimental Raman spectra of strontium hydrides synthesized from $SrH_2$/AB in DAC Sr1 at different temperatures (60–200 K) and pressures (105–125 GPa, the laser wavelength was 532 nm). Inset: Raman shift measured near the gasket (at the point marked by a bright green dot in the picture) shows that the signal at ~4513 $cm^{-1}$ does not belong to the sample. (c, d) Comparison of the experimental and calculated resonant Raman spectra for the excitation wavelengths of (c) 532 nm and (d) 650 nm. Inset: photo of the $SrH_{22}$ sample in transmitted light.

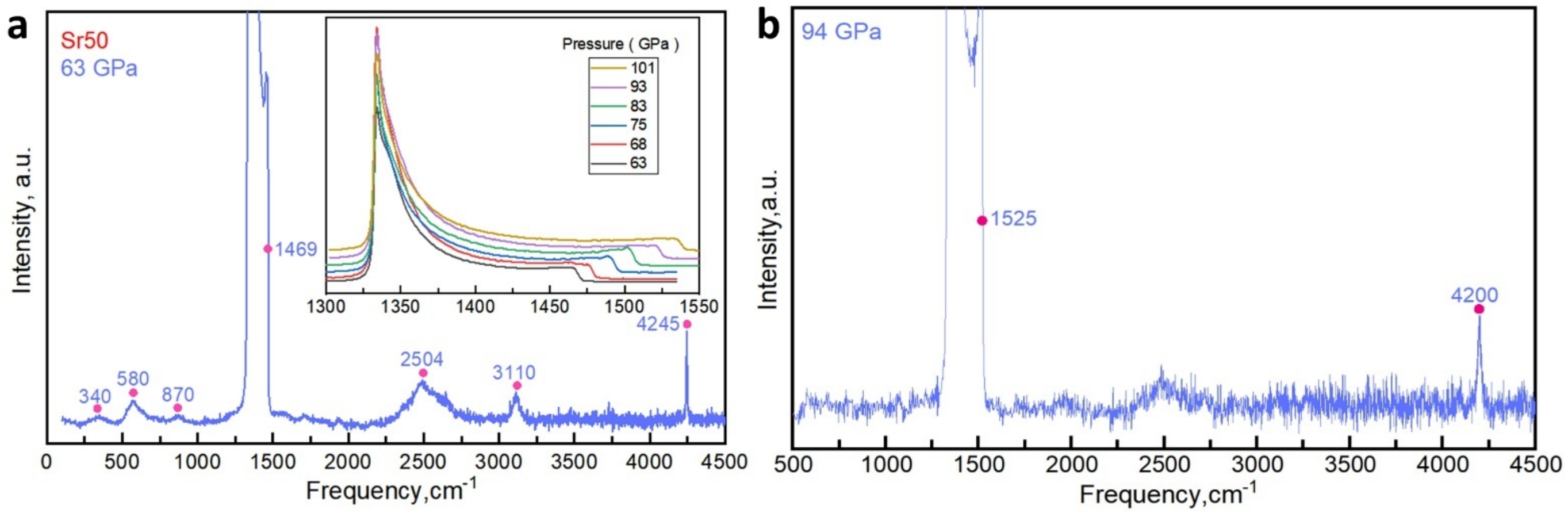

**Figure S29.** Raman spectra of strontium hydrides synthesized after the laser heating (the laser wavelength was 532 nm). In (a) DAC Sr50 at 63 GPa, the XRD analysis shows that the sample consists mainly of c-$SrH_9$ and $Sr_2H_3$. Besides the hydrogen vibron at 4245 $cm^{-1}$, there are peaks at 340 $cm^{-1}$, 580 $cm^{-1}$, 870 $cm^{-1}$, and 3110 $cm^{-1}$. The inset shows the Raman shift of the edge of the diamond during compression from 63 to 101 GPa. (b) Raman spectra of strontium hydrides synthesized after the laser heating in DAC Sr90 at 94 GPa.

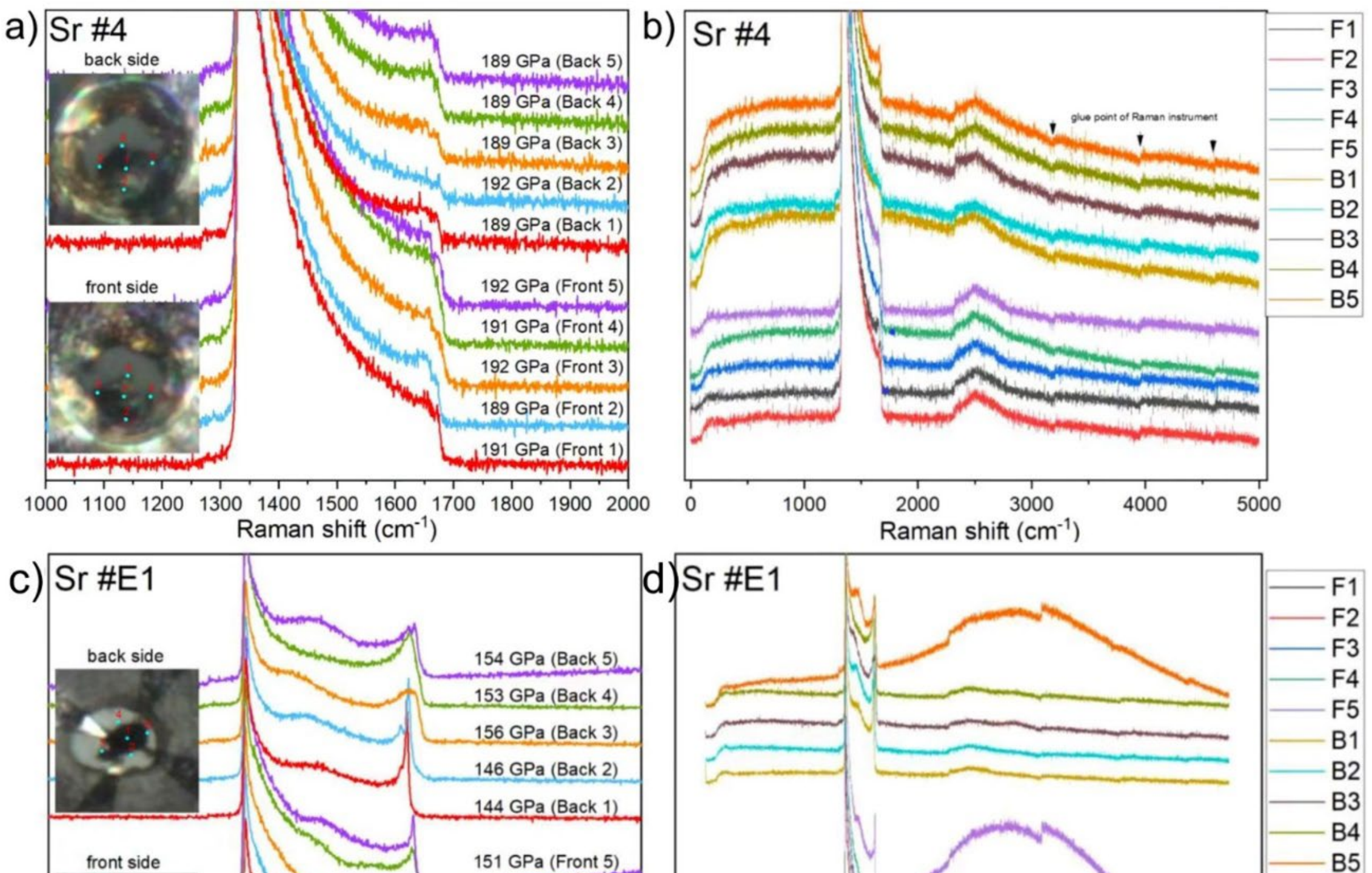

**Figure S30.** Raman spectra (633 nm excitation laser) of strontium hydrides in DAC Sr4 at 190 GPa (a, b) and in the electrical DAC E1 at ~148 GPa (c, d). (a) Pressure gradient in the sample area. (b) Full Raman spectra from the back (B) and front (F) sides of the DAC. (c) Pressure gradient in the sample area, ±6 GPa. (d) Full Raman spectra from the back (B) and front (F) sides of the DAC.

## *Optical properties*

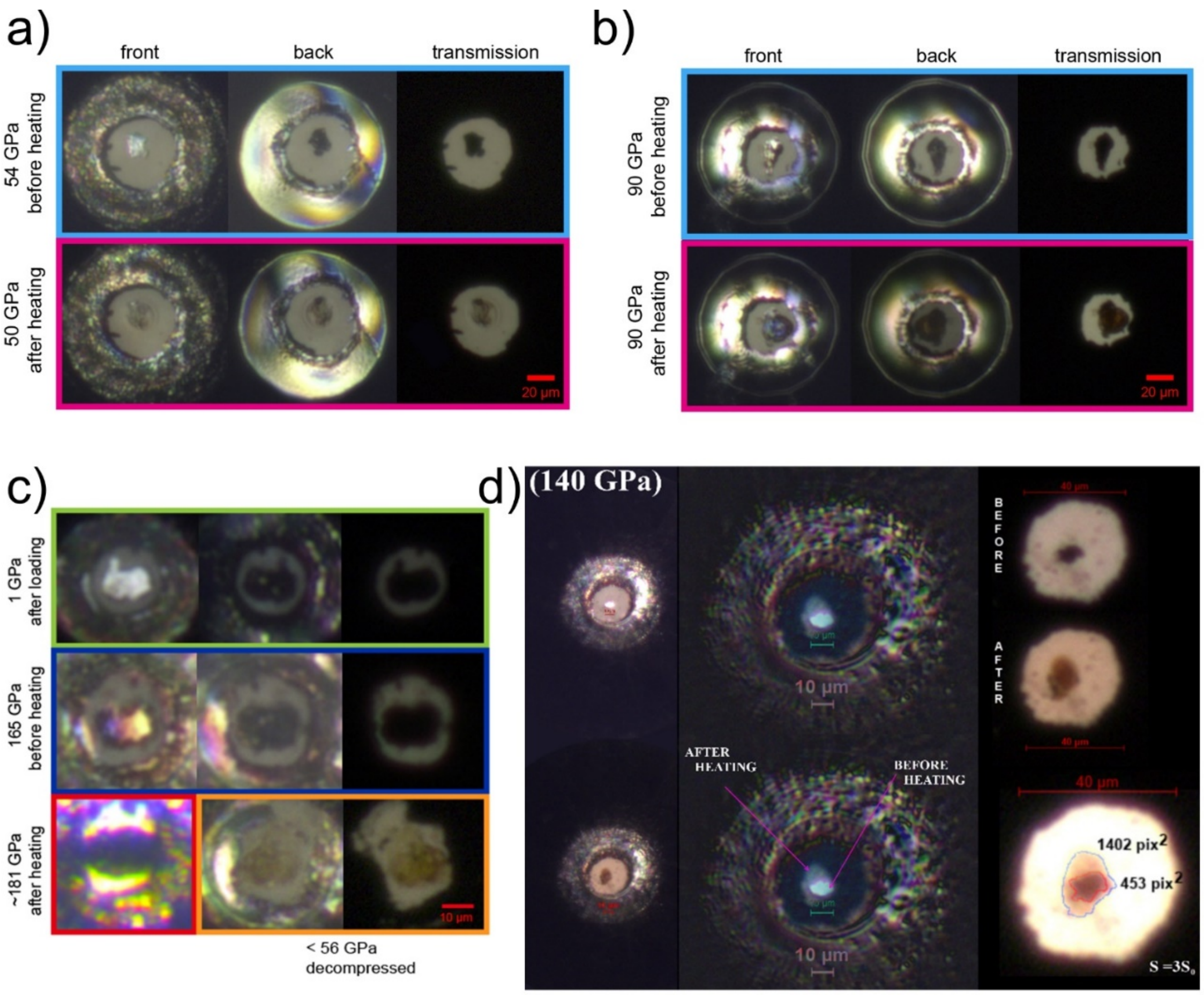

**Figure S31.** Photographs of samples (Sr/AB) before/after laser heating observed with reflected light and transmitted light. (a) The sample in DAC Sr50. (b) The sample in DAC Sr90. (c) The sample in DAC Sr165. Below 56 GPa the sample becomes translucent, which speaks in favor of its semiconducting properties. (d) The sample in DAC W2. A significant increase in the volume of the particle after the laser heating and a change in its color and transparency were observed.

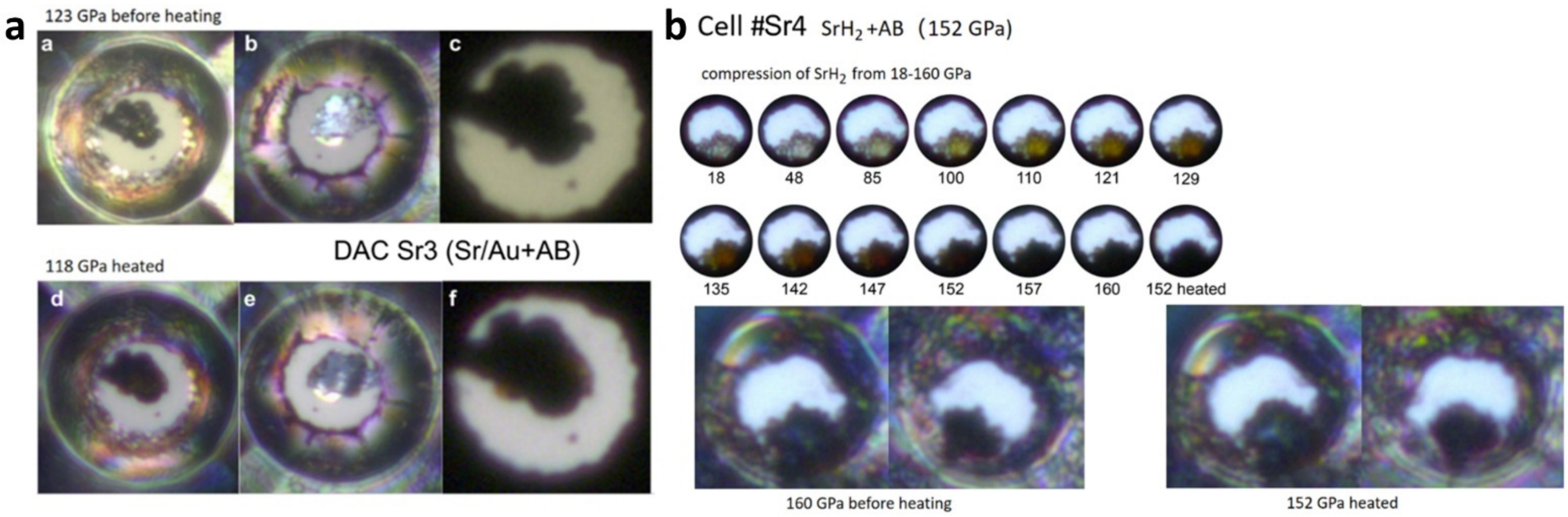

**Figure S32.** Sample photos in reflected and transmitted light in cells before and after the laser heating. (a) DAC Sr4 was loaded with $SrH_2$/AB at 152 GPa. According to ab initio calculations (PBE GGA), the bandgap in $P6_3/mmc$-$SrH_2$ narrows with increasing pressure: 3.12 eV at 0 GPa, 1.2 eV at 100 GPa, and 0.4 eV at 150 GPa. The observed change in the color and transparency of the sample is in qualitative agreement with the calculation results. A better quantitative agreement can be achieved in calculations using the TB09 functional. No additional Raman signals other than that of diamond were detected in the spectra of the sample. (b) The sample in DAC Sr3, loaded with Sr/Au/AB, before and after the laser heating, photographed in reflected and transmitted light.

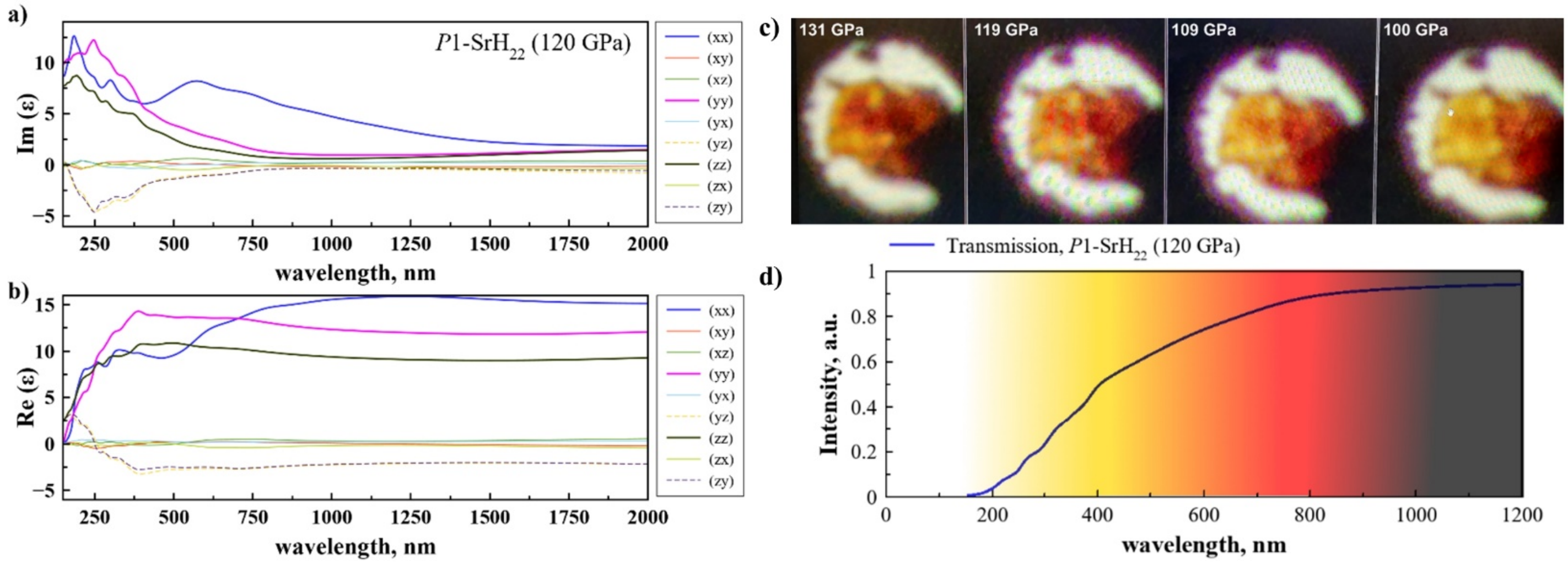

**Figure S33.** (a) Imaginary and (b) real parts of the dielectric function of strontium superhydride $P1$-$SrH_{22}$ calculated at 120 GPa. (c) A sample of $SrH_{22}$ in DAC Sr1 at various pressures (100–131 GPa). As the pressure increases, the sample becomes darker, which corresponds to the closure of the bandgap in this semiconductor. (d) Calculated transmission spectrum of the $SrH_{22}$ at 120 GPa. The color of the sample in DAC Sr1 observed in transmitted light corresponds to the calculated values.

## *Impedance spectroscopy*

**Table S26.** Parameters of the equivalent electrical circuit used to approximate the experimental behavior of the sample in DAC E1 at different pressures and a fixed temperature of 300 K. *C*1 and *R*1 correspond to the first semicircle, *R*2 and CPE1 — to the second half-ellipse and low-frequency tail.

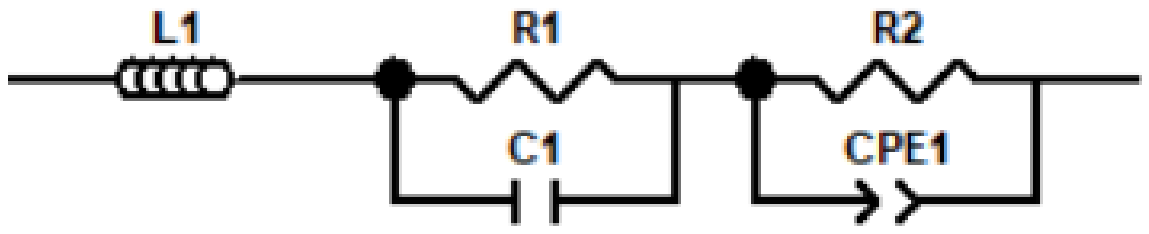

| Pressure, GPa | Temperature, K | Inductance, (*L*1), µH | *R*1, MΩ | *C*1, pF | *R*2, MΩ | CPE-T (Q), $10^{-6}$ | CPE-P (*n*) | $\chi^2$ |
|---|---|---|---|---|---|---|---|---|
| 147 | 300 | 86.373 | 0.344 | 2.13 | 1.081 | 0.910 | 0.477 | 0.04 |
| 144 | 300 | 73.199 | 0.604 | 2.47 | 3.1105 | 0.607 | 0.482 | 0.035 |
| 139 | 300 | 90.65 | 1.239 | 2.09 | 9.3872 | 0.344 | 0.500 | 0.046 |
| 136 | 300 | 83.881 | 1.669 | 2.24 | 11.665 | 0.255 | 0.508 | 0.046 |
| 130 | 300 | 67.121 | 3.120 | 2.73 | 13.091 | 0.108 | 0.553 | 0.04 |
| 126 | 300 | 284.16 | 3.244 | 0.80 | ∞ | 0.140 | 0.413 | 0.13 |

**Table S27**. Parameters of the equivalent electrical circuit used to approximate the experimental behavior of the sample in DAC E1 at different temperatures and a fixed pressure of 150 GPa. *C*1 and *R*1 correspond to the first semicircle, *R*2 and CPE1 — to the second half-ellipse and low-frequency tail.

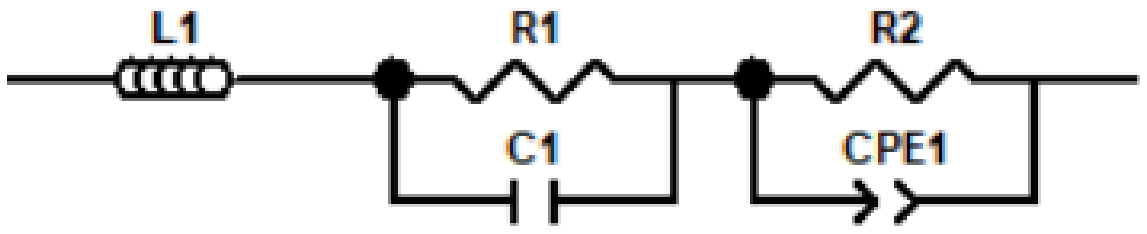

| Pressure, GPa | Temperature, K | Inductance, (*L*1), µH | *R*1, kΩ | *C*1, pF | *R*2, kΩ | CPE-T (Q), $10^{-6}$ | CPE-P (*n*) | $\chi^2$ |
|---|---|---|---|---|---|---|---|---|
| 150 | 300 | 48.43 | 251.2 | 3.385 | 672 | 1.174 | 0.489 | 0.03 |
| 150 | 320 | 40.53 | 117.9 | 3.827 | 411 | 1.889 | 0.470 | 0.024 |
| 150 | 340 | 39.42 | 67.82 | 3.997 | 276 | 2.354 | 0.470 | 0.024 |
| 150 | 360 | 38.52 | 42.27 | 4.005 | 179 | 2.699 | 0.486 | 0.027 |
| 150 | 380 | 35.81 | 27.58 | 4.072 | 120 | 3.134 | 0.501 | 0.028 |
| 150 | 400 | 33.67 | 20.09 | 4.056 | 86.5 | 3.510 | 0.519 | 0.03 |
| 150 | 420 | 28.03 | 14.24 | 4.181 | 58.8 | 3.953 | 0.536 | 0.03 |

**Table S28.** Parameters of the equivalent electrical circuit used to approximate the experimental behavior of the sample in DAC E1 at different temperatures and a fixed pressure of 126 GPa. *C*1 and *R*1 correspond to the first semicircle, *R*2 and CPE1 — to the second half-ellipse and low-frequency tail.

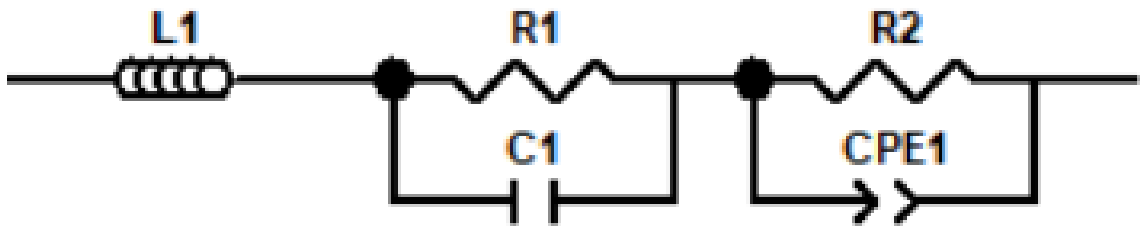

| Pressure, GPa | Temperature, K | Inductance, (*L*1), µH | *R*1, kΩ | *C*1, pF | *R*2, MΩ | CPE-T (Q), $10^{-6}$ | CPE-P (*n*) | $\chi^2$ |
|---|---|---|---|---|---|---|---|---|
| 126 | 300 | 415.52 | 3263 | 0.755 | 12.95 | 0.083 | 0.526 | 0.176 |
| 126 | 320 | 38.0 | 1242 | 4.03 | 19.65 | 0234 | 0.412 | 0.084 |
| 126 | 340 | 233 | 604.8 | 0.917 | ∞ | 0.449 | 0.416 | 0.1 |
| 126 | 360 | 210 | 308.6 | 0.979 | ∞ | 0.686 | 0.421 | 0.1 |
| 126 | 380 | 191 | 175.4 | 1.02 | ∞ | 1.00 | 0.436 | 0.1 |
| 126 | 400 | 150 | 112.1 | 1.236 | ∞ | 1.278 | 0.445 | 0.085 |
| 126 | 420 | 127 | 80.18 | 1.36 | ∞ | 1.676 | 0.455 | 0.083 |
| 126 | 440 | 120 | 65.95 | 1.33 | ∞ | 2.38 | 0.465 | 0.083 |

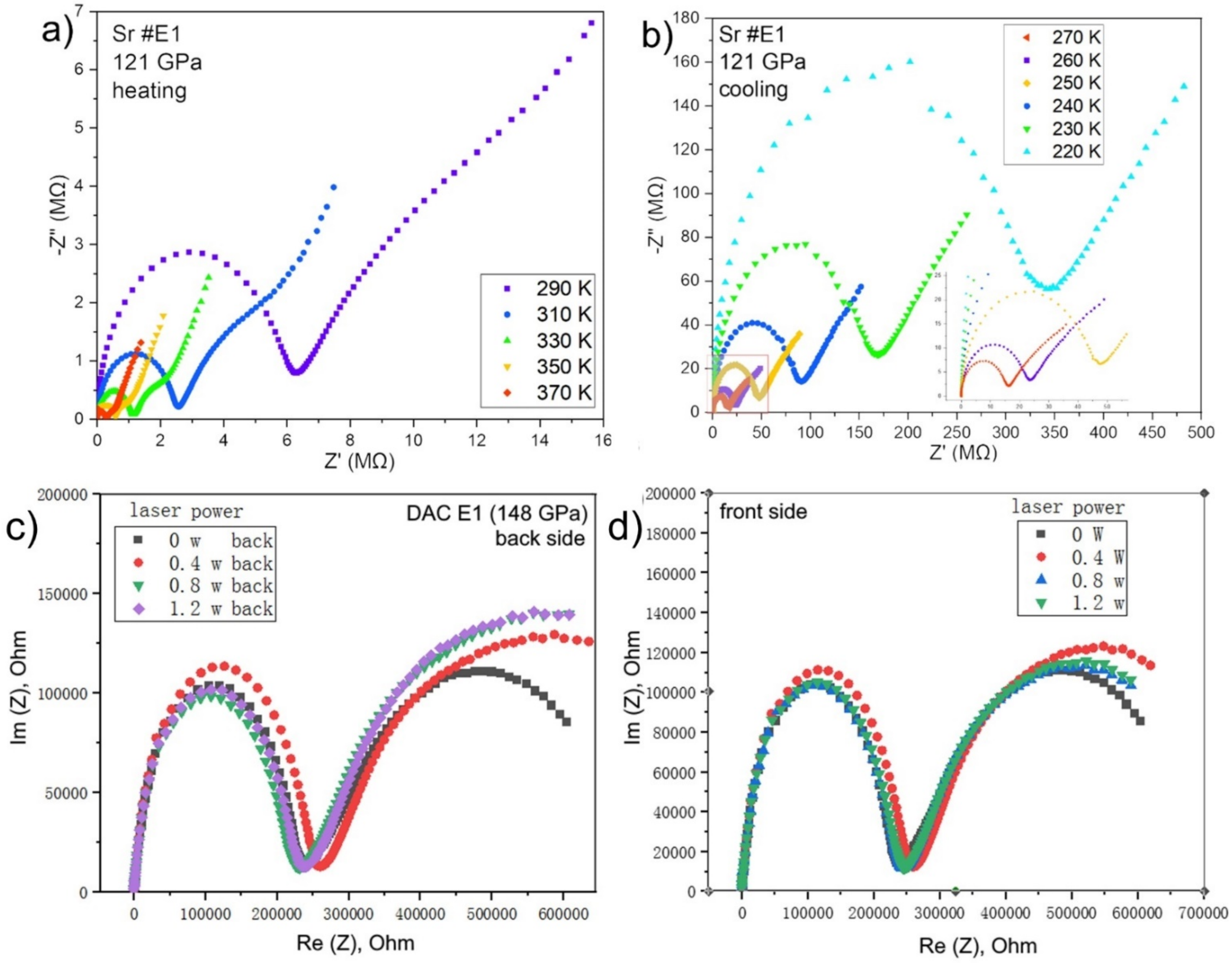

**Figure S34.** Impedance spectroscopy (Nyquist diagram) of the $P1$-$SrH_6$ sample (DAC E1) in the frequency range of 0.1–$10^7$ Hz at 121 GPa in the (a) heating and (b) cooling cycles. Impedance spectroscopy (Nyquist diagram) of the $P1$-$SrH_6$ sample in DAC E1 with (red, green, violet) and without (black) illumination by a 532 nm laser (power 0–1.2 W) at 300 K. Illumination from the (c) back and (d) front sides of the DAC leads to a significant increase in the size of the half-ellipse.

The illumination of the sample with a green laser (532 nm, power 0–1.2 W) practically does not change the parameters of the first semicircle in the hodograph but increases the size of the subsequent half-ellipse corresponding to the electrode processes. This suggests that the sample does not heat up during irradiation, whereas the resistance $R2$ and pseudocapacitance CPE-P increase. Irradiation significantly reduces the electrical resistance of the bulk $SrH_6$ phase because of photoconductivity. However, this change (~$10^2$–$10^3$ Ω) cannot be seen probably because the contribution of bulk resistance to the total resistance of the sample is very small.

**Table S29.** Parameters of the equivalent electrical circuit used to approximate the experimental behavior of the sample in DAC E1 at different powers of the light source (532 nm laser) and fixed pressure (148 GPa) and temperature (300 K). $C1$ and $R1$ correspond to the first semicircle, $R2$ and CPE1 — to the second half-ellipse and low-frequency tail.

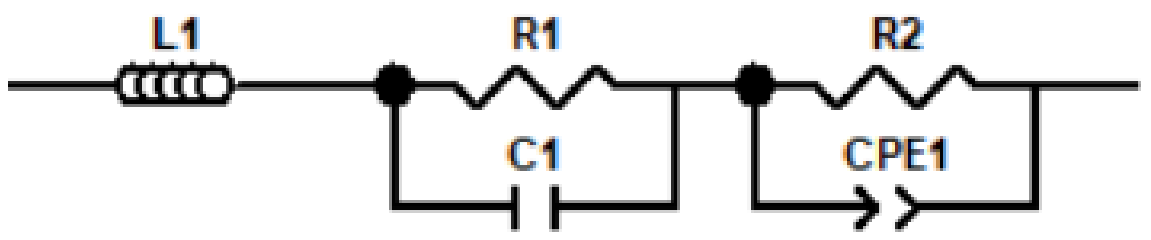

| Pressure, GPa | Laser power (532 nm), W | Inductance, ($L1$), μH | $R1$, kΩ | $C1$, pF | $R2$, kΩ | CPE-T (Q), $10^{-6}$ | CPE-P ($n$) | $\chi^2$ |
|---|---|---|---|---|---|---|---|---|
| 148 | 0 | 5.931E-5 | 225.8 | 2.865 | 537.0 | 1.14 | 0.463 | 0.027 |
| 148 | 0.8 | 5.259E-5 | 216.1 | 3.145 | 722.9 | 1.49 | 0.446 | 0.025 |
| 148 | 1.2 | 5.280 E-5 | 223.1 | 3.136 | 735.5 | 1.54 | 0.449 | 0.025 |

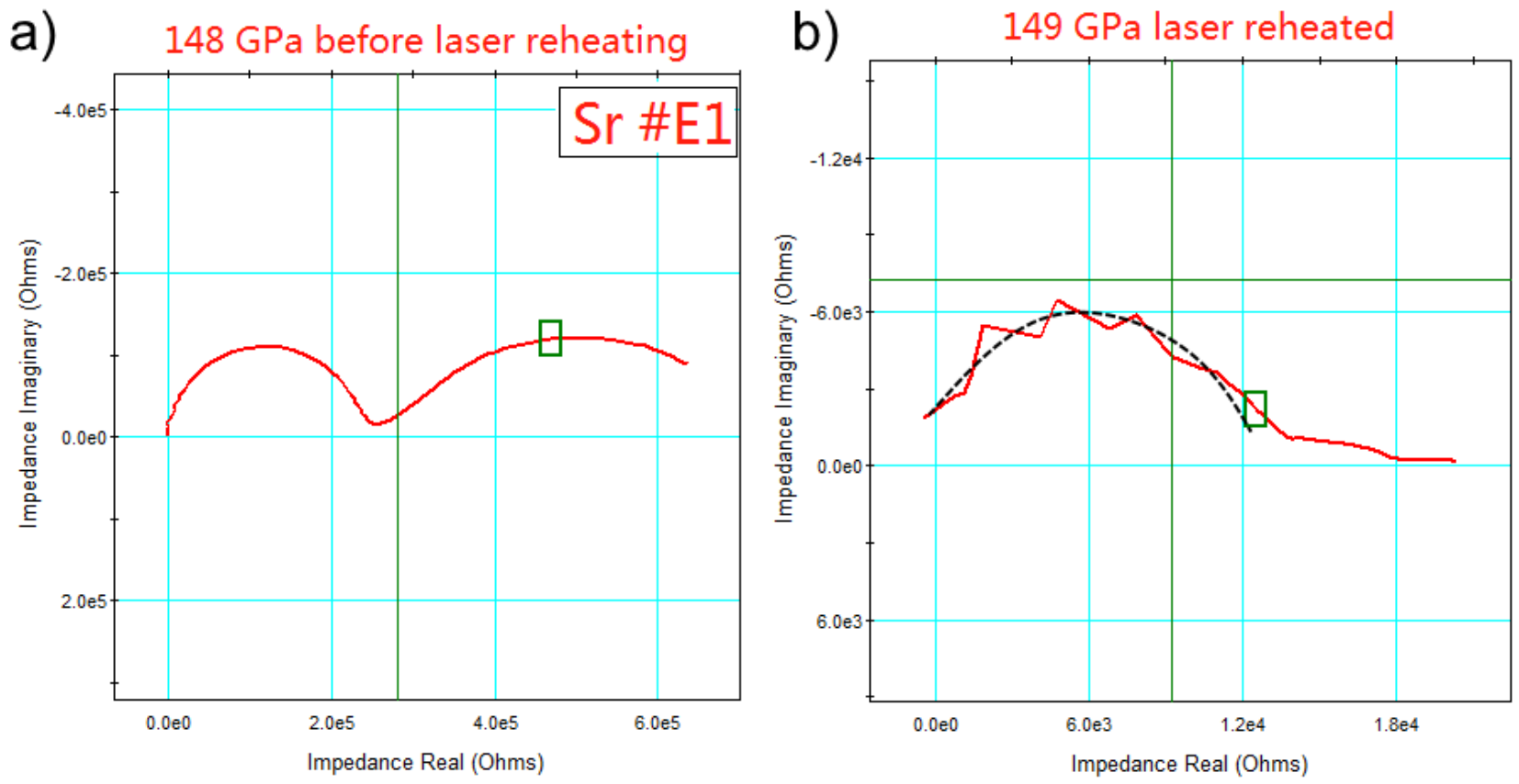

**Figure S42.** Impedance spectroscopy (Nyquist diagram) of the $P1$-$SrH_6$ sample in DAC E1 (a) before and (b) after the laser reheating. Strong laser heating at 148–149 GPa leads to transformation of the initial Sr hydride into a compound having about 10 times lower active resistance and no second half-ellipse in the impedance diagram.

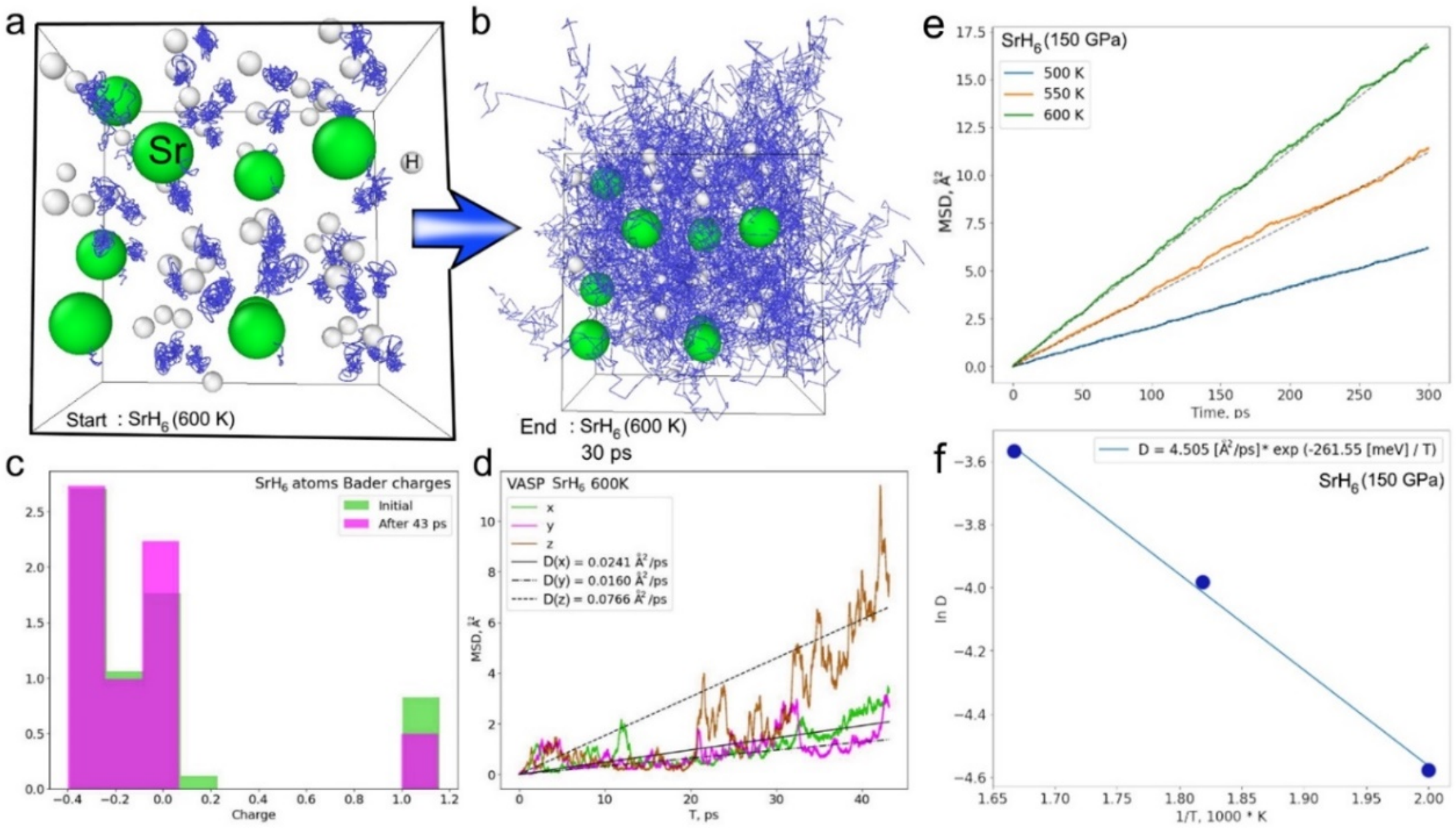

**Figure S35.** Thermal diffusion in $P1$-$SrH_6$ at 150 GPa and 600 K in an external electric field $E(z)$ of $10^6$ V/m (VASP, PBE GGA). (a, b) The structure of $SrH_6$ (a) before and (b) after the simulation. (c) Comparison histogram of the Bader charges of the Sr and H atoms before and after the molecular dynamics simulation shows that the charges do not change significantly. (d) Mean-square displacement (MSD) of the hydrogen atoms in different directions ($x$, $y$, $z$), averaged over all hydrogen atoms, calculated considering the shift of the center of gravity of the unit cell. Thermal diffusion in (e, f) $P1$-$SrH_6$ at 150 GPa in an external electric field $E(z)$ of $10^4$ V/m (MLIP, LAMMPS). (e) Diffusion coefficients obtained at 500, 550, and 600 K using the Einstein equation (mean-square displacement (MSD)-time) from the results of the molecular modeling with a duration of 300 ps. (f) Temperature dependence of the diffusion coefficients interpolated using the Arrhenius formula.

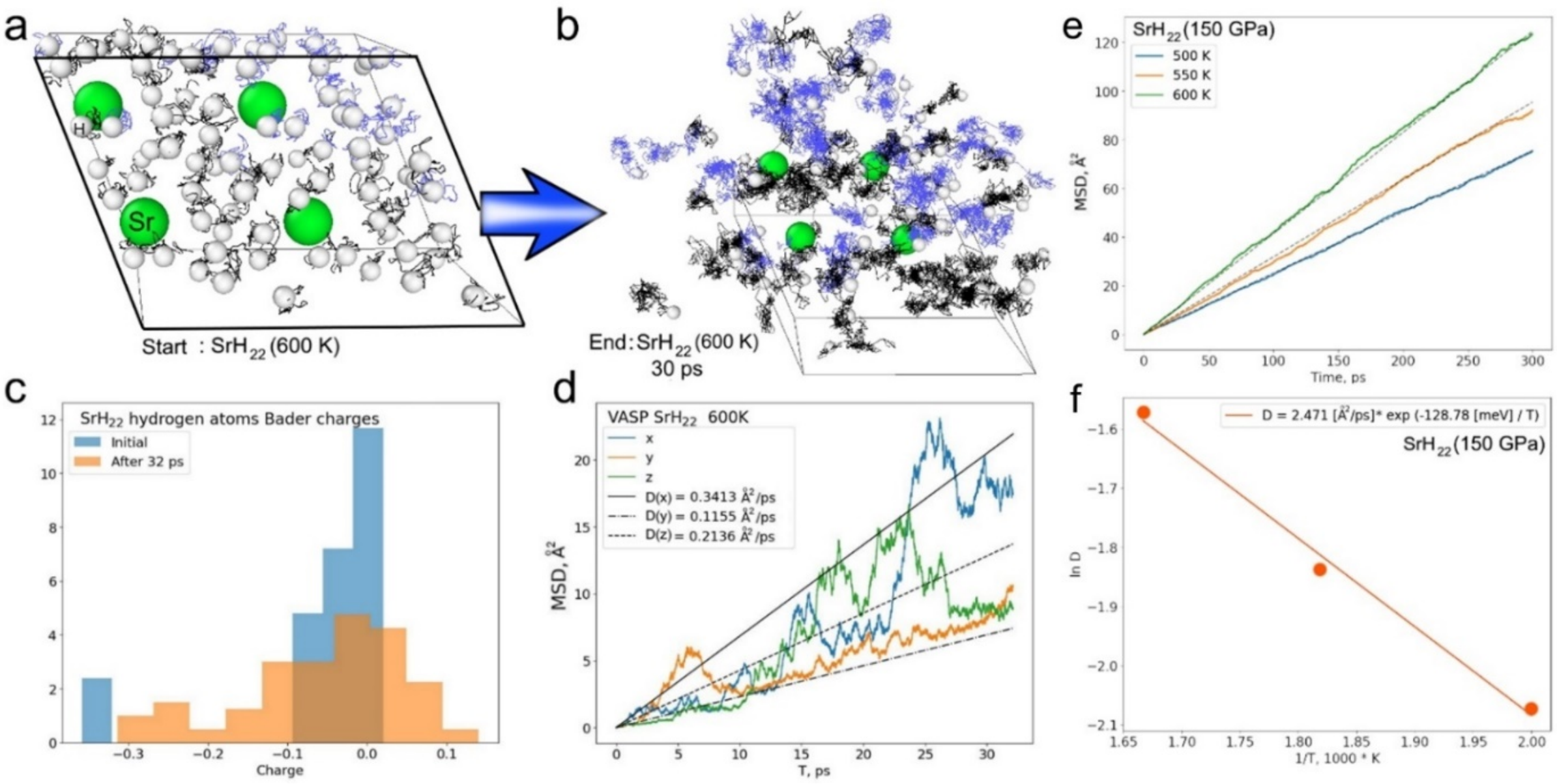

**Figure S36.** Thermal diffusion in $P1$-$SrH_{22}$ at 150 GPa and 600 K in an external electric field $E(z)$ of $10^6$ V/m (VASP, PBE GGA). (a, b) The structure of $SrH_{22}$ (a) before and (b) after the simulation. (c) Comparison histogram of the Bader charges of the H atoms before and after the molecular dynamics simulation shows that the charges do not change significantly. (d) Mean-square displacement (MSD) of the hydrogen atoms in different directions ($x$, $y$, $z$), averaged over all hydrogen atoms, calculated considering the shift of the center of gravity of the unit cell. Thermal diffusion in (e, f) $SrH_{22}$ at 150 GPa in an external electric field $E(z)$ of $10^4$ V/m (MLIP, LAMMPS). (e) Diffusion coefficients obtained at 500, 550, and 600 K using the Einstein equation (mean-square displacement (MSD)-time) from the results of the molecular modeling with a duration of 300 ps. (f) Temperature dependence of the diffusion coefficients interpolated using the Arrhenius formula.

# References for Supporting Information